\documentclass{cambridge6A}

\usepackage{natbib}
\usepackage{rotating}
\usepackage{floatpag}
\rotfloatpagestyle{empty}

\usepackage{amssymb}
\usepackage{amsmath}
\usepackage{amsthm}
\usepackage{mathrsfs}
\usepackage{graphicx}
\usepackage{multind}
\usepackage[dvipsnames]{xcolor}
\usepackage{hyperref}

\makeindex{authors}
\makeindex{subject}

\newcommand{\ie}{\emph{i.e.}}

\newcommand{\avg}[1]{\langle #1\rangle}

\theoremstyle{plain}

\theoremstyle{definition}

\theoremstyle{remark}

\begin{document}
\title[]{Reconstructing networks}

\author{Giulio Cimini, Rossana Mastrandrea, Tiziano Squartini\\[3\baselineskip]}

\frontmatter
\maketitle

\begin{abstract}
Complex networks datasets often come with the problem of missing information: interactions data that have not been measured or discovered, may be affected by errors, or are simply hidden because of privacy issues. This Element provides an overview of the ideas, methods and techniques to deal with this problem and that together define the field of network reconstruction. Given the extent of the subject, we shall focus on the inference methods rooted in statistical physics and information theory. The discussion will be organized according to the different scales of the reconstruction task, that is, whether the goal is to reconstruct the macroscopic structure of the network, to infer its mesoscale properties, or to predict the individual microscopic connections.\\

\noindent{\bf keywords}: network reconstruction, maximum-entropy inference, exponential random graphs, mesoscale structures, link prediction.

\null\vfill
\noindent {\bf Authors contributions.} GC wrote chapter 4. RM wrote chapter 3. TS wrote chapter 2. All authors wrote chapters 1 and 5, reviewed and approved the manuscript.
\end{abstract}

\tableofcontents

\mainmatter
\chapter{Introduction}
\label{intro}

\paragraph{Missing information: a general problem.} 
Laying at the heart of the scientific method, data analysis is about using data to validate models, acquire useful information and support decision-making. When the data is incomplete, so are the conclusions that can be drawn from it. Unfortunately, the problem of missing data is a common occurrence, both in science and in many practical situations. Even in our era of Big Data, data can be incomplete for a variety of reasons -- such as sheer lack of information, annotation errors, collection problems and privacy concerns. The problem is even more complicated when the data has a non-homogeneous structure, because it describes a system characterised by an irregular pattern of relations or connections, such as a {\em complex network}. To be more concrete, we draw a few examples. 

Consider a system biologist looking for the proteins in an organism that have a physical or functional pair interaction. The scientist would need to pick two candidate proteins and set up an experiment to determine whether they interact or not. Blindly considering all possible pairs is infeasible because experiments can be quite costly, and this is why interactions within the proteome are largely unknown. 
Therefore the need to pick good candidates for the experiment using the prior information on the protein interactions that have already been discovered \citep{redner2008teasing,guimera2009missing}.
As another example, consider a social scientist trying to extract a given social network. Two types of problems can arise in this context: i) the available data report only aggregated statistics on individuals (e.g. the total number of contacts) without disclosing sensible information such as the identities of friends; and ii) the network is extremely large to be explored by crawling algorithms, whence the need to consider subsamples that are representative \citep{leskovec2006sampling,liben-nowell2007link}.

Farther from the classical scientific domain, consider an entrepreneur running an e-commerce platform that sells books. In order to improve sales, it would be a good idea to set up a {\em recommender system} that shows customers the books they may be interested in buying. The algorithm works well if it is able to predict customer tastes (i.e., possible future purchases) using their buying records \citep{lu2012recommender}.
As a final example we take a regulator working in a central bank. Her job is to run stress tests to determine the ability of a given bank to deal with a crisis event. Since in a financial system losses and distress propagate through the various financial exposures banks have with each other and with other financial institutions, to accomplish her task properly the regulator should know the detailed network of exposures (who is exposed with whom, and to what extent). Unfortunately this information is confidential, and the regulator must resort to publicly disclosed information, i.e., the balance sheet of the banks containing only their aggregate exposures \citep{squartini2018reconstruction,anand2017missing}. 

The common theme of all these situations is that the system at hand is a {\em network}, namely a system that independently of its nature can be modelled by a complex pattern of interactions (the {\em links}) between its constituents (the {\em nodes}). When the network is known only partially, the task is to reconstruct the unknown part. 
The set of techniques that make up the field of \emph{network reconstruction} precisely aim at inferring the (unknown) structure of a network, making an optimal use of the partial knowledge about its properties \citep{squartini2018reconstruction,lu2011link}.

\paragraph{Approaching network reconstruction.} Generally speaking, the fundamental assumption grounding network reconstruction is \emph{statistical homogeneity}: the empirically observed network structures should be representative of the statistical properties concerning the network as a whole. The validity of such an assumption is the necessary condition for a reconstruction algorithm to work. Clearly, this approach limits the accuracy that can be achieved when reconstructing strongly heterogeneous structures. However, it prevents possible inference biases introduced by arbitrary assumptions not supported by the available information.

In order to deal with the problem of missing information, many different approaches have been attempted so far. Among the most successful ones there are those defined within the framework of information theory \citep{cover2006elements}. In a nutshell, these methods prescribe to 1) consider all configurations that are compatible with the available information (an \emph{ensemble}, in the jargon of statistical mechanics) and 2) assign a degree of plausibility to each of them. As it has been proven elsewhere, the least-biased way to do this rests upon the renowned \emph{entropy maximization} prescription \citep{cover2006elements,jaynes1957information}. Notably, this approach naturally leads to the Exponential Random Graphs (ERG) formalism \citep{park2004statistical,cimini2019statistical}. The importance of ERG models within the network reconstruction field is motivated by three desirable features they posses: \emph{analytical character}, \emph{general applicability} and \emph{versatility}. This is the reason why a large portion of the book is devoted to discuss the applications of such a powerful formalism.

\paragraph{A quick overview of the book.} The discussion of the network reconstruction problem is divided into three sections, according to the scale of the reconstruction task: \emph{macroscale}, \emph{mesoscale} and \emph{microscale}. This distinction is intended to provide a wide overview of the reconstruction techniques while presenting detailed results in some specific contexts. 

The chapter \textbf{Network reconstruction at the macroscale} focuses on the inference of global features of the network, such as \emph{assortativity} and \emph{hierarchical} patterns. In this case, reconstruction techniques are typically informed on node-specific properties (and possibly on trends that characterize the network as a whole), without considering any specific topological detail (i.e., the occurrence of a particular link). 
After describing the general ideas and results, particularly in the context of ERG, we will delve into the estimation of \emph{systemic risk} in a partially-accessible network. As already mentioned, this exercises is particularly relevant for financial networks, where the knowledge of the interconnections between financial institutions is required to run stress tests and assess the stability of the system.

The chapter \textbf{Network reconstruction at the mesoscale} instead deals with the detection and reproduction of network patterns like \emph{modular}, \emph{core-periphery} and \emph{bipartite} structures. The topic is of great interest for disciplines as diverse as epidemiology, finance, biology and sociology as it ultimately boils down to identify some sort of \emph{structural} or \emph{functional} similarity between nodes. The presence of mesoscale patterns then affects a wide range of dynamical processes \emph{on} networks (e.g. information and epidemic spreading, fake-news diffusion, etc.) whence the need to properly account for them. However, a fundamental point is to understand to what extent \emph{accessible} node properties are informative about the presence of mesoscale structures.

Finally, the chapter \textbf{Network reconstruction at the microscale} is devoted to the topic of \emph{single link} inference, a problem that is better known as \emph{link prediction}. Differently from the network reconstruction problem at the macro- and at the meso-scale, when considering the micro-scale many details of the network are known (typically a large number of connections) and the goal is to predict those links that are either not known---because the source data used to define the network is incomplete, or simply do not exist yet. We will review the link prediction techniques that build on the partial knowledge on the network, and not on any additional information like nodes features.

\chapter{Network reconstruction at the macroscale}
\label{chap2}

A network is defined as a set of constituent elements (the \emph{nodes}) and a set of connections (the \emph{links}) among them. Mathematically speaking, a network is a graph with non-trivial topological features. In practice, networks are the natural way to represent and model large class of very diverse systems, and thus we can speak of technological and information networks, social and economic networks as well as biological and brain networks.

\section{Macroscale properties: an overview}\label{sec1}

\subsection*{Binary properties}

Let us start by introducing the basic notation and the macroscale properties of \emph{binary}, \emph{undirected} (\emph{directed}) \emph{graphs} with $N$ nodes. Graphs of this kind are completely specified by a symmetric (generally asymmetric) $N\times N$ \emph{adjacency matrix} $\mathbf{A}$, whose generic entry is either $a_{ij}=0$ or $a_{ij}=1$, respectively indicating the absence or the presence of a connection between nodes $i$ and $j$ (from $i$ to $j$). As usual \emph{self-loops}, namely links starting and ending at the same node, will be ignored (in formulas, $a_{ii}=0,\:\forall\:i$). The description above also applies to \emph{bipartite} graphs, where nodes form two disjoint sets that are not connected internally.

\paragraph*{Connectance.} The simplest macroscopic characterization of a network is the connectance, or \emph{link density}, defined as

\begin{equation}
\rho(\mathbf{A})=\frac{2L}{N(N-1)}
\end{equation}
where $L(\mathbf{A})=\sum_{i<j}a_{ij}\equiv L$ is the total number of links in the network. Thus $\rho(\mathbf{A})$ is the fraction of node pairs that are connected by a link. 
For directed networks, $\rho(\mathbf{A})=\tfrac{L}{N(N-1)}$ with $L=\sum_{i\neq j}a_{ij}$. 

Notably, real-world networks are usually characterized by a very low density of links, i.e., they are sparse. 
Reproducing the network connectance is a sort of baseline requirement of any reconstruction method. 
The simplest model satisfying this requirement is the \emph{Erd\"os-Renyi} (ER) random graph \citep{erdos1960evolution,park2004statistical}. According to this model, the probability $p_{ij}$ of a connection between nodes $i$ and $j$ (that is, the average value of the adjacency matrix element $a_{ij}$ in the model, $\langle a_{ij}\rangle_{\text{ER}}$) reads
\begin{equation}
p_{ij}^{\text{ER}} = p = \frac{2L}{N(N-1)}=\rho,\:\forall\:i\neq j
\end{equation}
therefore any two nodes establish a connection with the same probability $p$. 

\paragraph*{Degrees.} The \emph{degree} of a node counts the number of its neighbors, or equivalently the number of its incident connections. In formulas, $k_i(\mathbf{A})=\sum_{j\neq i}a_{ij},\:\forall\:i$. An important and ubiquitous characterization of real-world networks is the \emph{heavy-tailed} shape of the degree distribution, with a few \emph{hub} nodes that are highly-connected ($k_{hub}=O(N)$) and the vast majority of other nodes (of the order $O(N)$) with a small degree. Although the mathematical nature of these heavy-tailed distributions is still debated, often they have been found to be \emph{scale-free}  \citep{caldarelli2007scale,barabasi2009scale}:
\begin{equation}
P(k)\sim k^{-\gamma},\:2<\gamma<3.
\end{equation}
for which the ``typical'' degree is simply missing. 
In any event, the strong heterogeneity of the degree distribution is the basic feature that makes networks different from homogeneous systems and regular lattices. Therefore, any good reconstruction algorithm should be able to reproduce it\footnote{Moreover, preserving the degrees automatically ensures that the link density is preserved.}. Notice that such a requirement rules out the ER model as a potentially good reconstruction model: in fact, although it ensures that the link density is reproduced, it fails in preserving the degree heterogeneity, since the model average $\langle k_i\rangle_{\text{ER}}=\sum_{j\neq i}\langle a_{ij}\rangle_{\text{ER}}=\sum_{j\neq i}p_{ij}^{\text{ER}}=2L/N,\:\forall\:i$. Such an evidence has motivated the definition of the Chung-Lu (CL) model \citep{chung2002connected}, according to which

\begin{equation}
p_{ij}^{\text{CL}}=\frac{k_ik_j}{2L},\:\forall\:i\neq j;
\end{equation}
by definition, then, $\langle k_i\rangle_{\text{CL}}=\sum_{j\neq i}\langle a_{ij}\rangle_{\text{CL}}=\sum_{j\neq i}p_{ij}^{\text{CL}}\simeq k_i,\:\forall\:i$.

In the directed case, there are two kinds of degree: the total number of links out-going from a node (the \emph{out-degree} $k^{out}_i(\mathbf{A})=\sum_{j\ne i}a_{ij},\:\forall\:i$) and the total number of links incoming to a node (the \emph{in-degree} $k^{in}_i(\mathbf{A})=\sum_{j\ne i}a_{ji},\:\forall\:i$). The directed extension of the Chung-Lu model (DCL) reads

\begin{equation}
p_{ij}^{\text{DCL}}=\frac{k_i^{out}k_j^{in}}{L},\:\forall\:i\neq j.
\end{equation}

\paragraph*{Assortativity.} Generally speaking, this term indicates the tendency of nodes to establish connections with other nodes having either similar (\emph{positive assortativity}) or different (\emph{negative assortativity} or \emph{disassortativity}) characteristics. Particularly relevant in the study of complex networks is the assortativity \emph{by degree}. In this case, assortativity can be studied by considering the \emph{average nearest neighbors degree} (ANND), which for generic node $i$ is defined as

\begin{equation}
k^{nn}_i(\mathbf{A})=\frac{\sum_{j\neq i}a_{ij}k_j}{k_i},\:\forall\:i.
\label{eq_bun_knn}
\end{equation}
ANND is a quadratic function of the adjacency matrix, and thus is a \emph{second-oder} network property. 
Plotting $k^{nn}_i$ versus $k_i$ reveals the two-points correlation structure of the network: 
an increasing trend corresponds to an assortative pattern (poorly connected nodes are connected to other poorly connected nodes, highly connected nodes are connected to other highly connected nodes), while a decreasing trend to the opposite disassortative pattern (poorly connected nodes connected to highly connected nodes and viceversa). 
Notice that assortativity is typically observed in social networks (where it is also known with the term \emph{homophily}), whereas, economic and technological networks are usually disassortative \citep{newman2002assortative}.

Assortativity acts as the testbench for the CL model. Since $\langle k^{nn}_i\rangle_{\text{CL}}\simeq\frac{\sum_{j\neq i}p_{ij}^{\text{CL}}k_j}{k_i}=\frac{\sum_{j\neq i}k_j^2}{2L},\:\forall\:i$, in this model $k^{nn}_i$ is weakly dependent on node $i$ - basically, the ANND is the same for all nodes \citep{squartini2011analytical}. As a consequence, the CL model is not capable to reproduce any (dis)assortativity, thus missing one of the characteristic features of real-world networks. The solution lies in the definition of a more refined model, i.e., the \emph{Configuration Model} (CM - see below).

When considering directed networks, the ANND can be generalized in five different ways \citep{squartini2011randomizing}.

\paragraph*{Hierarchy.} Assortativity and ANND accounts for second-order interactions, that is, interactions between nodes along patterns of length two. Third-order interactions (i.e., three-points correlations) are instead typically measured through the \emph{clustering coefficient}, which for any node $i$ is defined as the percentage of pairs of neighbors of $i$ that are also neighbors of each other:

\begin{equation}
c_i(\mathbf{A})=\frac{\sum_{j\neq i}\sum_{k\ne i,j}a_{ij}a_{ik}a_{jk}}{k_{i}(k_{i}-1)},\:\forall\:i;
\label{eq_bun_ck}
\end{equation}
otherwise stated, $c_i$ measures the fraction of potential triangles attached to $i$ (and defined by the product $a_{ij}a_{ik}$) that are actually realized (i.e., closed by the third link $a_{jk}$). A decreasing trend of $c_i$ as a function of $k_i$ indicates that neighbors of highly connected nodes are poorly interconnected, whereas neighbors of poorly connected nodes are highly interconnected. This behavior characterizes a \emph{hierarchical} network, i.e. a network of densely connected subgraphs that are poorly inter-connected. In real-world networks, the a scale-free degree distribution often coexists with a large value of the clustering coefficient \citep{albert2002statistical}.

As for the assortativity, the CL model predicts a value for the clustering coefficient that is only weakly dependent on $i$, thus calling for a more refined model to reproduce empirical patterns of real-world networks.

Generalizations to directed networks also exist for third-order quantities. Beside five different definitions of the clustering coefficient \citep{squartini2011randomizing}, there are thirteen possible patterns involving three nodes and all possible connections between them: these quantities are called \emph{motifs} and, as discussed in chapter \ref{chap3}, have been proven to play a fundamental role in the self-organization of biological, ecological and cellular networks: certain \emph{structures} have been, in fact, suggested to promote specific \emph{functions} \citep{milo2002network}.

\paragraph*{Higher-order patterns.} The presence of higher-order patterns can be inspected using the powers of the network adjacency matrix $\mathbf{A}$. Indeed, the entry indexed by $i$ and $j$ of  $\mathbf{A}^n$ (i.e. the $n$th power of $\mathbf{A}$) counts the number of paths of length $n$ existing between $i$ and $j$ (or from $i$ to $j$).\\

A very popular higher-order pattern is given by the \emph{shortest path length}, a concept entering into the definition of the well known \emph{small-world effect} \citep{watts1998collective}. Small-worldness refers to the evidence that, in many real-world networks, two (apparently) competing features co-exist: a large clustering coefficient and a small average shortest path length. More quantitatively, the small-world phenomenon is characterized by an average shortest path length typical of random graphs

\begin{equation}
\overline{d}\simeq d_\text{random}\propto\ln N
\end{equation}
(i.e. growing ``slowly'' with the size of the system) and by an average clustering coefficient typical of regular lattices (i.e., independent of the system size), much larger than that of a random graph

\begin{equation}
\overline{c}\gg\overline{c}_\text{random}\propto N^{-1}
\end{equation}
where, in both expressions, the term ``random'' refers to the ER model.

\paragraph*{Nestedness.} A pattern that has recently attracted much attention is the \emph{nestedness}. It quantifies how much the \emph{biadjacency} matrix of a bipartite network can be rearranged to let a triangular structure emerge \citep{johnson2013factors,mariani2019nestedness}. Several measures have been defined to quantify the nestedness, among which the NODF (an acronym for ``Nestedness metric based on Overlap and Decreasing Fill'') quantifies a matrix ``triangularity'' by measuring the overlap between rows and between columns \citep{almeida2008consistent}. 
Nestedness has been observed in ecological and economic systems alike. The classical example of nested ecological systems is given by the interactions between plants and pollinators, where nestedness emerges due to the presence of generalist pollinators (being attracted by all species of plants) co-existing with specialist pollinators (being attracted by only a small number of species of plants). Such a structure has been argued to promote the stability of the ecosystem \citep{bascompte2003nestedness}. For what concerns economic systems, nestedness is observed in the structure of countries' exports: a few very diversified countries have a large export basket, while others only export some simple products. Interestingly, this pattern contradicts classical economic theories, for which countries should specialize and export only those products in which they have a competitive advantage - this would imply a block-diagonal biadjacency matrix and not a nested one \citep{tacchella2012ec}.

\paragraph*{Centrality.} The concept of \emph{centrality} aims at quantifying the ``importance'' of a node in a network \citep{newman2018networks}. Besides \emph{degree centrality}, i.e., the centrality given by the degree, other well-known measures are the \emph{closeness centrality}, defined as

\begin{equation}
C_i(\mathbf{A})=\frac{1}{\overline{d}_i},\:\forall\:i
\end{equation}
i.e. as the reciprocal of the average topological distance of a node from the others, and the \emph{betweenness centrality}, defined as

\begin{equation}
B_i(\mathbf{A})=\sum_{j\neq i}\sum_{k\neq i,j}\frac{\sigma_{jk}(i)}{\sigma_{jk}},\:\forall\:i
\end{equation}
where $\sigma_{jk}$ is the total number of shortest paths from node $j$ to node $k$ and $\sigma_{jk}(i)$ is the number of these paths passing through $i$. 

Most of the proposed centrality measures are computable only on undirected networks. A notable exception is the \emph{PageRank centrality} \citep{page1999pagerank}, which can be computed by solving the iterative equation
\begin{equation}
P_i(\mathbf{A})=\frac{1-\alpha}{N}+\alpha\sum_{j\neq i}\left(\frac{a_{ji}}{k_j^{out}}\right)P_j(\mathbf{A}),\:\forall\:i.
\end{equation}

In general, it is very difficult to reconstruct the patterns of centrality of a network, unless these are strongly correlated with the degree centrality \citep{barucca2018tackling}.

\paragraph*{Reciprocity.} In the specific case of directed networks, it is of particular interest to measure the percentage of links having a counterpart pointing in the opposite direction. This quantity is known as \emph{reciprocity} and reads

\begin{equation}
r(\mathbf{A})=\frac{L^\leftrightarrow}{L}=\frac{\sum_i\sum_{j\neq i}a_{ij}a_{ji}}{\sum_i\sum_{j\neq i}\sum_ia_{ij}};
\end{equation}
remarkably, different classes of real-world networks are characterized by different values of reciprocity \citep{garlaschelli2004patterns}. For instance, reciprocity is a distinguishing feature of financial networks, where it proxies ``trust'' between banks \citep{squartini2013early}. 

\paragraph*{Spectral properties.} With this term one refers to the features of eigenvalues and eigenvectors of both the adjacency matrix $\mathbf{A}$ and of the \emph{Laplacian matrix} $\mathbf{L}=\mathbf{D}-\mathbf{A}$ of the network\footnote{The focus on undirected binary networks is justified by the easiness of treating symmetric matrices, a characteristic ensuring that eigenvalues are real, for example.} (here $\mathbf{D}$ is the diagonal matrix whose generic entry reads $d_{ii}=k_i,\:\forall\:i$). While Laplacian spectral properties provide information on macroscale network properties like the number of connected components (that matches the multiplicity of the zero eigenvalue of $\mathbf{L}$), spectral properties of $\mathbf{A}$ provide information on higher-order patterns like cycles \citep{estrada2015first} as well as on dynamical properties of spreading processes \citep{bardoscia2017pathways}. Notice that the reconstruction of spectral properties of empirical networks is still a largely underexplored topic, although a first result in this sense is provided by the Silverstein theorem \citep{silverstein1994spectral}.

\subsection*{Weighted properties}

While binary networks are characterized by an adjacency matrix whose entries assume only the values $0$ and $1$, \emph{weighted}, \emph{undirected} (\emph{directed}) graphs are specified by a symmetric (generally asymmetric) ${N\times N}$ matrix $\mathbf{W}$ whose generic entry $w_{ij}$ quantifies the intensity of the link connecting nodes $i$ and $j$: in the most general case, $w_{ij}$ is a real number; however, in many cases $w_{ij}$ assumes integer values. Naturally, $\mathbf{A}$ and $\mathbf{W}$ are related by the position $a_{ij}=\Theta[w_{ij}],\:\forall\:i,j$, simply stating that any positive weight between $i$ and $j$ carries the information that $i$ and $j$ are indeed connected.

\paragraph*{Weight distribution.} When links are characterized by ``magnitudes'', the first step is to inspect the distribution of these magnitudes. When considering real-world networks, weight distributions are often found to be \emph{fat-tailed}. 

\paragraph*{Strengths.} The weighted analogue of the degree is the so-called \emph{strength}. It is defined as $s_i(\mathbf{W})=\sum_{j\neq i} w_{ij},\:\forall\:i$, i.e. as the sum of the weight of the links connected to node $i$. Similarily to the case of degrees, strength distributions are often found to be \emph{fat-tailed}. When directed networks are considered, one speaks of \emph{out-strength} and \emph{in-strength}, respectively defined as $s_i^{out}(\mathbf{W})=\sum_{j\neq i} w_{ij},\:\forall\:i$ and $s_i^{in}(\mathbf{W})=\sum_{j\neq i} w_{ji},\:\forall\:i$.

From a network reconstruction perspective, strengths play an important role, since they often represent the only kind of information available for the system under consideration. The typical example is that of financial networks, where only the total \emph{assets} and \emph{liabilities} of each bank (respectively the out- and in-strengths of the respective node) are accessible. This has motivated the definition of the weighted analogue of the Chung-Lu model, also known as the \emph{MaxEnt} (ME) recipe. Its directed version, reading

\begin{equation}
\hat{w}_{ij}^{\text{ME}}=\frac{s_i^{out}s_j^{in}}{W},\:\forall\:i\neq j
\end{equation}
(with $W=\sum_is_i^{out}=\sum_is_i^{in}$) is extensively used to estimate the magnitude of links in economic and financial networks \citep{mistrulli2011assessing,upper2011simulation,squartini2018reconstruction}.

\paragraph*{Weighted assortativity.} The concept of assortativity can be easily extended to the weighted case. The weighted counterpart of the average nearest neighbors degree of node $i$ is the \emph{average nearest neighbors strength} (ANNS):

\begin{equation}
s^{nn}_i(\mathbf{W})=\frac{\sum_{j\neq i}a_{ij}s_j}{k_i},\:\forall\:i.
\label{eq_wun_snn}
\end{equation}

Analogously to the binary case, the correlation between strengths can be inspected by plotting $s^{nn}_i$ versus $s_i$. Note that since $\langle a_{ij}\rangle_\text{ME}=p_{ij}^\text{ME}=\Theta[\hat{w}_{ij}^\text{ME}]$, the weighted version of the CL model always generates a very densely connected network and, as a consequence, a value for the ANNS of node $i$ that is weakly dependent on $i$ itself, i.e. $\langle s^{nn}_i\rangle_{\text{ME}}\simeq \frac{\sum_{j\neq i}p_{ij}^{\text{ME}} s_j}{\langle k_i\rangle_{\text{ME}}}\simeq\frac{\sum_{j\neq i}s_j}{N-1}\simeq\frac{2W}{N-1},\:\forall\:i$ \citep{squartini2011analytical}. The weighted CL model thus suffers from the same limitations affecting the binary CL model\footnote{As we will see in what follows, the weighted counterpart of the CM, namely the \emph{Weighted Configuration Model} (WCM), does not represent the solution to this problem. We will need to consider degrees and strengths together as in the \emph{Enhanced Configuration Model} (ECM).}.

\paragraph*{Weighted hierarchy.} A \emph{weighted clustering coefficient} (WCC) can be defined to capture the ``intensity'' of the triangles in which node $i$ participates \citep{squartini2011randomizingII}:

\begin{equation}
c_i^w(\mathbf{W})=\frac{\sum_{j\neq i}\sum_{k\ne i,j}(w_{ij}w_{jk}w_{ki})^{1/3}}{k_{i}(k_{i}-1)}.
\label{eq_wun_c}
\end{equation}

Contrarily to what is observed for the vast majority of binary networks, plotting $c_i^w$ versus $s_i$ reveals an increasing trend for many real-world networks, indicating that nodes with larger total activity participate in more ``intense'' triangles.

For extensions of ANNS and WCC to directed networks, see \citet{squartini2011randomizingII}.

\paragraph*{Weighted reciprocity.} A weighted version of link reciprocity can be defined as

\begin{equation}
r^w(\mathbf{W})=\frac{W^\leftrightarrow}{W}=\frac{\sum_i\sum_{j\neq i}\min[w_{ij},w_{ji}]}{\sum_i\sum_{j\neq i}w_{ij}}
\end{equation}
a quantity whose numerator accounts for the ``minimum exchange'' between any two nodes \citep{squartini2013reciprocity}.

\paragraph*{Higher-order patterns and Centrality.} Differently from the purely binary case, higher-order patterns in weighted networks are rarely inspected. An attempt to define \emph{weighted motifs} has been done in \citep{onnela2005intensity}, while \emph{weighted centrality measures} have been defined in \citep{opshal2010node}.

\section{Macroscale reconstruction of economic and financial networks: a quick historical survey}
\label{appa}

The network reconstruction problem can be formulated as follows. Let us consider the most general case of a weighted, directed network, represented by an $N\times N$ asymmetric matrix $\mathbf{W}$ with entries $w_{ij}\in\mathbb{R}$ $\forall\:i,j$. When considering financial networks, the generic entry $w_{ij}$ may represent the value of the exposure of $i$ towards $j$; in the case of economic networks, it may represent the value of exports from country $i$ to country $j$.

In the well-studied case of economic and financial networks, the available information is represented by the out-strength and in-strength sequences, i.e. $s_i^{out}(\mathbf{W})=\sum_{j\neq i} w_{ij},\:\forall\:i$ and $s_i^{in}(\mathbf{W})=\sum_{j\neq i} w_{ji},\:\forall\:i$: the network reconstruction goal, thus, becomes the estimation of the generic entry $w_{ij}$ via the aforementioned, aggregate information \citep{squartini2018reconstruction,anand2017missing}.\\

One of the first-ever proposed reconstruction algorithms was born with the aim of inferring the value of direct exposures between financial institutions and is known as \emph{MaxEnt} algorithm \citep{wells2004financial,mistrulli2011assessing}. The method prescribes to maximize the entropic functional

\begin{equation}\label{me}
S=-\sum_i\sum_j w_{ij}\ln w_{ij}
\end{equation}
under the constraints represented by the out- and in-strengths (respectively assets and liabilities, in the jargon of finance). The solution to the aforementioned constrained maximization problem reads

\begin{equation}\label{maxe}
\hat{w}_{ij}^{\text{ME}}=\frac{s_i^{out}s_j^{in}}{W},\:\forall\:i\neq j
\end{equation}
where $W=\sum_{i}\sum_{j\neq i}w_{ij}$ denotes the total economic value of the system at hand. The recipe above is simple and allows the constraints to be satisfied: in fact, $\hat{s}_i^{out}=\sum_j\hat{w}_{ij}^{\text{ME}}=s_i^{out}$ and analogously for $s_i^{in}$. However, it suffers from two major drawbacks. Firstly, constraints are satisfied only if the summation index runs over all values $j=1\dots N$, including the ones corresponding to the diagonal entries: the method needs self-loops to work. Second, the generated network topologies are unrealistically densely-connected - in fact, no entry can be predicted to be zero, unless either $s_i^{out}=0$ or $s_i^{in}=0$ for some nodes. 
This is quite a problem for financial networks, since employing too dense configurations for running the stress tests typically leads to underestimate the systemic risk \citep{mistrulli2011assessing}. On the other hand, too sparse configurations lead to systemic risk overestimation \citep{anand2014filling}. 
Nevertheless, the MaxEnt prescription provides quite accurate estimates of the \emph{magnitude} of empirical weights \citep{squartini2017stock,mazzarisi2017methods,almog2019enhanced}.\\

A first attempt to build more realistic configurations is represented by the \emph{iterative proportional fitting} (IPF) algorithm \citep{deming1940least,bacharach1965estimating,fienberg1970iterative}. This is a simple recipe to obtain a matrix that 1) lies at the ``minimum distance'' from the MaxEnt matrix $\hat{\mathbf{W}}^{\text{ME}}$, 2) satisfies the constraints represented by the available information, 3) admits the presence of a set of zero entries (in the simplest case, the diagonal ones). The main drawback of the IPF, however, is that of requiring the knowledge of the position of zeros \emph{in advance}, a piece of information that is (practically) never accessible.\\

The evidence that the outcome of systemic risk estimation strongly depends on the link density has pushed many researchers to devise a way to tune the density of the reconstructed network \citep{mastromatteo2012reconstruction,drehmann2013measuring,halaj2013assessing,cimini2015systemic}. This is typically achieved though a free parameter in the algorithm, intended to enforce\footnote{The attractiveness of these methods lies in the possibility of generating different topologies still satisfying the (weighted) constraints: in fact, since the link density is \emph{chosen} rather than \emph{reproduced}, these methods are well suited for defining possible scenarios over which running stress tests.} the desired density. As an ``extreme'' example, let us consider the \emph{minimum density algorithm} \citep{anand2014filling}, intended to find the network structure with minimum link density that still satisfies the weighted constraints. Although its main limitation is that of overestimating the impact of shocks on the predicted configuration (intuitively, the few admitted links carry the maximum possible load, thus propagating the largest possible shocks), its combined use with the MaxEnt algorithm allows one to provide an upper and a lower bound to the systemic risk.

\section{The Exponential Random Graphs framework}

We now introduce the popular framework of Exponential Random Graphs that will be used to carry out the network reconstruction at the macroscale. The idea is that of building a network model able to reproduce a bunch of chosen quantities (the {\em constraints}) while ensuring that everything else is kept as random as possible. The core quantity of the method is \emph{Shannon entropy}, defined as 

\begin{equation}
S=-\sum_{\mathbf{W}\in\mathcal{W}}P(\mathbf{W})\ln P(\mathbf{W})
\end{equation}
where $P(\mathbf{W})$ is the probability distribution of the specific network configuration $\mathbf{W}$ output by the model - defined over the ensemble $\mathcal{W}$ of allowable configurations. The constrained maximization of $S$ represents an inference procedure that has been proved to be maximally non-committal with respect to the missing information \citep{jaynes1957information}: in other words, it allows one to derive a probability distribution that minimizes the number of unjustified assumptions about the unavailable data.

More quantitatively, it can be implemented by defining the Lagrangean function

\begin{eqnarray}\label{cons}
\mathscr{L}[P]&=&S-\sum_{m=0}^M\theta_m\left(\sum_ {\mathbf{W}\in\mathcal{W}}P(\mathbf{A})C_m(\mathbf{W})-C_m^*\right)\nonumber\\
&=&S-\sum_{m=0}^M\theta_m\left(\langle C_m\rangle-C_m^*\right)
\end{eqnarray}
with $\theta_m$ representing the lagrange multiplier associated to the $m$-th constraint, $C_m(\mathbf{W})$ the value of the $m$-th constraint measured on the configuration $\mathbf{W}$, $\langle C_m\rangle$ its average value over the ensemble $\mathcal{W}$ and $C_m^*$ the value we impose for it. 
$C_0(\mathbf{W})=C_0^*=1$ sums up the normalization condition. Upon solving the equation $\frac{\delta\mathscr{L}[P]}{\delta P(\mathbf{W})}=0$ we find the expression $P(\mathbf{W}|\vec{\theta})=e^{-1-\vec{\theta}\cdot\vec{C}(\mathbf{W})}$ that can be further re-written as

\begin{equation}\label{erg}
P(\mathbf{W}|\vec{\theta})=\frac{e^{-H(\mathbf{W},\vec{\theta})}}{Z(\vec{\theta})}
\end{equation}
a formula defining the \emph{Exponential Random Graphs} (ERG) formalism in its full generality, with the quantity $H(\mathbf{W},\vec{\theta})=\vec{\theta}\cdot\vec{C}(\mathbf{W})=\sum_{m=1}^M\theta_m C_m(\mathbf{W})$ usually called \emph{Graph Hamiltonian} \citep{park2004statistical}.

As $P(\mathbf{W}|\vec{\theta})$ depends on the vector of parameters $\vec{\theta}$, we need a recipe to estimate them. 
Such a recipe comes from the \emph{likelihood maximization principle}, prescribing to solve the system of equations

\begin{equation}\label{eqlik}
\frac{\partial\mathcal{L}(\vec{\theta})}{\partial\vec{\theta}}=\frac{\partial\ln P(\mathbf{W}^*|\vec{\theta})}{\partial\vec{\theta}}=\vec{0}
\end{equation}
with respect to the unknowns. Here $\mathbf{W}^*$ indicates an empirical network configuration that meets the values of the imposed constraints. Notice that substituting eq. (\ref{erg}) into eq. (\ref{eqlik}) and solving it leads to the system of equations $\langle C_m\rangle(\vec{\theta})=C_m^*,\:\forall\:m$, guaranteeing that the expected values of the constraints match the imposed ones \citep{squartini2011analytical}.

\paragraph*{The Configuration Model (CM).} 
The most popular model of the ERG family is the \emph{configuration model}, where the constraints imposed are the node degrees \citep{park2004statistical,cimini2019statistical}. The Hamiltonian thus reads

\begin{equation}
H(\mathbf{A},\vec{\theta})=\sum_i\theta_ik_i(\mathbf{A})
\end{equation}
and the probability of a network in the ensemble is
\begin{equation}
P(\mathbf{A})=\prod_{i<j} p^{a_{i\alpha}}_{ij}(1-p_{ij})^{1-a_{i\alpha}}
\end{equation}
where $p_{ij}=\frac{x_ix_j}{1+x_ix_j}$ stands for the probability that a link exists between nodes $i$ and node $j$.
The parameters are numerically determined using the likelihood maximization equation

\begin{equation}
k_i(\mathbf{A}^*)=\sum_{j(\neq i)}\frac{x_ix_j}{1+x_ix_j}=\langle k_i\rangle_\text{CM},\:\forall\:i\\
\end{equation}

\paragraph*{Constraining linear vs non-linear quantities.} The ERG framework allows considering as a constraint any function of the adjacency matrix of the network. In what follows, we will focus on ERG models defined by linear constraints (i.e., linear functions of the adjacency matrix elements) for at least two reasons. Firstly, as we will see, linear models perform remarkably well when employed to reconstruct several networks of interest. Secondly, they are not affected by the (theoretical and practical) limitations characterizing models with non-linear constraints. Examples of the latter ones are provided by the models studied in \citep{park2004solution} and \citep{park2005solution}, constraining the total number of \emph{links} and \emph{two-stars} and the total number of \emph{links} and \emph{triangles}, respectively. What the authors find for these models is the presence of \emph{phases} and phase transitions, analogous to the ones that characterize classical disordered systems of spins. Although interesting from a purely theoretical perspective, these models cannot be easily used for reproducing the properties of real-world networks. Other examples are provided by models specifying the degree distribution and degree-degree correlations (and clustering, to some extent), which however can be approached mainly through numerical Monte Carlo sampling methods \citep{coolen2009constrained,annibale2009tailored,orsini2015quantifying}

\section{The best-performing reconstruction method}

We now show how to apply the ERG framework in the representative case of financial networks, where nodes' out- and in-strengths are the only information on the network. To overcome the problems affecting the ME recipe, one may be tempted to solve eq. (\ref{erg}) by imposing the out- and in-strength sequences as constraints. This model, known as \emph{Directed Weighted Configuration Model} (DWCM) \citep{squartini2011analytical}, is characterized by a geometric weight-specific distribution

\begin{equation}
q_{ij}^\text{DWCM}(w_{ij})=(x_iy_j)^{w_{ij}}(1-x_iy_j),\:\forall\:i\neq j
\end{equation}
whose unknowns can be estimated via the constraints equations

\begin{eqnarray}
s_i^{out}(\mathbf{W}^*)&=&\sum_{j\neq i}\frac{x_iy_j}{1-x_iy_j}=\sum_{j\neq i}\langle w_{ij}\rangle_\text{DWCM}=\langle s_i^{out}\rangle_\text{DWCM},\:\forall\:i\\
s_i^{in}(\mathbf{W}^*)&=&\sum_{j\neq i}\frac{x_jy_i}{1-x_jy_i}=\sum_{j\neq i}\langle w_{ji}\rangle_\text{DWCM}=\langle s_i^{in}\rangle_\text{DWCM},\:\forall\:i.
\end{eqnarray}

An alternative estimation procedure characterizes the \emph{Maximum-Entropy Capital Asset Pricing Model} (MECAPM) \citep{digangi2018assessing}, which prescribes to equate the expression for the expected weights under the DWCM to the MaxEnt estimate, i.e.

\begin{equation}
\langle w_{ij}\rangle_\text{DWCM}=\frac{x_iy_j}{1-x_iy_j}=\hat{w}_{ij}^\text{ME},\:\forall\:i\neq j.
\end{equation}
Solving this system of equations then leads to recover the numerical value of the expressions $x_iy_j,\:\forall\:i\neq j$.\\

Unfortunately, both the DWCM and MECAPM generate very dense network configurations and thus, like the ME recipe, perform poorly in reproducing the topological structure of the network \citep{squartini2018reconstruction,mazzarisi2017methods}. This drawback can be solved only by \emph{imposing} some kind of topological information, beside the weighted one represented by the sequences $\{s_i^{out}\}_{i=1}^N$ and $\{s_i^{in}\}_{i=1}^N$. The need of adding some topological information leads to the definition of two broad classes of algorithms: those simultaneously imposing binary and weighted constraints and those acting iteratively on \emph{ad-hoc} topologies. Among the algorithms belonging to the first group, a special mention is deserved by the \emph{Enhanced Configuration Model} (ECM)\footnote{The ECM is particularly interesting from a theoretical viewpoint. In fact, it generates mixed Bose-Fermi statistics in the case of integer weights \citep{garlaschelli2009generalized}, whereas, in the case of continuous weights it is equivalent to an Ising model on a lattice gas \citep{,gabrielli2019grand}.} \citep{mastrandrea2014enhanced}, defined (in the simpler undirected case) by the recipe

\begin{equation}\label{eq:weight_ecm}
q_{ij}^\text{ECM}(w_{ij})=\left\{ \begin{array}{ll}
1-p_{ij}^\text{ECM} & \textrm{if $w_{ij}=0$},\\
p_{ij}^\text{ECM}(y_iy_j)^{w_{ij}-1}(1-y_iy_j) & \textrm{if $w_{ij}>0$}
\end{array} \right.
\end{equation}
with $p_{ij}^\text{ECM}=\frac{x_ix_jy_iy_j}{1-y_iy_j+x_ix_jy_iy_j}$ being the binary connection probability. 
Examples of algorithms belonging to the second group are those using the ME recipe and then iteratively adjusting the link weights (e.g. via the IPF recipe \citep{bacharach1965estimating}) on top of some previously-determined topological structure, in such a way to satisfy the strengths constraints \emph{a posteriori}.

As will be shown below, the knowledge of both the degrees and the strengths allows the ECM to achieve a very good reconstruction of many different kinds of networks \citep{mastrandrea2014enhanced,mastrandrea2014reconstructing}. Degrees, however, are rarely accessible, whence the need to find an alternative recipe to estimate them. As shown in \citep{squartini2017network}, the basic information encoded into the link density of the network can be successfully used to this aim. Its use leads to the definition of a two-step approach to reconstruction\footnote{As opposed to the algorithms employing the IPF recipe to adjust weights, whose second step is \emph{deterministic}, both steps of these ERG-based approaches are \emph{probabilistic} in nature.}. This approach has been tested in four different horse races \citep{anand2017missing,mazzarisi2017methods,ramadiah2020reconstructing,lebacher2019search}, resulting the method consistently performing the best (or among the best).

The first step of such an ERG-based approach consists in estimating the network topology, resting upon the following three hypotheses.

\begin{itemize}
\item[$\text{I.}$] The binary topology of the empirical network $\mathbf{W}^*$ is drawn from the ensemble induced by the \emph{Directed Configuration Model} (DCM) \citep{squartini2011analytical,park2004statistical}. The DCM induces a set of configurations that are maximally random, except for the (ensemble) averages of the out- and in-degrees. This amounts at considering the entries $a_{ij}=\Theta[w_{ij}],\:\forall\:i\neq j$ of the binary adjacency matrix as independent random variables, fully described by the link-specific probability coefficients reading

\begin{equation}\label{eq:prob}
p_{ij}^\text{DCM}=\langle a_{ij}\rangle_\text{DCM}=\frac{x_iy_j}{1+x_iy_j},\:\forall\:i\neq j.
\end{equation}

The Lagrange multipliers $\{x_i\}_{i=1}^N$ and $\{y_i\}_{i=1}^N$ can be numerically determined by solving the system of equations

\begin{eqnarray}
k_i^{out}(\mathbf{A}^*)&=&\sum_{j\neq i}\frac{x_iy_j}{1+x_iy_j}=\langle k_i^{out}\rangle_\text{DCM},\:\forall\:i\\
\label{eq.DCM1}
k_i^{in}(\mathbf{A}^*)&=&\sum_{j\neq i}\frac{x_jy_i}{1+x_jy_i}=\langle k_i^{in}\rangle_\text{DCM},\:\forall\:i
\label{eq.DCM2}
\end{eqnarray}
where $\mathbf{A}^*=\Theta[\mathbf{W}^*]$. Degrees, however, are rarely accessible, whence the need of a second assumption.

\item[$\text{II.}$] The Lagrange multipliers $\{x_i\}_{i=1}^N$ and $\{y_i\}_{i=1}^N$ controlling for the ensemble average of the degrees are assumed to be linearly correlated with accessible quantities, generally called \emph{fitnesses} \citep{caldarelli2002scale,iori2008network}. The typical approach is to use the out- and in-strengths themselves (i.e. the only available information) as fitnesses, because of the strong correlation between degrees and strengths observed in several real-world networks \citep{cimini2015systemic}. Whence the position

\begin{equation}\label{eq:lagra}
x_i=\sqrt{z}\cdot s_i^{out},\:\forall\:i\:\:\:\mbox{and}\:\:\:y_i=\sqrt{z}\cdot s_i^{in},\:\forall\:i.
\end{equation}

Since we cannot make a direct use of the DCM, we can resort to the ansatz above, which assumes the network topology to be determined by intrinsic node properties. This approach, also known as \emph{fitness-induced Configuration Model} (fiCM), has been successfully employed to model financial networks, where fitnesses are nothing but assets and liabilities \citep{cimini2015systemic,cimini2015estimating}. For economic networks, node fitness have been identified for instance with the countries GDP \citep{garlaschelli2004fitness,almog2017double}.

\item[$\text{III.}$] Besides the heterogeneity induced by the fitness-induced degrees, the network is assumed to be \emph{homogeneous}, so that its (global) link density can be estimated by sampling subsets of nodes. To this aim, the best recipe is represented by the \emph{random-nodes sampling scheme}, any other procedure being biased towards unrealistically large, or small, link density values \citep{squartini2017network}.
\end{itemize}

The assumptions above leave us with the task of determining only one (global) proportionality constant, which is obtained by equating the ensemble average of the total number of links to the (known or estimated) $L(\mathbf{W}^*)$ value, i.e.

\begin{equation}\label{eq:L}
\langle L\rangle_\text{fiCM}=\sum_i\sum_{j\neq i}\frac{zs_i^{out}s_j^{in}}{1+zs_i^{out}s_j^{in}}=L(\mathbf{W}^*);
\end{equation} 
once $z$ has been determined, the numerical value of the linking probabilities

\begin{equation}
p_{ij}^\text{fiCM}=\langle a_{ij}\rangle_\text{fiCM}=\frac{zs_i^{out}s_j^{in}}{1+zs_i^{out}s_j^{in}},\:\forall\:i\neq j
\end{equation}
can be straightforwardly estimated. Notice that the fiCM can be easily generalized to address the problem of bipartite networks reconstruction (see also Appendix \ref{appb}) \citep{squartini2017stock}.\\

The second step of the reconstruction procedure then concerns the estimation of link weights. To this aim, one can extend the traditional MaxEnt prescription by defining the following Bernoulli-like recipe:

\begin{equation}\label{eq:weight}
q_{ij}^\text{dcGM}(w_{ij})=\left\{ \begin{array}{ll}
\:\:\:0 & \textrm{with probability $1-p_{ij}^\text{fiCM}$},\\
\frac{\hat{w}_{ij}^{\text{ME}}}{p_{ij}^\text{fiCM}} & \textrm{with probability $p_{ij}^\text{fiCM}$}.
\end{array} \right.
\end{equation}

Since the MaxEnt weight $\hat{w}_{ij}^{\text{ME}}$ is placed between nodes $i$ and $j$ with probability $p_{ij}^\text{fiCM}$, both the node strengths and the link density are correctly reproduced\footnote{In order for node strengths to be correctly reproduced, the probability distribution defined by eq. (\ref{eq:weight}) must include self-loops; in \citep{squartini2017network}, the authors have proposed a slightly modified version of this recipe, to deal with the more realistic case in which self-loops are absent.} whatever the underlying topology of the network. In fact, $\langle L\rangle_\text{fiCM}=L(\mathbf{W}^*)$ and

\begin{equation}
\langle w_{ij}\rangle_\text{dcGM}=0\cdot (1-p_{ij}^\text{fiCM})+\frac{\hat{w}_{ij}^{\text{ME}}}{p_{ij}^\text{fiCM}}\cdot p_{ij}^\text{fiCM}=\hat{w}_{ij}^{\text{ME}},\:\forall\:i\neq j.
\end{equation}

The reconstruction method described here is known as \emph{density-corrected Gravity Model} (dcGM) \citep{cimini2015systemic}. In a sense, by disentanging the binary and weighted statistics of the ensemble, it constitutes a simplified version of the \emph{Directed Enhanced Configuration Model} (DECM), yet retaining the same accuracy in reconstructing real-world networks. Note that the dcGM approach can be ``enriched'' with an \emph{exponential} weight-specific distribution that overcomes the limitations affecting the simpler Bernoulli-like recipe defined in eq. (\ref{eq:weight}) \citep{parisi2020faster}. 

An alternative efficient approach to the dcGM consists in using degrees estimated through the first step together with empirical strengths to inform an ECM \citep{cimini2015estimating}.

\section{Testing reconstruction at the macroscale}

This second section of the chapter is devoted to answering the question \emph{which of the aforementioned network patterns can be reconstructed and by which model?} As it will be shown, specifying only \emph{local} information (i.e., the one encoded into the degree and strength sequences) is often enough to achieve a satisfactory reconstruction of the network under consideration. In what follows, we will mainly focus on economic systems. In particular, we will cite results concerning the WTW, which is the network of trade exchanges between world countries. The WTW data is publicly available, allowing reconstruction models to be testable and comparable.

\begin{figure}[t!]
\includegraphics[width=0.49\textwidth]{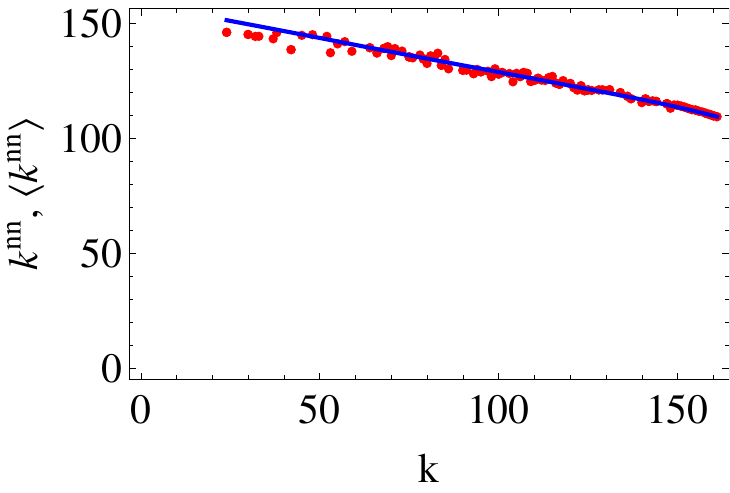}
\includegraphics[width=0.49\textwidth]{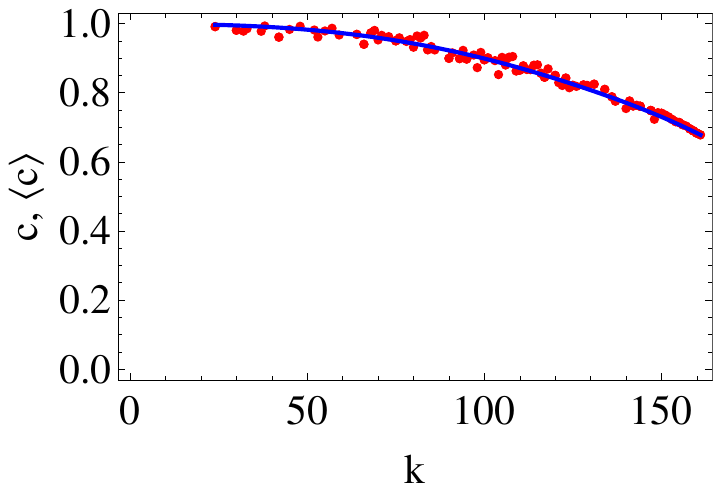}
\caption{Comparison between the observed (red points) and the reconstructed (blue solid curve) values of the average nearest neighbors degree (left panel) and of the clustering coefficient (right panel) plotted as functions of the degree, for the 2002 snapshot of the binary undirected WTW. The fact that the reconstruction is done through the CM we can  conclude that the information encoded into the degree sequence is enough to reproduce the disassortative and the hierarchical character of the WTW network. Source: \citep{squartini2011randomizing}.}
\label{fig1}
\end{figure}

\paragraph*{Assortativity and hierarchy.} Let us start by inspecting if and how assortativity can be reconstructed. Figure \ref{fig1} shows the empirical ANND of the WTW for the year 2002 \citep{squartini2011randomizing}: the network is disassortative, i.e. countries with large degree preferentially connect with countries with low degree and viceversa; from a macroeconomic point of view, this reflects the evidence that the partners of richer countries are (preferentially) poorer countries and viceversa. Figure \ref{fig1} also shows that constraining the degrees allows degree-degree correlations to be reproduced quite well: in other words, the CM allows one to achieve an accurate reconstruction of the second-order properties of the WTW. A similar conclusion can be drawn about third-order properties like the clustering coefficient: the CM correctly reconstruct the WTW as a hierarchical network.

\begin{figure}[t!]
\includegraphics[width=0.76\textwidth]{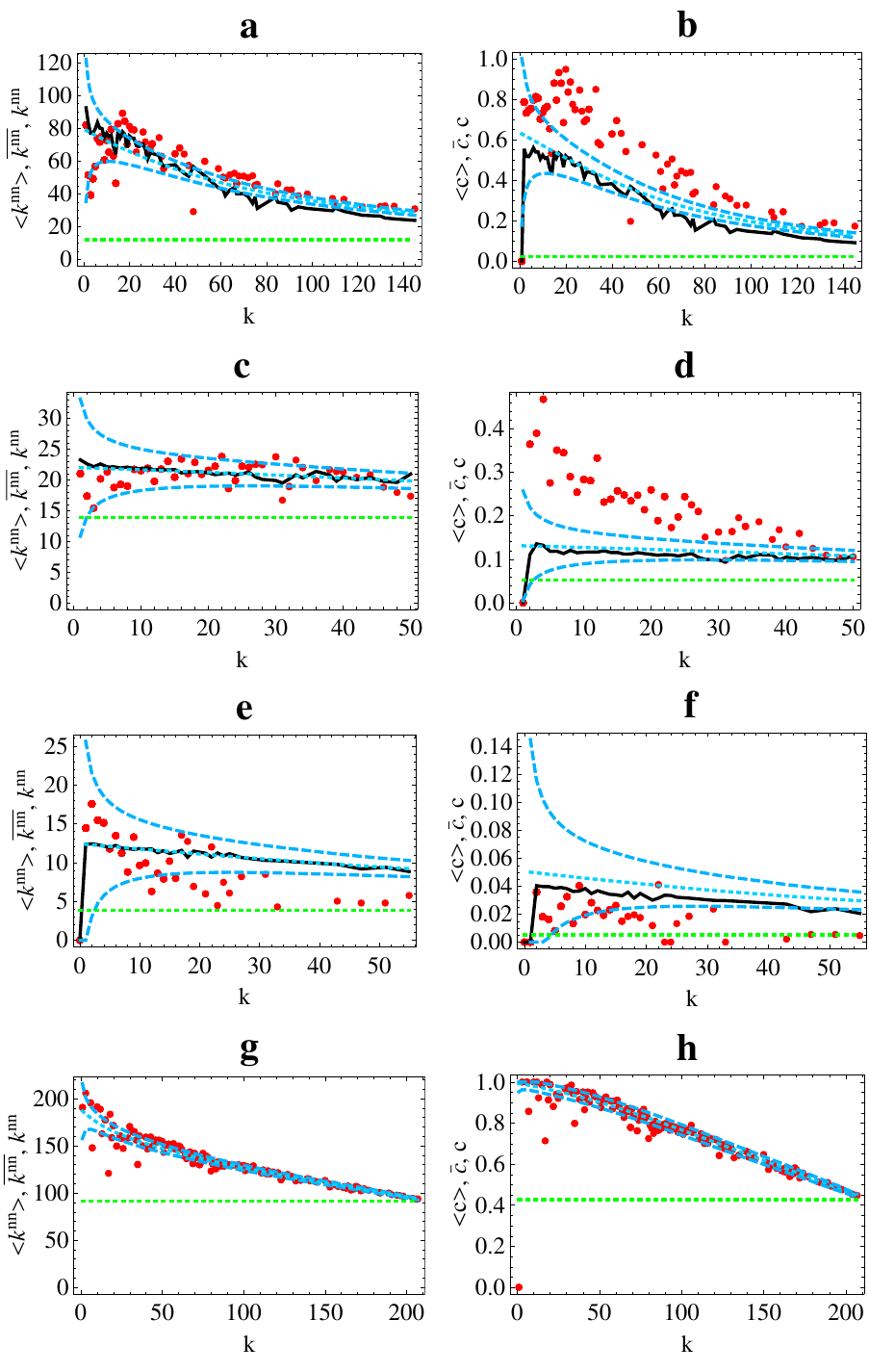}
\caption{Average nearest neighbors degree (left panel) and clustering coefficient (right panel) of various real-world networks. Blue dashed curves are the expectations under the CM and green dashed curves are the expectations under the ER model. Source: \citep{squartini2011analytical}.}
\label{fig2}
\end{figure}

Figure \ref{fig2} shows the performance of the CM in reproducing the ANND and the clustering coefficient for a wide variety of networks (technological, neural, cellular and financial ones): while the former seems to be reproduced overall quite satisfactorily, the latter is not; the best agreement is observed for a financial system, i.e. e-MID (electronic Market for Interbank Deposits), the Italian unsecured interbank network \citep{cimini2015systemic}. The comparison between the CM and the ER model is also shown: as anticipated, the ER model performs quite poorly in reproducing the empirical patterns considered here. This is readily seen by calculating $\langle k_i^{nn}\rangle_\text{ER}=p(N-1),\:\forall\:i$ and $\langle c_i\rangle_\text{ER}=p,\:\forall\:i$, a result confirming that the ER model is not able to account for the heterogeneity of nodes characterizing any real-world network.

\begin{figure}[t!]
\includegraphics[width=0.46\textwidth]{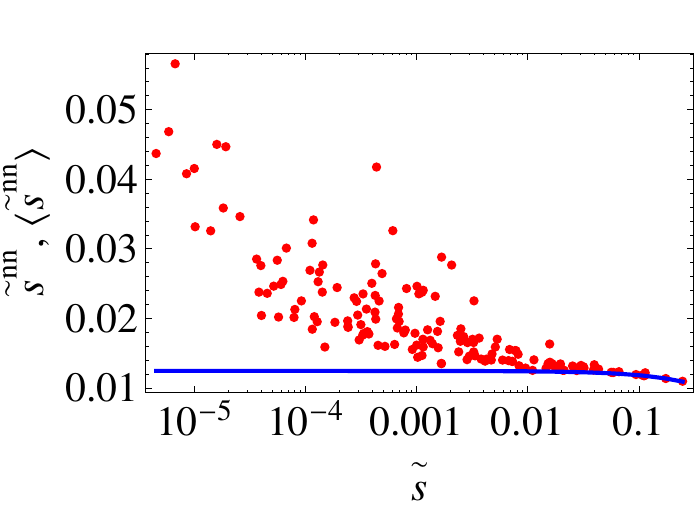}
\includegraphics[width=0.53\textwidth]{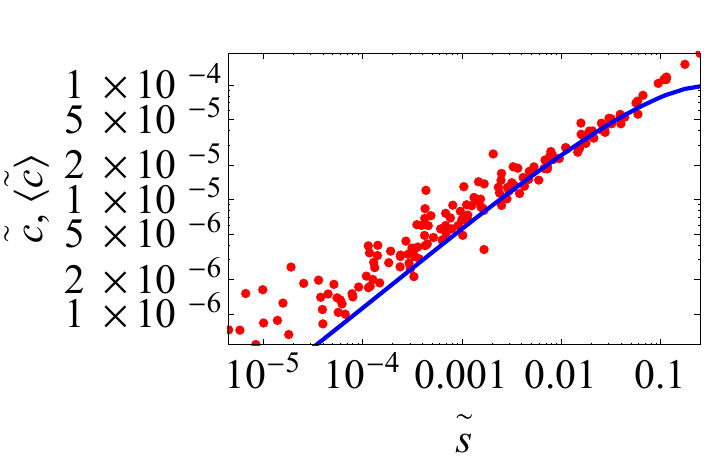}
\caption{Comparison between the empirical (red points) and expected (blue solid curve) values of the average nearest neighbors strength (left panel) and of the weighted clustering coefficient (right panel) plotted as functions of the strength, for the 2002 snapshot of the weighted, undirected WTW (red points). The expectations are computed under the WCM: while the latter fails in reproducing the trend of the ANNS, it seems to capture the rising trend of the WCC. Source:  \citep{squartini2011randomizingII}.}
\label{fig3}
\end{figure}

It is interesting to notice how the weighted counterparts of the ANND and the clustering coefficient are, instead, badly reproduced by the WCM. As fig. \ref{fig3} seems to suggest, the information encoded into the strength sequence is not enough to satisfactorily reconstruct the weighted WTW structure. The reason lies in the poor performance of the WCM in reproducing the purely topological structure of the WTW: it fact, it predicts a very dense network, i.e. $\langle k_i\rangle_\text{WCM}\simeq(N-1),\:\forall\:i$, a result inducing a flat trend of higher-order properties like the ANNS. Under this respect, the ME model and the WCM perform similarly: $\langle s_i^{nn}\rangle_\text{ME}\simeq\langle s_i^{nn}\rangle_\text{WCM}\simeq\frac{2W}{N-1},\:\forall\:i$.\\

\begin{figure}[t!]
\includegraphics[width=\textwidth]{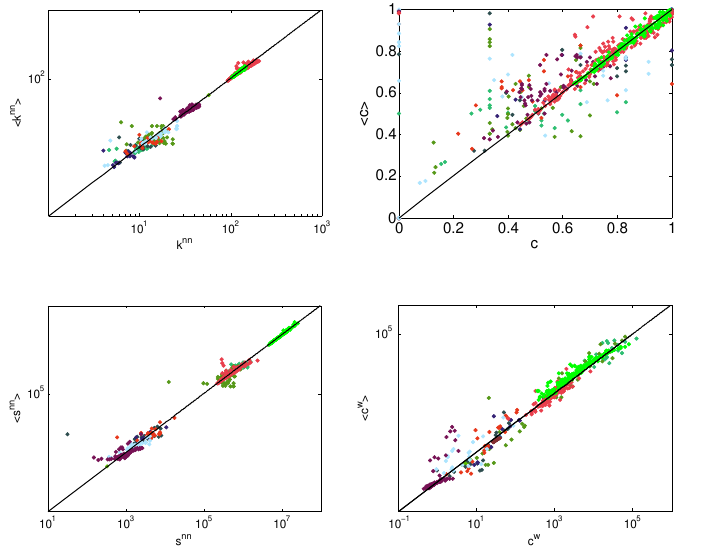}
\caption{Enhanced network reconstruction from strengths and degrees (ECM). Each panel shows the comparison between the reconstructed (y axis) and the empirical (x axis) value of a node-specific network property, for several real-world networks: top left, ANND; top right: clustering coefficient; bottom left: ANNS; bottom right: WCC. Source:  \citep{mastrandrea2014enhanced}.}
\label{fig4}
\end{figure}

As mentioned in the previous section, the solution to this problem lies in constraining some kind of topological information beside the (weighted) one represented by strengths. In the ideal case, both degrees and strengths are available: constraining them simultaneously leads to the definition of the ECM \citep{mastrandrea2014enhanced,gabrielli2019grand}, whose good performance in reproducing a wide range of real-world networks is shown in fig. \ref{fig4}.

\begin{figure}[t!]
\includegraphics[width=0.8\textwidth]{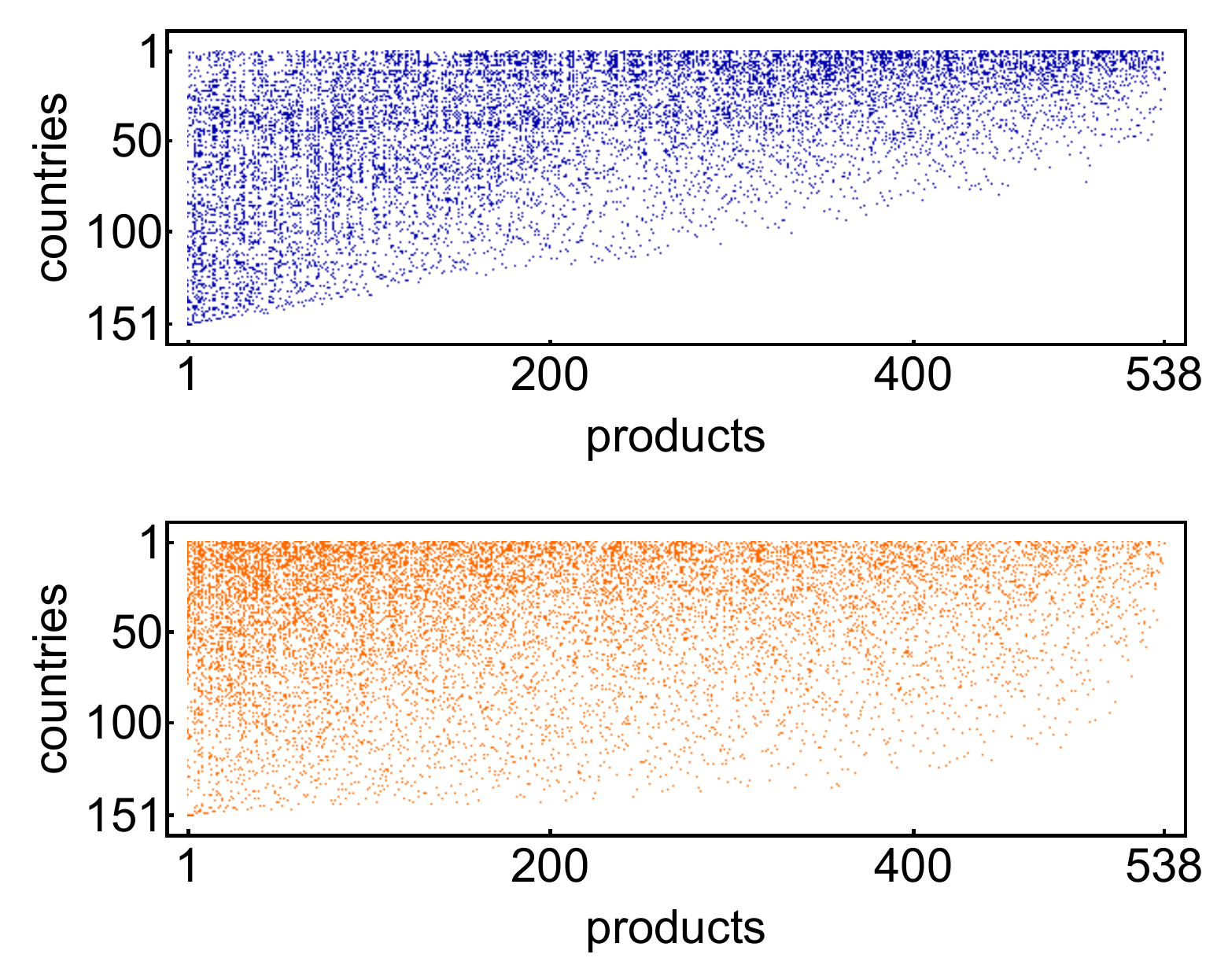}\\
\includegraphics[width=0.75\textwidth]{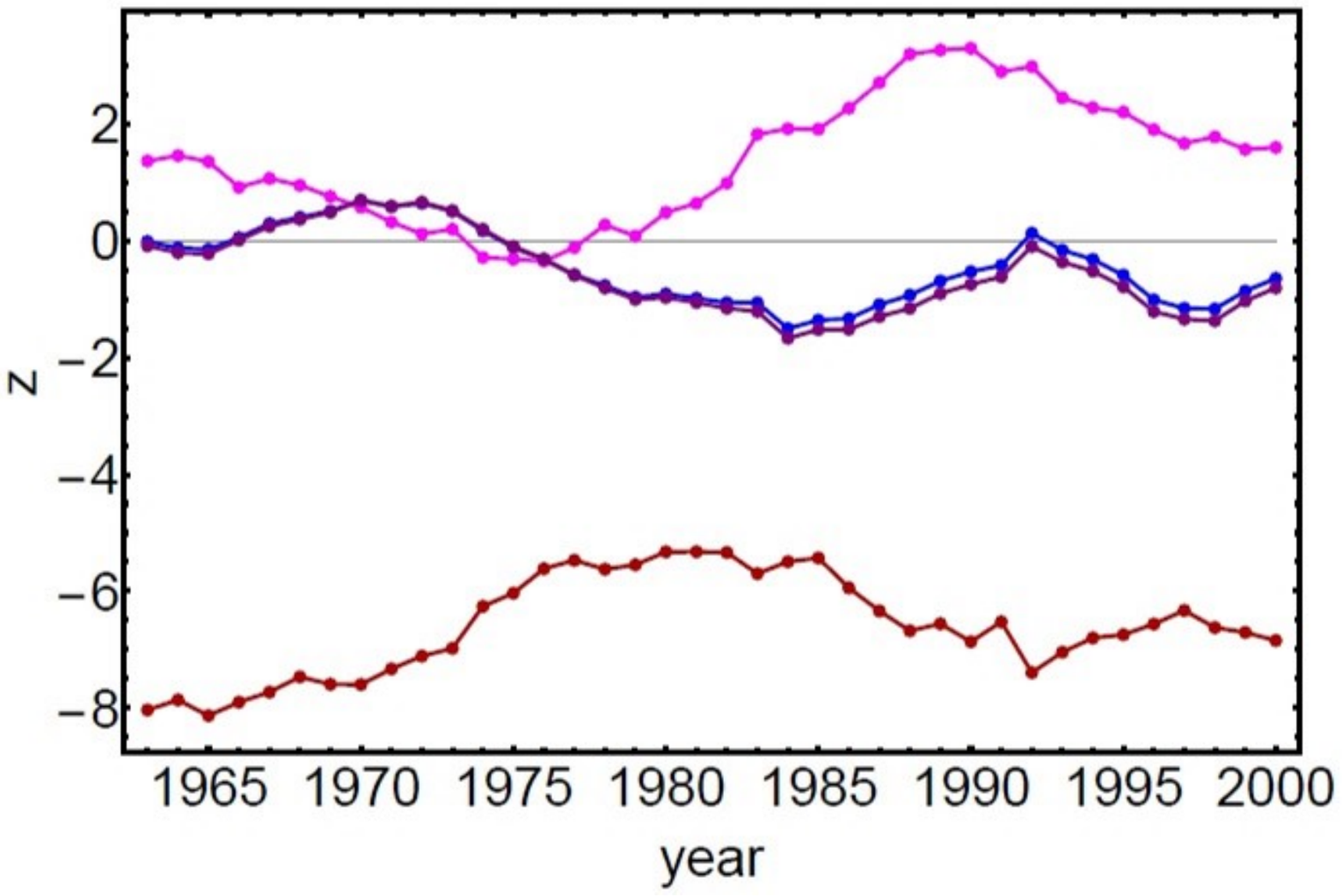}
\caption{Upper panel: the binary, undirected, bipartite representation of the WTW in the year 2000. Upon reordering the rows and columns according to the fitness-complexity algorithm introduced in \citep{tacchella2012ec}, a triangular pattern clearly emerges. Middle panel: matrix drawn from the ensemble induced by the BiCM for the same year and ordered according to the same criterion. Lower panel: evolution of the z-score for the assortativity index (brown), the NODF (blue) and its ``reduced'' versions along rows (magenta) and columns (purple). Source:  \citep{saracco2015randomizing}.}
\label{fig5}
\end{figure}

\paragraph*{Nestedness.} The upper panel of figure \ref{fig5} shows the nestedness (measured by NODF) of the bipartite WTW in the year 2000; the lower panel, instead, shows the performance of the \emph{Bipartite Configuration Model} (BiCM - see also Appendix \ref{appb}) in reproducing it \citep{saracco2015randomizing}. The empirical and the expected value of the NODF are compared via a z-score, defined as

\begin{equation}
z_\text{NODF}=\frac{\text{NODF}-\langle\text{NODF}\rangle_\text{BiCM}}{\sigma_\text{NODF}}
\end{equation}
and quantifying the difference between them in units of standard deviation\footnote{An empirical value $X^*$ corresponding to a largely positive (negative) value of $z_{X^*}$ is assumed to indicate that the quantity $X$ is over(under)-represented in the data, hence not explained by the model itself. More precisely, for a quantity that is normally distributed under a given model, values falling outside the intervals $z_X=\pm1$, $z_X=\pm2$, $z_X=\pm3$ occur with a probability of 32\%, 5\%, 1\%, respectively.}. As it can be appreciated, $-1<z_\text{NODF}<1$, even if values $z_\text{NODF}\simeq \pm2$ are, sometimes, reached. Similar results are found in \citep{borras2019breaking} where ecological networks are considered. These results seem to indicate that the information encoded into the degree sequence is indeed enough to explain the empirical nestedness observed in economic and ecological systems -- see however \citep{bruno2020nested}.

\paragraph*{Reciprocity.} Contrarily to what happens for other binary properties, the DCM often fails in reproducing the empirical values of reciprocity. An example is provided by the Dutch Interbank Network (DIN), whose observed and expected reciprocity are compared in \citep{squartini2013early}. While during the first seven years covered by the dataset, the reciprocity structure of the network (inspected via its \emph{dyadic} structure - see also chapter \ref{chap3}) is still consistent with the DCM prediction, the remaining three years are characterized by an increasing difference between the values $r$ and $\langle r\rangle_\text{DCM}$.

Although this signals that the structure of the DIN cannot be fully explained by the degree heterogeneity, it also highlights the versatility of the models defined within the ERG framework: in case the chosen amount of information is not capable of reproducing the observations, it can still be used to define a \emph{null model}, i.e. a benchmark against which comparing the empirical patterns \citep{cimini2019statistical}. Alternatively, a more refined reconstruction model can be defined; in this case, explicitly constraining reciprocity together with degrees defines the \emph{Reciprocal Configuration Model} (RCM) \citep{squartini2011analytical,garlaschelli2006multispecies}.

\section{Quantifying systemic risk}

As an additional test of the goodness of the discussed reconstruction methods, let us consider the problem of quantifying systemic risk. This problem became extremely relevant since the aftermath of the financial crisis. Systemic risk is rooted in the evidence that the complex patterns of interconnections between financial institutions have the potential to make the system \emph{as a whole} extremely fragile, as these connections constitute the channels through which financial distress can spread \citep{gai2010contagion,haldane2011systemic,acemoglu2015systemic,battiston2016complexity,bardoscia2017pathways}. As a consequence, both researchers and regulators have paid increasing attention to inferring the structural features of financial systems, with the aim of properly estimating the \emph{systemicness} of an institution \citep{squartini2018reconstruction}: intuitively, a systemically-important institution adversely affects a large number of other institutions in case of default \citep{komatsu2012distributed}. In what follows, we will provide two illustrative examples of systemic risk estimators.\\

\paragraph*{Systemic risk for monopartite networks.} In the case of monopartite interbank networks, stress tests have been typically performed using the propagation of a shock as a consequence of the default of an institution. This is simulated by 1) deleting the defaulted institution and its connections from the network, 2) checking the impact of such an event on the other nodes, 3) repeating the deletion step if other defaults have happened as a consequence of the previous step \citep{furfine2003interbank}.

It is, however, of greater interest to check the level of \emph{distress} of an institution, i.e. its ``closeness'' to default. A compact measure in this direction is provided by the \emph{DebtRank} (DR) indicator \citep{battiston2012debtrank}. 
To obtain this indicator the starting point is the balance sheet equation governing the financial situation of a bank $i$, namely $E(i)=a(i)-l(i)$ where $E(i)$ is the value of $i$'s equity, $a(i)$ the value of its assets and $l(i)$ the value of its liabilities. The DR of bank $i$, denoted as $h(i)$, is equal to $h(i)=0$ if the bank is ``healthy'', i.e. its equity is positive and has not suffered any losses; if $h(i)=1$, bank $i$ is defaulted, i.e. its equity is zero; the intermediate values $0<h(i)<1$ correspond to different levels of distress (banks are not defaulted yet, but are ``closer'' to default as a consequence of a propagating shock).

Given the (known or reconstructed) weighted, directed adjacency matrix $\mathbf{W}$ of the network, let us call $\mathbf{E}_0$ the vector of banks equities at time $t=0$, $\mathbf{E}_1$ the vector of banks equities at time $t=1$ and $\tau$ the total amount of time during which the system dynamics is observed. The algorithm to calculate the DR index works as follows \citep{bardoscia2015debtrank}:

\begin{itemize}
\item[$\bullet$] the equity of all banks is assumed to be affected by an external shock: as a consequence, $E_1(i)<E_0(i),\:\forall\:i$;
\item[$\bullet$] the \emph{relative equity loss} of each bank is $h_1(i)=\frac{E_0(i)-E_1(i)}{E_0(i)}>0$, which measures its level of distress (even if the bank is not defaulted, it has become ``closer'' to default as a consequence of the equity reduction);
\item[$\bullet$] a distressed bank $j$ is less likely to meet its obligations; thus, the distress of bank $j$ becomes a distress for each bank $i$ that lent money to $j$, i.e., for which $w_{ij}>0$.
\item[$\bullet$] the overall distress bank $i$ receives at a generic time $t$ can be calculated as

\begin{equation}
\Delta_t(i)=\sum_{j\neq i}\Lambda_{ij}\cdot(h_t(j)-h_{t-1}(j))
\end{equation}
where $\Lambda_{ij}=\frac{w_{ij}}{E_0(i)},\:\forall\:i\neq j$ is the so-called \emph{leverage matrix} and $h_t(i)=\frac{E_{t-1}(i)-E_t(i)}{E_{t-1}(i)}$. As a consequence, the state of bank $i$ is updated according to the rule

\begin{equation}
h_{t+1}(i)=\min\{1,h_t(i)+\Delta_t(i)\};
\end{equation}
when $h(i)\geq1$ bank $i$ is defaulted (and remains in the ``default'' state at all subsequent time steps).
\end{itemize}

The DR algorithm outputs the list $h_t(i),\:\forall\:i$ for each time step $t=1\dots\tau$; in addition, it also outputs a global index, i.e. the ``group'' DebtRank, which is defined as the weighted average of the relative equity losses, i.e.

\begin{equation}
\text{DR}_t=\sum_ie(i)\cdot(h_t(i)-h_1(i)),\:\forall\:t
\end{equation}
with $e(i)=\frac{E_0(i)}{\sum_{j}E_0(j)}$. In other words, $h(i)$ measures the economic value of node $i$ that is potentially lost because of distress: notice that this is done in a recursive fashion, to properly account for reverberation effects. \\

\begin{figure}[t!]
\includegraphics[width=\textwidth]{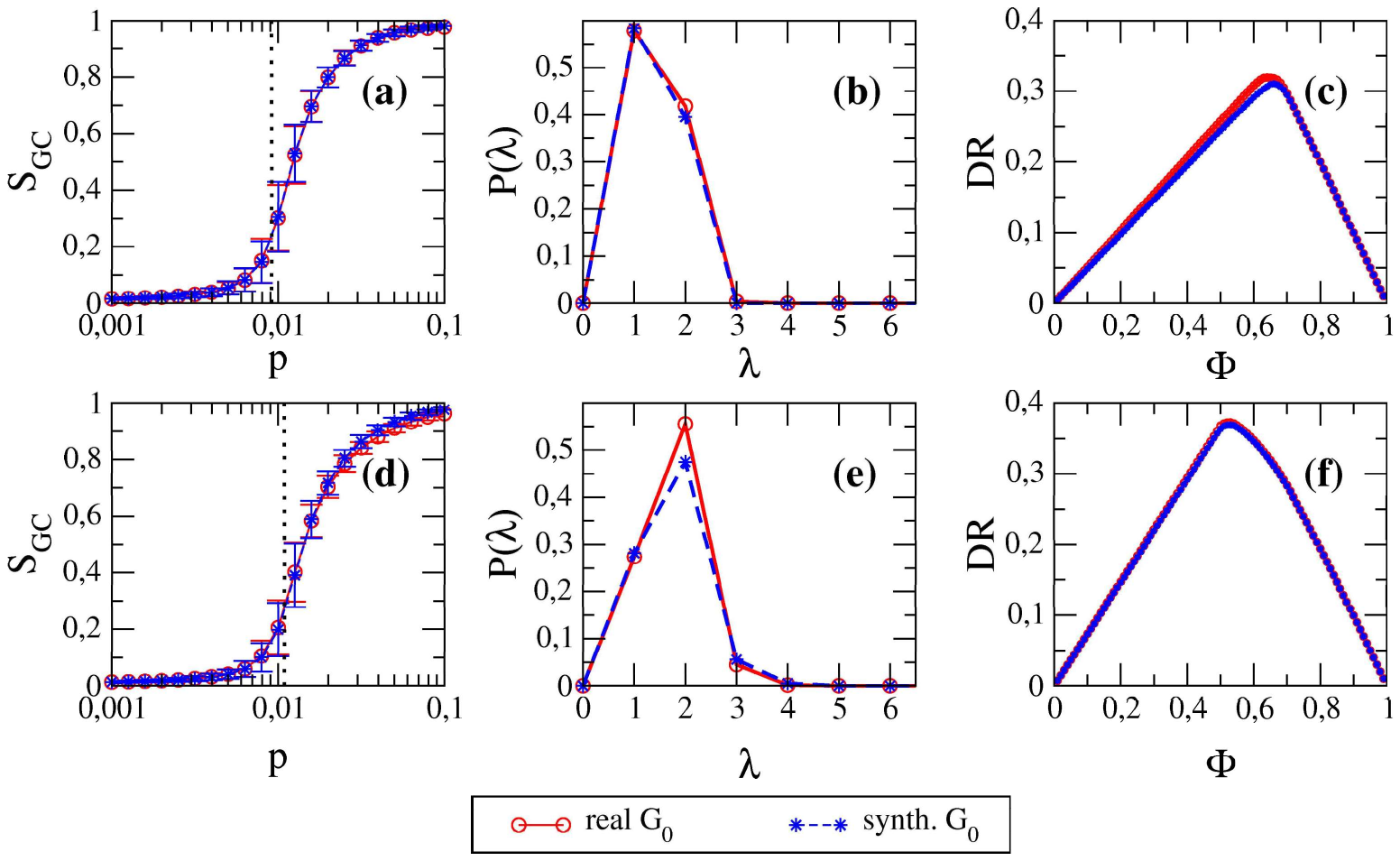}
\caption{Properties of real and reconstructed networks (obtained by implementing the dcGM). Left plots (a,d): evolution of the size of the giant component as a function of the occupation probability. Central plots (b,e): empirical probability distribution of the directed shortest path length. Right plots (c,f): dependence of the DR on the initial distress $\Phi$. Top panels (a,b,c) refer to WTW, bottom panels (d,e,f) to e-MID. Source: \citep{cimini2015systemic}.}
\label{fig6}
\end{figure}

\begin{figure}[t!]
\centering
\includegraphics[width=0.75\textwidth]{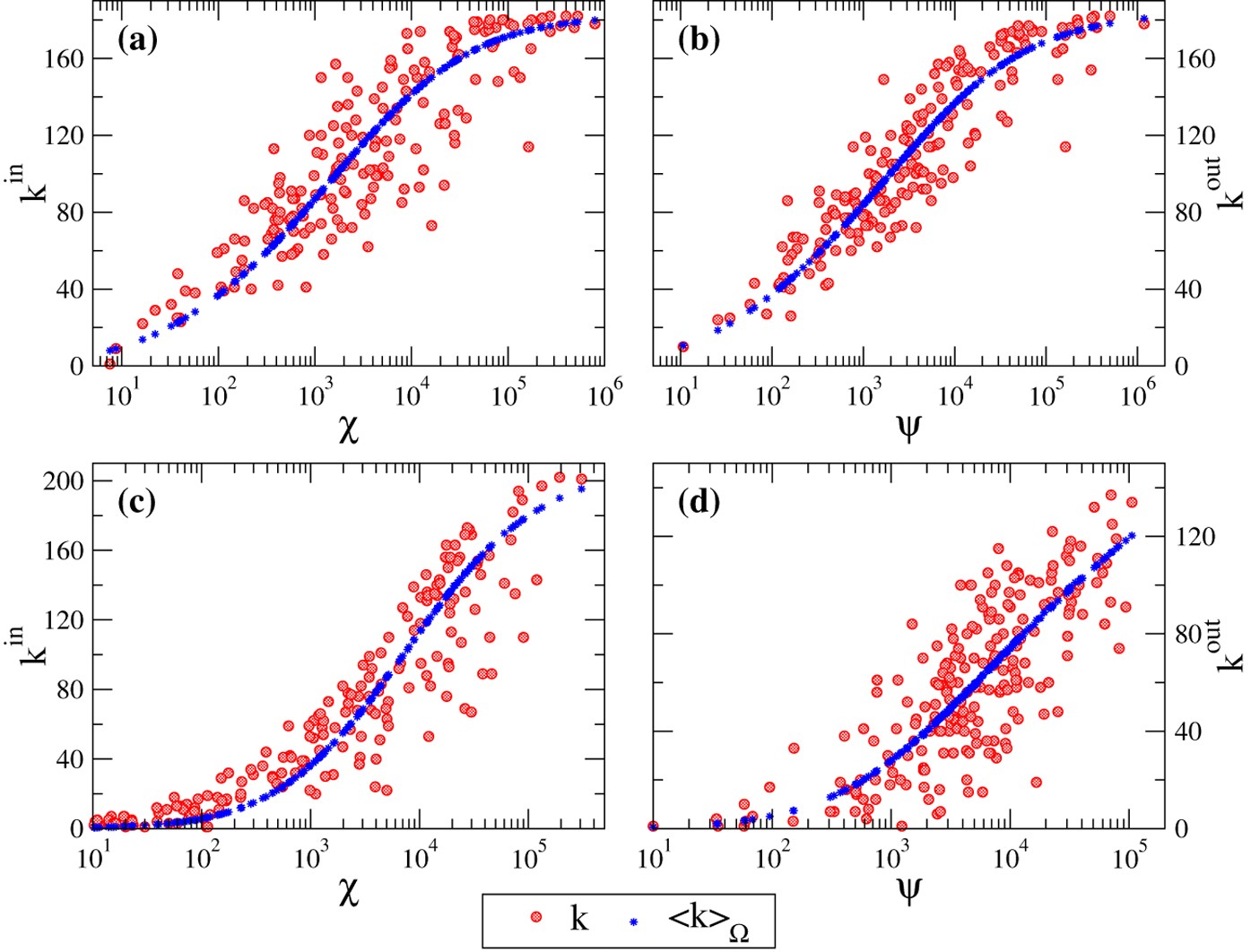}
\caption{Qualitative assessment of the fiCM ansatz: Scatter plots of node fitnesses $\chi$, $\psi$ (the strengths) versus real node in- and out-degrees (red circles) and their fiCM ensemble averages (blue asterisks). Upper panels (a,b) refer to WTW, lower panels (c,d) to eMID. Source: \citep{cimini2015systemic}.}
\label{fig7}
\end{figure}

Figure \ref{fig6} shows the performance of the dcGM in reproducing the DR index on a snapshot of the WTW and e-MID (beside other two important macroscale properties): the agreement between the dcGM-induced trend and the observed one is remarkable. As can be seen from Figure \ref{fig7}, this happens because of the high correlation between node degrees and strengths (used as fitnesses in the fiCM). Further analyses also show that the dcGM effectively estimates the DR also when the available information is minimal (i.e. a small percentage of nodes is used to estimate the overall density of the network) \citep{musmeci2013bootstrapping,squartini2017network}.\\

\paragraph*{Systemic risk for bipartite networks.} When considering bipartite networks as those of portfolio holdings by financial institutions, the focus is on the risk related to the sale of illiquid assets and the subsequent losses during fire-sales \cite{shleifer2011fire,caccioli2014stability,cont2016fire,gualdi2016statistically}. The estimation of this kind of risk can be done by adopting the \emph{systemicness} index $S_i$ as a measure of the impact of a financial institute $i$ on the whole system \citep{greenwood2015vulnerable}:

\begin{equation}\label{eq.systemicness}
S_i(\mathbf{W})=\frac{\Gamma_iV_i}{E}B_ir_i,\:\forall\:i.
\end{equation}

The systemicness index is a function of $\Gamma_i=\sum_j\sum_\alpha l_\alpha(w_{i\alpha}w_{j\alpha}),\:\forall\:i$, that quantifies the overlap of portfolio $i$ with other portfolios, the illiquidity parameter $l_\alpha$ of asset $\alpha$, the leverage $B_i$ and the portfolio return $r_i$, the total equity of the system $E$ and the portfolio value (i.e. the strength) of $i$, $V_i=\sum_\alpha w_{i\alpha}$. In order to simplify the estimation of the systemicness index, we can assume the presence of homogeneous shocks, as well as identical illiquidity parameter for all assets. Upon doing so, the ratio between the expected and the observed value of systemicness becomes solely defined in terms of quantities that can be readily estimated, i.e. the link weights:

\begin{equation}\label{eq.systemicness2}
\frac{\langle S_i\rangle}{S_i(\mathbf{W})}=\frac{\sum_j\sum_\alpha\langle w_{i\alpha}\rangle\langle w_{j\alpha}\rangle}{\sum_j\sum_\alpha w_{i\alpha}w_{j\alpha}},\:\forall\:i.
\end{equation}

\begin{figure}[t!]
\includegraphics[width=0.75\textwidth]{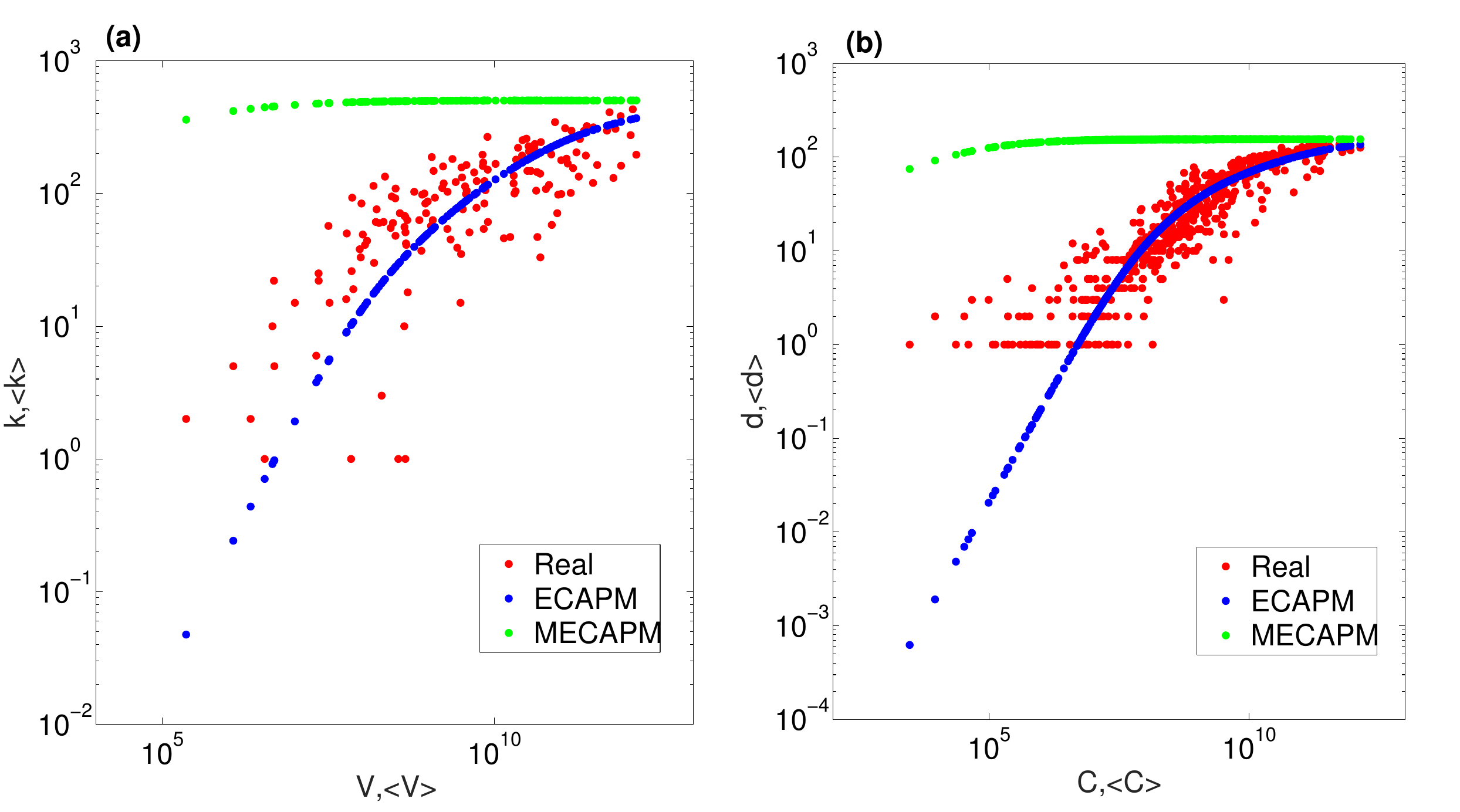}
\caption{Performance of the (bipartite version of the) fiCM  and MECAPM in reproducing the degrees of the nodes defining the Security Holding Statistics (SHS) network. Source: \citep{squartini2017stock}.}
\label{fig8}
\end{figure}

Not surprisingly, the (bipartite versions of the) dcGM \citep{squartini2017stock} and the MECAPM  \citep{digangi2018assessing} reconstruct the same expected value of systemicness, which is a natural consequence of the fact that the expected values of weights, under the two models, coincide. This happens despite the fiCM can well reproduce degrees while MECAPM cannot - see fig. \ref{fig8}.

However, results again differ when single network instances are drawn from the corresponding ensembles. Let us, in fact, replace the ensemble averages in eq. (\ref{eq.systemicness2}) with the single-instance values of the weights (i.e. the weights of \emph{a} particular configuration drawn from the dcGM or MECAPM ensembles), i.e.

\begin{equation}
\frac{\tilde S_i}{S_i(\mathbf{W})}=\frac{\sum_j\sum_\alpha\tilde{w}_{i\alpha}\tilde{w}_{j\alpha}}{\sum_j\sum_\alpha w_{i\alpha}w_{j\alpha}},\:\forall\:i
\end{equation}

\begin{figure}[t!]
\includegraphics[width=0.75\textwidth]{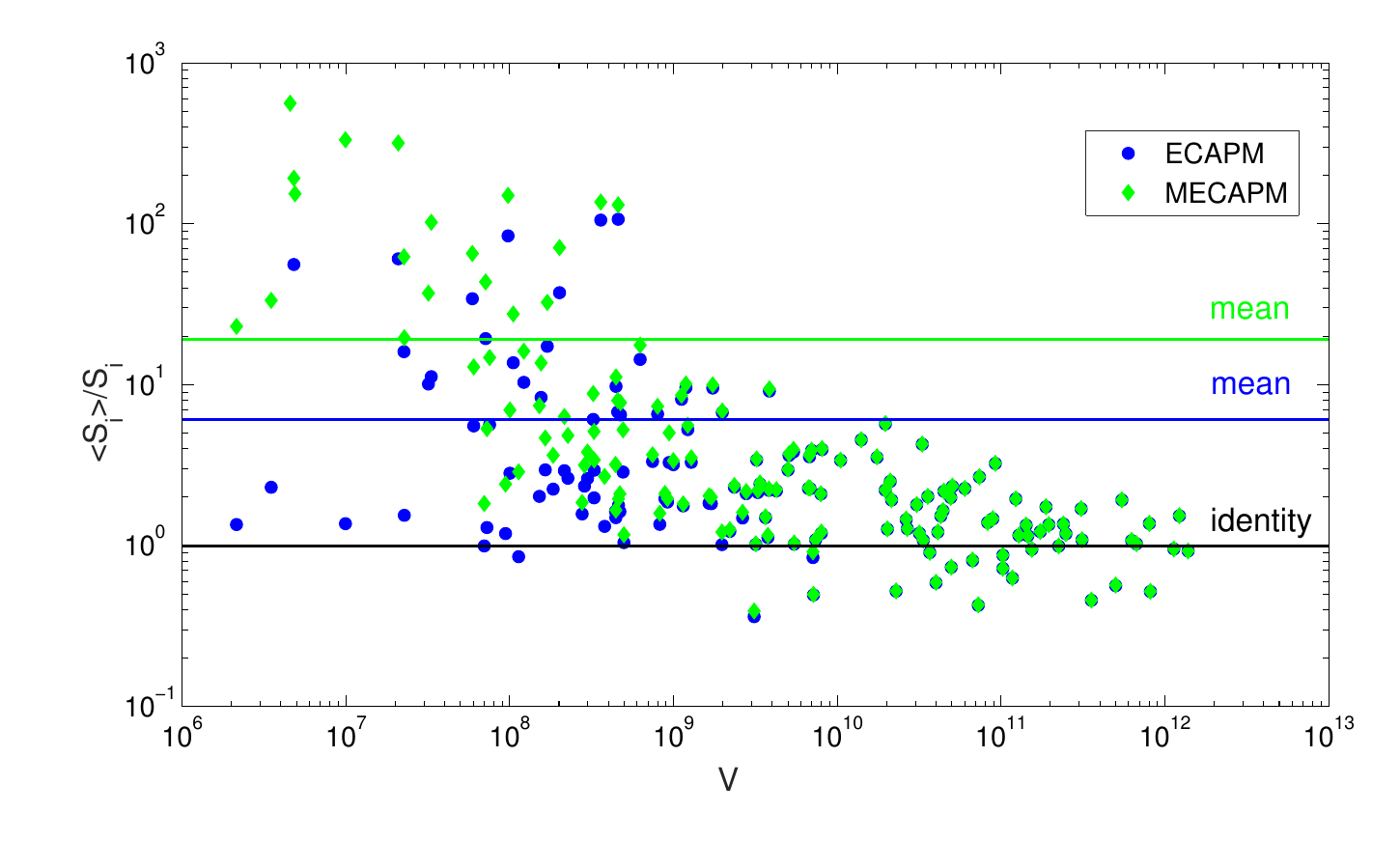}
\caption{Reconstruction of systemicness in the SHS dataset: each point represents the relative systemicness $\frac{\tilde S_i}{S_i(\mathbf{W})}$ of country-sector $i$ scattered versus its strength (here indicated with $V_i$), for a particular configuration drawn from the (bipartite) dcGM ensemble (blue dots) and the (bipartite) MECAPM ensemble (green dots). Averages are shown for both methods as horizontal solid lines. Source: \citep{squartini2017stock}.}
\label{fig9}
\end{figure}

As fig. \ref{fig9} shows, while the estimates provided by the two methods coincide for the largest nodes, the MECAPM tends to overestimate the systemicness of small nodes. As an additional test, let us consider the ratio of the ensemble standard deviations of the systemicness index, i.e. $r_{S_i}=\frac{\sigma_{S_i^{\text{dcGM}}}}{\sigma_{S_i^{\text{MECAPM}}}},\:\forall\:i$: it is found to be smaller than 1 for the 90\% of nodes. 

The explanation for such a difference between dcGM and MECAPM resides in the different errors affecting their estimates of a generic link weight $w_{ij}$: in formulas, the ratio $r_{ij}$ between the two errors is

\begin{equation}
r_{ij}=\frac{\sigma_{\hat{w}_{ij}^{\text{dcGM}}}}{\sigma_{\hat{w}_{ij}^
{\text{MECAPM}}}}\simeq\sqrt{\frac{1}{p_{ij}^\text{fiCM}}-1},\:\forall\:i\neq j.
\end{equation}

Notice the key role played by topology in lowering the uncertainty affecting the estimation of weights: requiring $r_{ij}<1$ is, in fact, equivalent at requiring $p_{ij}^\text{fiCM}>1/2$, further implying that the estimation of larger weights provided by the dcGM is less affected by uncertainty \citep{squartini2017network}. 

These results are particularly relevant for the estimation of systemic risk: since larger weights drive larger shocks, it is desirable to employ a model providing accurate estimations for them; hence, employing the (bipartite version of the) dcGM to generate possible scenarios seems to allow obtaining configurations over which the estimation of systemic risk provides closer values to the actual one \citep{squartini2017stock}.

\chapter{Network reconstruction at the mesoscale}
\label{chap3}

\paragraph{Reconstructing vs testing.} When studying the mesoscale structure of a network, it is fundamental to distinguish two different, yet complementary, methodological approaches: the one seeking for the best \emph{reconstruction} model and the one seeking for the best \emph{benchmark} model. In fact, testing the statistical significance of a network quantity requires: (i) to build the benchmark by choosing some (usually local) properties of the real network to be preserved while randomizing everything else; (ii) to compare the empirical value of the chosen quantity with its value according to the benchmark. Finding that the considered property is \emph{not statistically significant} ultimately means that the (local) information that has been preserved to define the benchmark is enough to explain that property: therefore, we can conclude that it is possible to \emph{reconstruct} such a property by just knowing the aforementioned local information. On the other hand, finding \emph{a statistically significant} discrepancy between the empirical value of the considered quantity and its randomized counterpart implies that additional information is required to explain it. As we observed in the previous chapter, the knowledge of \emph{binary local information} (i.e. the degree of nodes) allows one to reconstruct several higher-order properties, while the same kind of information in the weighted case (i.e. the strength of nodes) fails. In this section we will focus only on the binary part by asking \emph{to what extent the knowledge of binary local information allows one to reconstruct the mesoscale structure of networks}

\section{Motifs: the building blocks of networks}

In the previous chapter we have introduced the important concept of \emph{motif}, i.e. a specific interaction pattern involving a small number of nodes. 
The first work about motifs appeared in a paper by \citet{shen2002network} studying the gene regulation network of the bacteria \emph{E. Coli}. In the same year  \citet{milo2002network} presented a detailed study of the motifs  in different types of real-world networks: the authors showed that these subgraphs could help in identifying \emph{classes} of networks. Since then, an increasing body of literature has been devoted to the study of motifs in biological and neural networks \citep{lee2002transcriptional,yeger2004network,sporns2004motifs,cloutier2011dynamic,stouffer2012evolutionary,lim2013design,chen2013identification,messe2018toward}, economic systems  \citep{ohnishi2010network,squartini2012triadic,saracco2015randomizing,saracco2016detecting} and, more in general, to quantitatively characterize networks of different nature via the analysis of sub-structures  \citep{watts1998collective,sinatra2010networks,jiang2013review,schneider2013unravelling,stone2019network}.

\paragraph*{Dyadic motifs.} In a directed binary network the simplest motifs are represented by all possible subgraphs of dimension 2. There are four possibilities according to the presence and the direction of links between the two nodes.
Given the adjacency matrix of the network, it is possible to compactly write the occurrence of each dyadic pattern as follows:
\begin{eqnarray}\label{trimot}
a^\rightarrow_{ij}&\equiv&a_{ij}(1-a_{ji}),\\
a^\leftarrow_{ij}&\equiv&a_{ji}(1-a_{ij}),\\
a^\leftrightarrow_{ij}&\equiv&a_{ij}a_{ji},\\
a^\nleftrightarrow_{ij}&\equiv&(1-a_{ij})(1-a_{ji})
\label{trimot2}
\end{eqnarray}

\begin{figure}[t!]
\centering
\includegraphics[width=\textwidth]{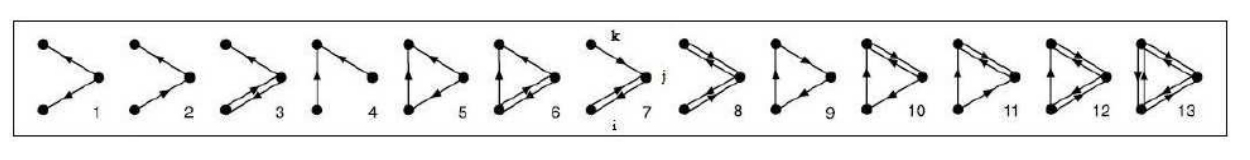}
\includegraphics[width=\textwidth]{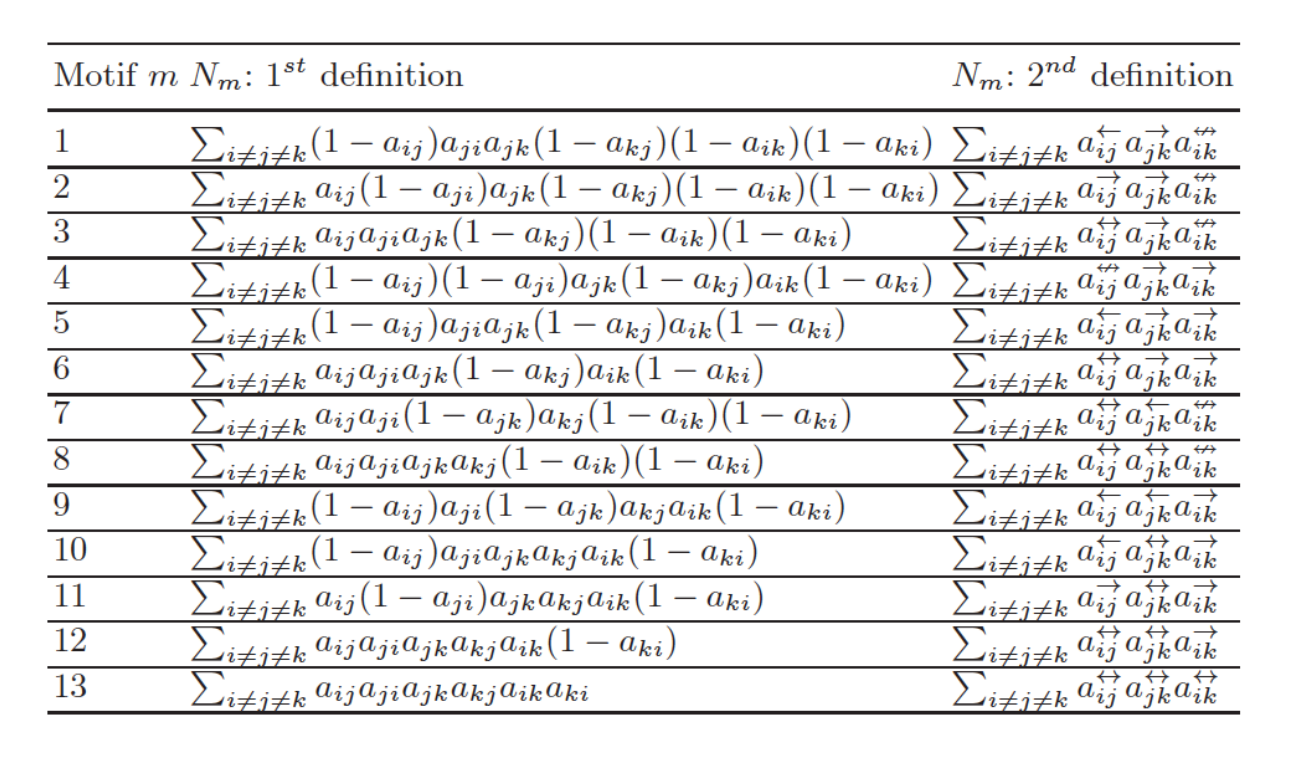}
\caption{Top: graphical representation of the triadic, binary, directed motifs. Bottom: classification and definition of triadic motifs. Source: \citep{squartini2012triadic}.}
\label{Motifs}
\end{figure}

Dyadic motifs thus refine the concept of reciprocity, a measure that simply counts the number of bilateral links with respect to the total number of links. And, despite their simplicity, these patterns can well characterize strongly symmetric networks such as the WTW as well as asymmetric networks such as the investments network of countries \citep{duenas2017spatio}.

\paragraph*{Triadic motifs.} Evidently, motifs with more than 2 nodes are of higher interest, but are also more complex to count as the increasing number of node implies an exponential increase of the possible patterns according to the presence/absence of links and their directionality. Indeed, most of the studies in literature focus on motifs of size 3 or 4. Despite their small size, triadic motifs offer interesting insights on network organization. Indeed, they can be considered as the natural extension of the directed clustering coefficient (number of closed triangles over all possible triplets of nodes) and represent a first step towards the deepest exploration of the network organization in communities \citep{kashtan2005spontaneous}: the table depicted in fig. \ref{Motifs} shows how to compute the occurrence of the 13 triadic motifs with the classical and the compact representation given by eqs. (\ref{trimot})-(\ref{trimot2}).

By their own nature, real networks do not necessarily contain all the triadic motifs described in the table. Indeed, \citet{milo2002network} defined the motifs as ``simple building blocks'' of networks, closely related to their specific  functioning: the ability of processing information.

\paragraph*{Motifs in biological networks.} \citet{milo2002network} identified classes of real networks by looking at the occurrence of motifs of size 3 and 4 with respect to properly-defined null-models. In particular, they tested the significance of such occurrences  by employing null models preserving (i) the out- and in-degrees of nodes and (ii) the occurrence of all subgraphs of size $n-1$ when testing the significance of subgraphs of size $n$. This second requirement amounts at preserving the number of diadic motifs when testing the significance of triadic motifs, and the occurrence of all the 13 triadic motifs when testing the significance of motifs of size 4. This choice allows to filter out the effect of  possible significant occurrences of smaller order motifs.
The authors found that different kinds of networks are characterized by different patterns. They considered two transcriptional regulation gene networks (nodes are genes, links are directed from a gene that encodes a transcriptional factor to a gene regulated by it), finding only 2 significant motifs over the 13 triadic motifs and the 199 4-size motifs (directed networks): the feed-forward loop and the bi-fan. 
The outcome for foodwebs (nodes are species and links are directed from predators to preys) of seven different ecosystems revealed the different nature of these networks, with the feed-forward loop always underrepresented while the three chain and the bi-parallel motifs being significantly shared among them. The neuronal network (nodes are neurons connecting through synapsis) of the \emph{C. Elegans} nematode shared two motifs with the gene networks (the feed forward loop and the bi-fan) and one with the foodwebs (the bi-parallel one). 
These results stress the great variability of patterns observable in real networks and their fundamental role in shaping their structural organization and functioning.

\begin{figure}[t!]
\centering
\includegraphics[width=\textwidth]{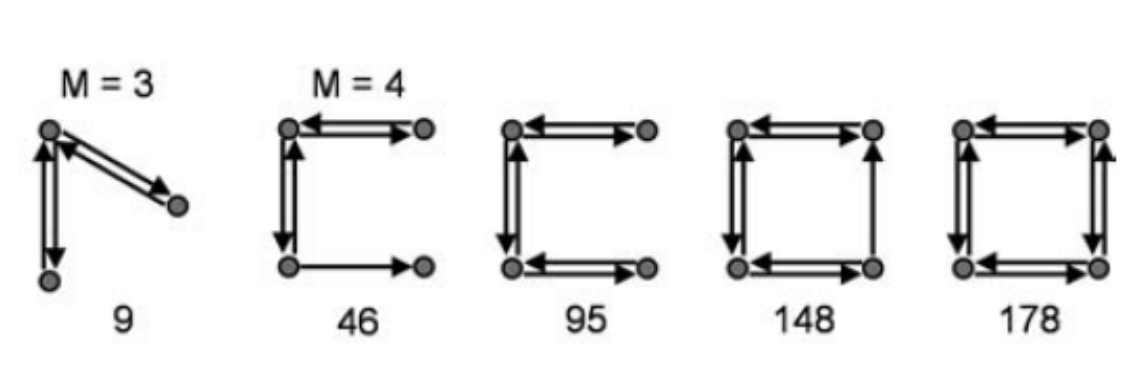}
\caption{Significant motifs in the structural brain network. Numbers indicate the type of motif. Source: \citep{sporns2004motifs}.}
\label{MotRes}
\end{figure}

Another relevant work in this direction was developed in the field of neuroscience. \citet{sporns2004motifs} studied the structural motifs in  brain networks (nodes are brain regions while directed links represent anatomical connections) of macaques, cats and \emph{C. Elegans}. 
They assessed the occurrence of motifs against a null model preserving the degree distribution \citep{maslov2002specificity} and lattice networks\footnote{The \citet{maslov2002specificity} randomization procedure is based on link swaps, which occur only if there are non zero entries closer to the main diagonal in the resulting adjacency matrix: this approach produces network structures which are similar to rings or lattices.}. The second null model washes away the effect of high local clustering on motifs occurrence. For large-scale cortical networks (macaque, cat), the authors found only one statistically significant 3-nodes subgraph (motif 9 in fig. \ref{MotRes}) and its expanded versions for 4-nodes patterns. These motifs are relevant for explaining the cortical functional organization as they combine the two main principles of \emph{segregation} and \emph{integration}: indeed, they contain reciprocal chains with unconnected end nodes. The result is different for the invertebrate \emph{C. Elegans}: other motifs appear statistically significant instead of motif 9. This outcome allows classifying the large-scale cortical networks in a separate family of brain networks with respect to the neuronal network of the nematode. In reconstruction terms, these findings imply that local node properties are not able to explain the principles driving the functional organization of brain networks also in the simplest known neuronal system (i.e. \emph{C. Elegans}): these networks cannot be reconstructed using local information only.

\paragraph{Motifs in economic and financial networks.} More recently, \citet{squartini2012triadic} and \citet{squartini2013early} tackled the problem of motifs reconstruction in economic networks using the ERG framework. They used the \textit{z-score} $z_X=\frac{X-\langle X\rangle}{\sigma[X]}$ where $X$ is the occurrence of (dyadic and triadic) motifs, while $\langle X\rangle$ and $\sigma[X]$ are its expected value and standard deviation over the ensembles induced by either the \emph{Directed Configuration Model} (DCM) and the \emph{Reciprocal Configuration Model} (RCM). 

\begin{figure}[t!]
\centering
\includegraphics[width=\textwidth]{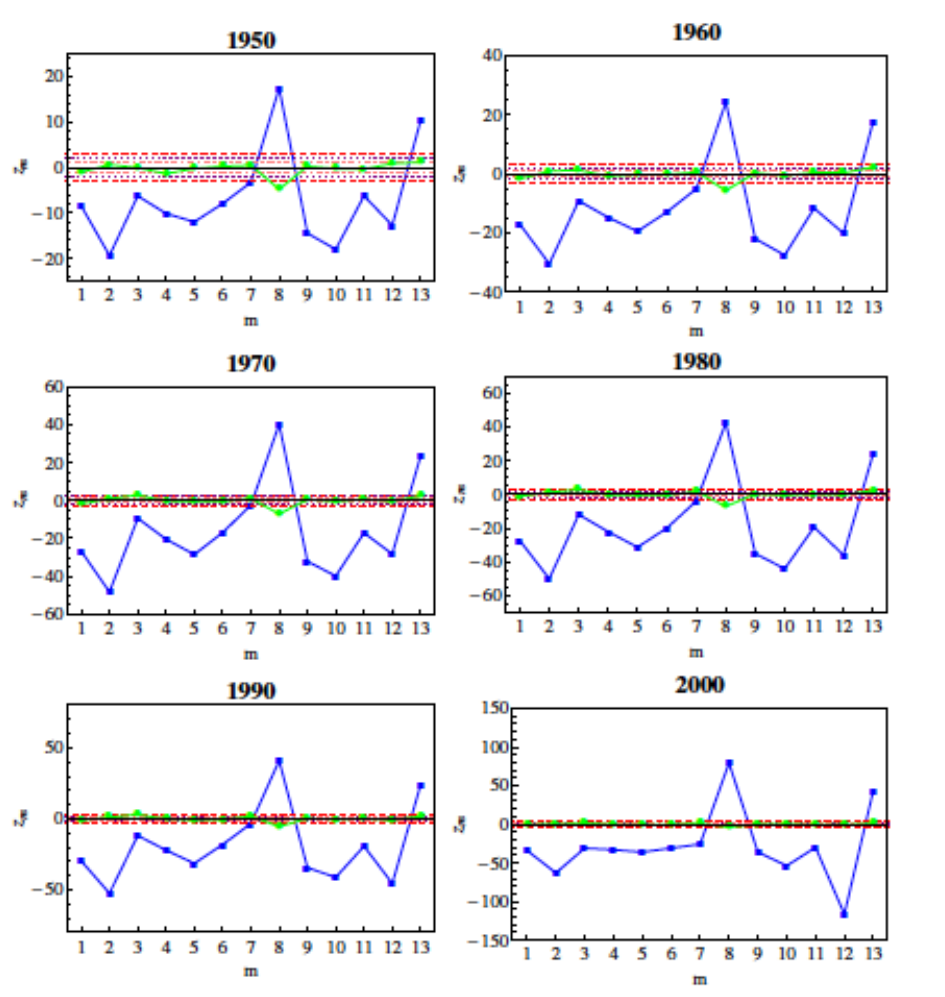}
\caption{Evolution of the z-scores profiles for the 13 triadic motifs defined in fig. \ref{Motifs}, under the DCM (blue line) and the RCM (green line): while the DCM is not able to reproduce the abundance of motifs, the RCM succeeds in predicting a number of triadic motifs that is \emph{not} significant under it. Source: \citep{squartini2012triadic}.}
\label{MotRes2}
\end{figure} 

\begin{figure}[t!]
\centering
\includegraphics[width=\textwidth]{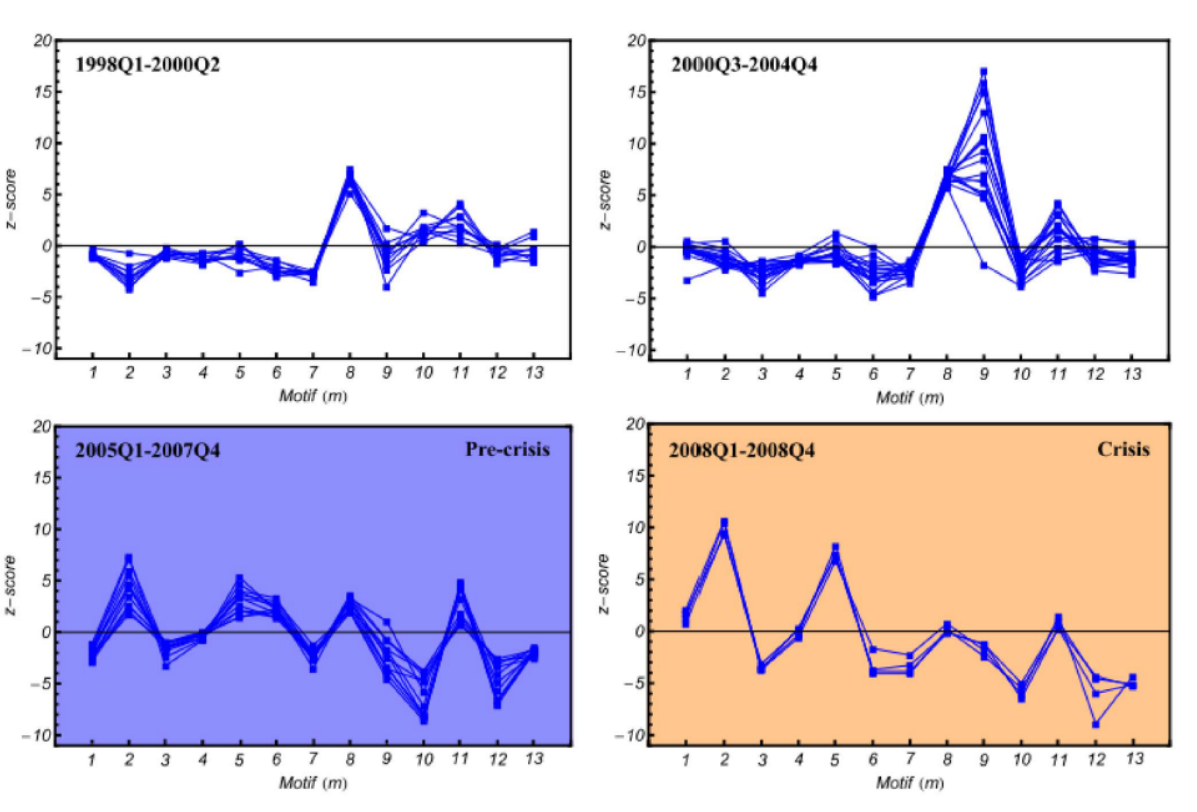}
\caption{Evolution of the z-scores profiles for the 13 motifs defined in fig. \ref{Motifs}, under the DCM: notice the presence of 4 different, temporal profiles characterizing the DIN. Source: \citep{squartini2013early}}
\label{MotRes4}
\end{figure}

The authors started with the WTW, an interesting case-study being the result of a self-organization process driven by the global economy. As fig. \ref{MotRes2} shows, the number of triadic motifs cannot be reproduced by simply knowing local information (i.e. under the DCM). Indeed, independently on the year under study, the number of all motifs, except two, is underestimated in the DCM reconstructed networks. It is not surprising that the two overestimated patterns are those characterized by reciprocated links only (motifs 8 and 13 in fig. \ref{Motifs}). In the attempt to find the minimum amount of information able to control for the triadic properties, the authors then considered the RCM: remarkably, the dyadic information seems enough to reproduce all motifs except one (which in this case appears slightly overestimated).

The same authors found a different result when considering a financial system, i.e. the Dutch Interbank Network (DIN) (nodes are banks and links represent loans among them). As fig. \ref{MotRes4} shows, a first result concerns the existence of four different temporal profiles describing the triadic motifs occurrence and identifying important periods related to the 2008 global financial crisis. Looking at the temporal evolution of z-scores, clear early-warning signals of the topological collapse can be appreciated both under the DCM and the RCM. An interesting interpretation for the temporal evolution of motif 9 is also provided: its over-representation during the period 2000-2004 signals a sort of ``cyclic anomaly''. This non-reciprocated loop, in fact, represents a risky configuration that may have destabilized the system well before the crisis, leading to a lack of trust between banks during the years 2005-2007. According to this viewpoint, the crisis represented the ending point of a process that was involving the entire Dutch system for years.

\paragraph*{Bipartite motifs.} The concept of motif can be also extended to bipartite networks. As explained in the previous chapter, in a bipartite network there are two disjoint and independent sets of nodes, and links can exist only between and not within the two sets. It is evident that odd cycles of any length are absent, therefore nor the clustering coefficient neither the standard triadic motifs can be observed in such systems. However, in the same spirit of the monopartite case, it is possible to define a class of motifs able to capture the higher-order correlations between nodes in bipartite networks.

There are examples in different fields of study. For instance, \citet{baker2015species} identified 44 motifs in a bipartite host-parasitoid foodweb made up of 2-6 species and uniquely-identified positions. \citet{saracco2015randomizing} considered the bipartite version of the WTW (i.e. the set of world countries and the set of products exported), and introduced the V- and $\Lambda$-motifs, respectively counting how many couple of countries export the same product and how many couple of products are in the basket of the same country. In other words, the V-motif measures the correlation among producers, while the $\Lambda$-motif focuses on correlations between products (see fig. \ref{BipMot2}, top):

\begin{figure}[t!]
\centering
\includegraphics[width=0.4\textwidth]{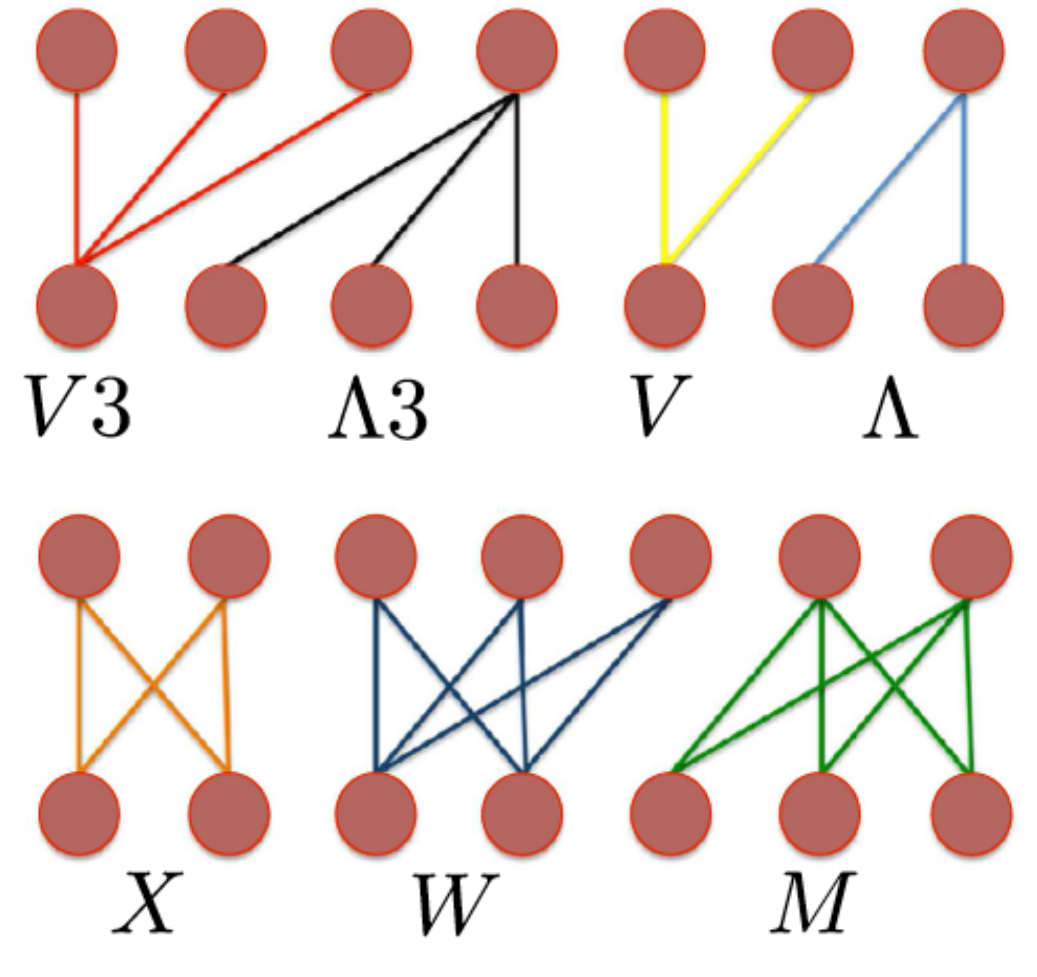}
\caption{Undirected bipartite motifs containing up to 5 nodes. Top: $V$- and $\Lambda$-motifs and their generalizations to 3 links. Bottom: $X$, $W$ and $M$ motifs. Source: \citep{saracco2015randomizing}.}
\label{BipMot2}
\end{figure}

\begin{eqnarray}\label{Vm}
N_\text{V}(\mathbf{B})&=&\sum_i\sum_{j>i}C_{ij}=\sum_i\sum_{j>i}\sum_\alpha b_{i\alpha}b_{j\alpha}=\sum_\alpha \binom{h_\alpha}{2},\\
N_\Lambda(\mathbf{B})&=&\sum_\alpha\sum_{\beta>\alpha}P_{\alpha\beta}=\sum_\alpha\sum_{\beta>\alpha}\sum_i b_{i\alpha}b_{i\beta}=\sum_i\binom{k_i}{2}
\label{Lm}
\end{eqnarray}
where $b_{i\alpha}$ is the generic entry of the biadjacency matrix (see Appendix \ref{appb}). While $C_{ij}$ counts the number of products exported by both countries $i$ and $j$, $P_{\alpha\beta}$ counts the number of countries exporting both products $\alpha$ and $\beta$. Notice that the abundance of V- and $\Lambda$-motifs can be compactly written in terms of node degrees  $k_i(\mathbf{B})=\sum_\alpha b_{i\alpha}$ (also called \textit{diversification} and measuring the number of products exported by each country) and $h_\alpha(\mathbf{B})=\sum_i b_{i\alpha}$ (also called \textit{ubiquity} and measuring the number of countries exporting each product).

V- and $\Lambda$-motifs can be generalized to include more than two products/countries: for example, V3- and $\Lambda3$-motifs quantify, respectively, the number of country triplets that export the same products and how many triplets of products are in the same basket of a producer; even more generally, formulas \ref{Vm} and \ref{Lm} can be extended to compute the occurrence of $n$-tuples of countries and products, i.e.

\begin{equation}\label{VmLm}
N_{\text{V}_n}(\mathbf{B})=\sum_\alpha\binom{h_\alpha}{n}\quad\text{and}\quad N_{\Lambda_n}(\mathbf{B})=\sum_i\binom{k_i}{n}.
\end{equation}

More complicated motifs capturing higher-order correlations among nodes can be defined by increasing the size of the subgraph. This is for example the case of X, W and M motifs (see fig. \ref{BipMot2}, bottom). X-motifs measure the co-occurrence of 2 countries in producing the same pair of products and the co-existence of 2 products in the basket of the same two countries. This allows quantifying the competitiveness among countries for different market segments. M- and W- motifs allow the competitiveness between countries to be measured with respect to larger baskets of products:

\begin{eqnarray}\label{X}
N_\text{X}(\mathbf{B})&=&\sum_{i<j}\sum_{\alpha<\beta}b_{i\alpha}b_{i\beta}b_{j\alpha}b_{j\beta}=\sum_{i<j}\binom{C_{ij}}{2}=\sum_{\alpha<\beta}\binom{P_{\alpha\beta}}{2},\\
N_\text{M}(\mathbf{B})&=&\sum_{i<j}\sum_{\alpha<\beta<\gamma}b_{i\alpha}b_{i\beta}b_{i\gamma}b_{j\alpha}b_{j\beta}b_{j\gamma}=\sum_{i<j}\binom{C_{ij}}{3},\\
N_\text{W}(\mathbf{B})&=&\sum_{\alpha<\beta}\sum_{i<j<k}b_{i\alpha}b_{i\beta}b_{j\alpha}b_{j\beta}b_{k\alpha}b_{k\beta}=\sum_{\alpha<\beta}\binom{P_{\alpha\beta}}{3}
\end{eqnarray}

\section{Community structure}

A network is characterized by a community structure if it contains groups of nodes (the communities, or modules) clearly identifiable/recognizable for sharing common properties. The simplest possible classification looks at the number of links within and between communities. In some sense, this generalizes the concept of motifs focusing on subgraphs of higher size and looking at the number of links more than at their directions or at the tendency to form specific patterns.

\begin{figure}[t!]
\centering
\includegraphics[width=0.75\textwidth]{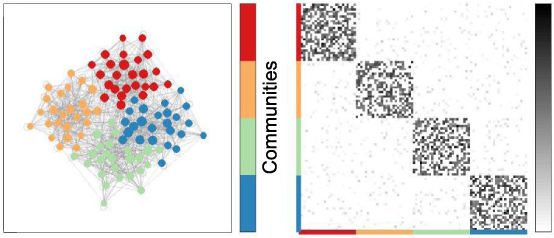}
\caption{Examples of network community structure and its adjacency matrix representation. Source: \citep{faskowitz2018weighted}.}
\label{cor2}
\end{figure}

Let us consider a binary, undirected graph with $N$ nodes and a subgraph $\mathcal{C}$ of $N_C$ nodes. The internal density $\delta_{int}(\mathcal{C})$ and the external density $\delta_{ext}(\mathcal{C})$ of $\mathcal{C}$ are defined as

\begin{equation}
\delta_{int}(\mathcal{C})=\frac{\#\:\text{internal links of $\mathcal{C}$}}{N_C(N_C-1)/2};\quad\delta_{ext}(\mathcal{C})=\frac{\#\:\text{external links of $\mathcal{C}$}}{N_C(N-N_C)};
\end{equation}
A first, intuitive, requirement to identify $\mathcal{C}$ as a community is that $\delta_{int}(\mathcal{C})\gg\rho$ and $\delta_{ext}(\mathcal{C})\ll\rho$ where $\rho$ is the density of the whole network (see fig. \ref{cor2} for an example illustration). Searching for the best trade-off between the two constraints is at the basis of most community detection techniques. 

Investigating the community structure of a network can be relevant for a number of reasons. First of all, grouping nodes sharing some properties leads to a meta-scale (coarse-grained) representation of the network, allowing a simplified analysis \citep{palla2005uncovering}. Moreover, the presence of communities exhibiting different properties can reveal novel features of the network with respect to its average properties, because the macroscale level does not necessarily reflect the mesoscale one. Communities could also provide insights on networks functioning as they can act as different specialized units, as in protein-to-protein interactions \citep{chen2006detecting} or metabolic cycles and pathways \citep{guimera2005functional}. Furthermore, they can help to classify vertices according to their role within and between modules with a consequent effect on network control and stability \citep{csermely2008creative}. They can relevantly shape diffusion patterns on networks (contagion, rumor spreading, innovation adoption). Finally, they can be used  to identify nodes sharing similar interests or opinions and thus to set up efficient recommendation systems \citep{reddy2002graph}.

The community detection problem still represents an open issue. Many algorithms have been introduced in the last two decades - see \citep{fortunato2010community,fortunato2016community} for extensive reviews on existing approaches.
It is possible to distinguish two main kinds of definitions to describe communities: the \emph{local} and the \emph{global} ones. 

Local definitions are based on internal properties of communities such as degree, mutuality, reachability and internal/external cohesion. Cliques (i.e, fully-connected subgraphs) or clique-like modules belong to this group of definition. In the same spirit, it is possible to associate to each module a \textit{fitness measure} to quantify how much it is well defined: for example, requiring that $\mathcal{C}$ is a community if its internal density is larger than a fixed threshold.

Global definitions are used when the communities forming a network cannot be considered as independent entities, and a measure taking into account the whole network structure is necessary. In this case, the standard procedure prescribes to compare the empirical features of a given network partition with the value of the same features provided by a benchmark: the larger the deviation from the benchmark, the stronger the presence of a community structure. The popular \textit{modularity} functional belongs to this class of definitions, since it compares the link density of a given partition and the expected link density of the same partition under a null model:

\begin{equation}\label{mod}
Q=\frac{1}{2L}\sum_i\sum_{j\neq i}(a_{ij}-p_{ij})\delta_{g_ig_j};
\end{equation}
in the definition above, $L$ is the total number of links, $p_{ij}$ represents the probability that nodes $i$ and $j$ are connected under the chosen null model, $g_i$ represents the group to which node $i$ belongs to and analogously for $g_j$, and the $\delta_{xy}$ function stands for the Kronecker delta, which is equal to 1 if $x=y$, i.e. if $i$ and $j$ belong to the same community, and to 0 otherwise.

When binary networks are considered, the most popular choice of benchmark is the Chung-Lu (CL) model \citep{chung2002connected}, which preserves the degrees of nodes. Eq. (\ref{mod}), thus, becomes
 
\begin{equation}\label{eq:2meso}
Q=\frac{1}{2L}\sum_i\sum_{j\neq i}\left(a_{ij}-\frac{k_ik_j}{2L}\right)\delta_{g_ig_j}.
\end{equation}
An alternative choice of benchmark is the \emph{Configuration Model} (CM) which, as we have seen in the previous chapter, is the model preserving the degrees in a proper and unbiased way. Notice that this kind of local information allows a wide set of other higher-order properties to be reconstructed \citep{squartini2011randomizing,mastrandrea2014enhanced,squartini2015unbiased}, hence representing a non-trivial benchmark against which comparing a network partition. Indeed, the modularity maximization identifies the best partition as the one that maximally deviates from the benchmark, i.e. the modular structure that is least likely to be reconstructed by knowing only local information: in other terms, a modularity value close to 0 indicates the presence of a community structure that can be inferred by just knowing the degrees of nodes\footnote{Notice that $Q$ is always smaller than 1 and can have negative values - for example for the trivial partition considering each node as a module.}\footnote{It is worth to notice that the maximum modularity grows with the network size and the number of partitions, therefore it cannot be used to compare partitions of networks having a different number of nodes.}.

\subsection{The Stochastic Block Model} 

From a network reconstruction perspective, the problem of inferring communities can be restated as the problem of finding the model \emph{best fitting a given community structure}. This approach requires a certain amount of information concerning the partition of nodes into modules to be explicitly included into the model \emph{ab initio}. Luckily, this can be done within the ERG framework  \citep{fronczak2013exponential}. To this aim, one of the most used benchmarks is the so-called \emph{Stochastic Block Model} (SBM) \citep{fienberg1981categorical,holland1983stochastic,snijders1997estimation}, prescribing that the probability of connection between nodes $i$ and $j$ only depends on the modules (or groups) they belong to:

\begin{equation}
p_{ij}\equiv p_{g_ig_j}.
\end{equation}
The Hamiltonian to consider, thus, becomes

\begin{equation}\label{sbmh}
H(\mathbf{A})=\sum_i\sum_{j>i}\theta_{g_ig_j}a_{ij}=\sum_r\sum_{s\geq r}\theta_{rs}L_{rs}(\mathbf{A})
\end{equation}
where $L_{rs}$ represents the number of links between groups $r$ and $s$ (or within the same group, in case $r=s$). The partition function is thus

\begin{equation}
Z(\vec{\theta})=\prod_r\left(1+e^{\theta_{rr}}\right)^{\binom{N_r}{2}}\prod_t\prod_{s>t}\left(1+e^{\theta_{ts}}\right)^{N_tN_s}
\end{equation}
where $N_r$ is the number of nodes in group $r$. 
The likelihood maximization prescription allows calculating the parameters as follows:

\begin{equation}\label{expW}
p_{rr}=\frac{L_{rr}(\mathbf{A^*})}{\binom{N_r}{2}},\:\forall\:r
\end{equation}
\begin{equation}\label{expB}
p_{ts}=\frac{L_{ts}(\mathbf{A^*})}{N_tN_s},\:\forall\:t<s.
\end{equation}

In other words, solving the SBM amounts at solving the ER model \emph{within} each block and \emph{between} blocks: in fact, eq. (\ref{expB}) is nothing else that the prescription defining the bipartite ER model. Despite its simplicity, this model allows reconstructing different network structures, e.g. networks with disconnected components, core-periphery, hierarchical or traditional modular structures.\\

The SBM, however, suffers from a limitation that is similar to the one affecting the ER model: in fact, the degree distribution predicted by the SBM is homogeneous and deviates from what observed in real-world networks. As nodes heterogeneity is fundamental for correctly understanding important network properties as their resilience to external shocks, the threshold of the percolation transition or the outcome of an epidemic spreading, \citet{karrer2011stochastic} proposed to incorporate in the model the information about the node degrees \emph{beside} the one concerning their group membership. In this way, they introduced a variation of the SBM, i.e. the \emph{degree-corrected Stochastic Block Model} (dcSBM). In a nutshell, while the SBM is defined by connection probabilities reading

\begin{equation}
p_{g_ig_i}=\frac{e^{-\theta_{g_ig_i}}}{1+e^{-\theta_{g_ig_i}}}\equiv\frac{\chi_{g_ig_i}}{1+\chi_{g_ig_i}},\:\forall\:i<j
\end{equation}
i.e. encoding information only about the group membership of each node, the dcSBM adds the degree information by considering probability coefficients reading

\begin{equation}
p_{ij}=\frac{e^{-(\theta_i+\theta_j+\theta_{g_ig_i})}}{1+e^{-(\theta_i+\theta_j+\theta_{g_ig_i})}}\equiv\frac{x_ix_j\chi_{g_ig_i}}{1+x_ix_j\chi_{g_ig_i}},\:\forall\:i<j
\end{equation}
according to which the probability that nodes $i$ and $j$ are linked depends \emph{both} on their group membership \emph{and} on their degree. In fact, the Hamiltonian can now be written as:

\begin{equation}\label{dcH1}
H(\mathbf{A})=\sum_i\theta_ik_i+\sum_i\sum_{j>i}\theta_{g_ig_j}a_{ij}
\end{equation}
and the likelihood maximization prescribes to solve the following system of equations to determine the unknown parameters 

\begin{eqnarray}
k_i(\mathbf{A}^*)&=&\sum_{j\neq i}\delta_{g_ir}\delta_{g_js}\frac{x_ix_j\chi_{rs}}{1+x_ix_j\chi_{rs}}=\langle k_i\rangle_\text{dcSBM},\:\forall\:i\\
L_{rs}(\mathbf{A^*})&=&\sum_{i}\sum_{j>i}\delta_{g_ir}\delta_{g_js}\frac{x_ix_j\chi_{rs}}{1+x_ix_j\chi_{rs}}=\langle L_{rs}\rangle_\text{dcSBM},\:\forall\:r\leq s
\end{eqnarray}

\begin{figure}[t!]
\centering
\includegraphics[width=\textwidth]{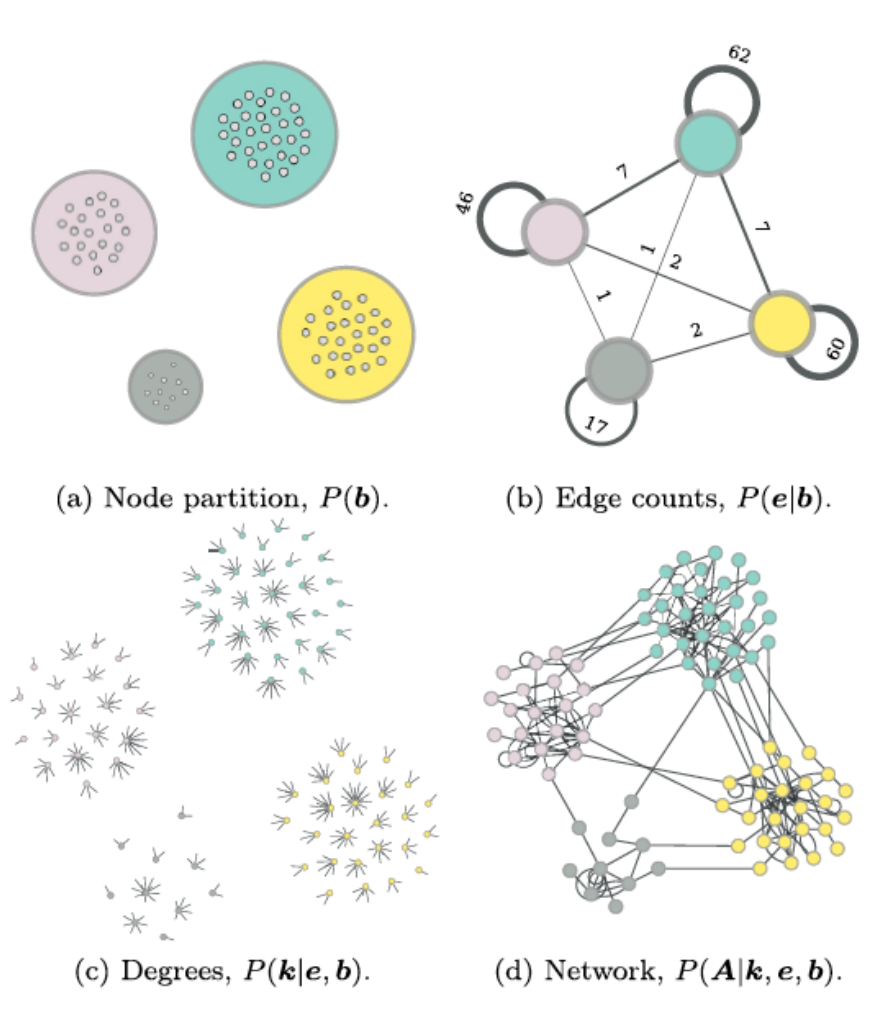}
\caption{Graphical representation of the non-parametric generative process for the degree-corrected SBM: (a) partition sampled; (b) link-counts between groups; (c) node degrees; (d) the network itself. Source: \citep{peixoto2017nonparametric}.}
\label{SBMcom}
\end{figure}

Notice that posing

\begin{equation}
\theta_i+\theta_j+\theta_{g_ig_j}=\left(\theta_i+\frac{\theta_{g_ig_j}}{2}\right)+\left(\theta_j+\frac{\theta_{g_ig_j}}{2}\right)\equiv\theta_i^{g_ig_j}+\theta_j^{g_ig_j}
\end{equation}
leads to degree-informed model where \emph{block-specific degrees} are constrained. In fact, the above position induces the following probability coefficients

\begin{equation}
p_{ij}=\frac{e^{-\left(\theta_i^{g_ig_j}+\theta_j^{g_ig_j}\right)}}{1+e^{-\left(\theta_i^{g_ig_j}+\theta_j^{g_ig_j}\right)}}\equiv\frac{x_i^{g_ig_j}x_j^{g_ig_j}}{1+x_i^{g_ig_j}x_j^{g_ig_j}},\:\forall\:i<j
\end{equation}
defining what is known as \emph{Block Configuration Model} (BCM). The latter, in turn, defines the following recipe for likelihood maximization:

\begin{equation}
k_i^{rs}(\mathbf{A}^*)=\sum_{j\neq i}\delta_{g_ir}\delta_{g_js}\frac{x_i^{rs}x_j^{rs}}{1+x_i^{rs}x_j^{rs}}=\langle k_i^{rs}\rangle_\text{BCM},\:\forall\:i,\:\forall\:r\leq s.
\end{equation}
In other words, solving the BCM amounts at solving the undirected version of the CM \emph{within each diagonal block} and the \emph{Bipartite Configuration Model} (BiCM - also Appendix \ref{appb}) \emph{within each off-diagonal block}.\\

In the same context of the SBM, \citet{peixoto2017nonparametric}  introduced a \emph{non-parametric} Bayesian method to infer the modular structure of a network without any \emph{a priori} information. Indeed, like any other parameter of the model as the node membership, also the optimal number of communities is inferred by the data themselves. The procedure (see a schematic representation in fig. \ref{SBMcom}) is based on the \emph{microcanonical} formulation of the degree-corrected SBM (i.e. the degree sequence is fixed exactly and not on average).

\section{The core-periphery organization}

The notion of \textit{core-periphery}, a structure consisting of a bulk of densely connected nodes and a periphery of weakly linked nodes, has a long tradition in social studies \citep{laumann2013networks,doreian1985structural} and was first formalized by \citet{borgatti2000models}. The core-periphery organization
has been detected in different kinds of networks: economic \citep{smith1992structure,van2014finding,fricke2015core,ma2015rich,barucca2018organization,kojaku2018structural,dejeude2019reconstructing,dejeude2019detecting}, social \citep{everett1999centrality,holme2005core,boyd2006computing,della2013profiling,csermely2013structure,zhang2015identification,rombach2017core,kojaku2017finding}, biological \citep{yang2014overlapping,bruckner2015graph}, neural \citep{bassett2013task,tuncc2015unifying} and transportation networks \citep{lee2014density,xiang2018unified}.

\begin{figure}[t!]
\centering
\includegraphics[width=\textwidth]{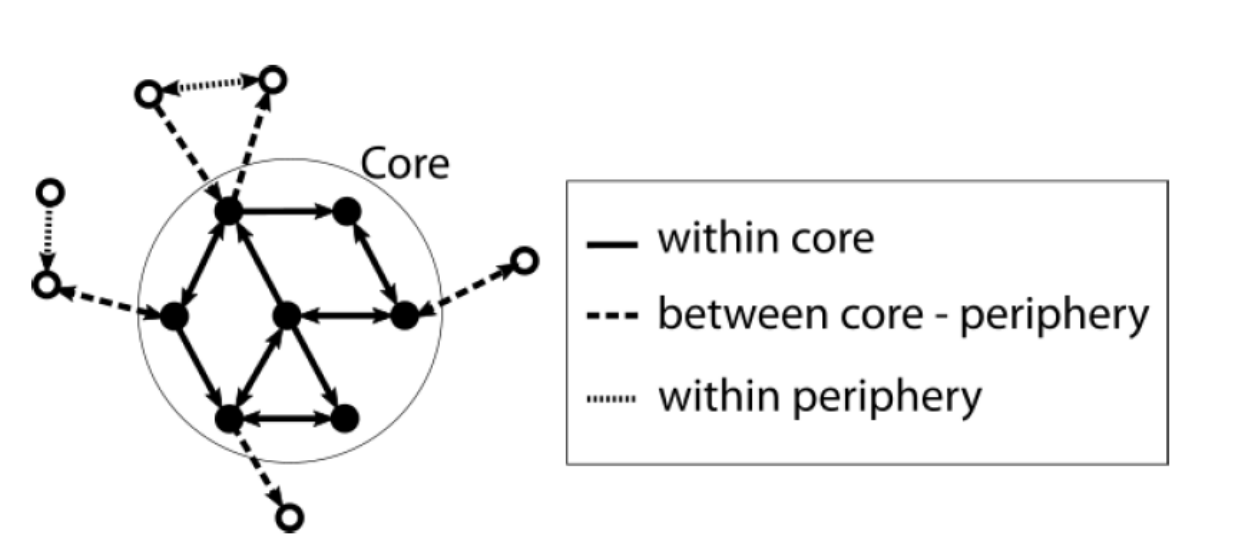}
\caption{Three kinds of links characterize a core-periphery structure: the ones within the core, the ones between the core and the periphery and the ones within the periphery.}
\label{cor3}
\end{figure}

The adjacency matrix of a core-periphery network can be rearranged as a 4 blocks matrix \citep{rombach2014core}

\begin{equation}
\mathbf{A}=\begin{pmatrix} 
\mathbf{A}^{\bullet} & \mathbf{A}^{\top}\\
\mathbf{A}^{\bot} & \mathbf{A}^{\circ}
\end{pmatrix}
\end{equation}
where $\mathbf{A}^{\bullet}$ is associated to the core subgraph, $\mathbf{A}^{\circ}$ to the periphery subgraph and $\mathbf{A}^{\top}$ and $\mathbf{A}^{\bot}$ contain all the connections between the two groups of nodes. It is worth to notice that while the two diagonal blocks are squared matrices, the off-diagonal ones are generally rectangular. Furthermore, generally the densities of the blocks satisfy the following chain of inequalities:
 
\begin{equation}
\rho(\mathbf{A}^{\bullet})>\rho(\mathbf{A}^{\top})\simeq \rho(\mathbf{A}^{\bot})>\rho(\mathbf{A}^{\circ})
\end{equation}
with the core being much denser than the periphery subgraph. Even if $\mathbf{A}^{\top}$ and $\mathbf{A}^{\bot}$ have a similar density, they give different information about the network structure, as the former contains all links pointing from the core to the periphery, while the latter takes into account the inverse direction. It is, thus, evident that three kinds of links are required to fully describe such a network structure: connections \emph{within} the core, \emph{within} the periphery and \emph{between} them (fig. \ref{cor3}).\\
 
The issue of identifying core-periphery structures has been tackled using three different approaches: 1) minimizing the distance with respect to an ideal core-periphery structure \citep{borgatti2000models,van2014finding}; 2) defining a proper benchmark against which detecting a statistically-significant topology \citep{holme2005core,della2013profiling,kojaku2017finding}; 3) finding the model best fitting a given mesoscale structural organization \citep{zhang2015identification,barucca2016disentangling,barucca2018organization}. As for the issue of community detection, the third approach is the one that can be better framed within the network reconstruction perspective.\\

\begin{figure}[t!]
\centering
\includegraphics[width=\textwidth]{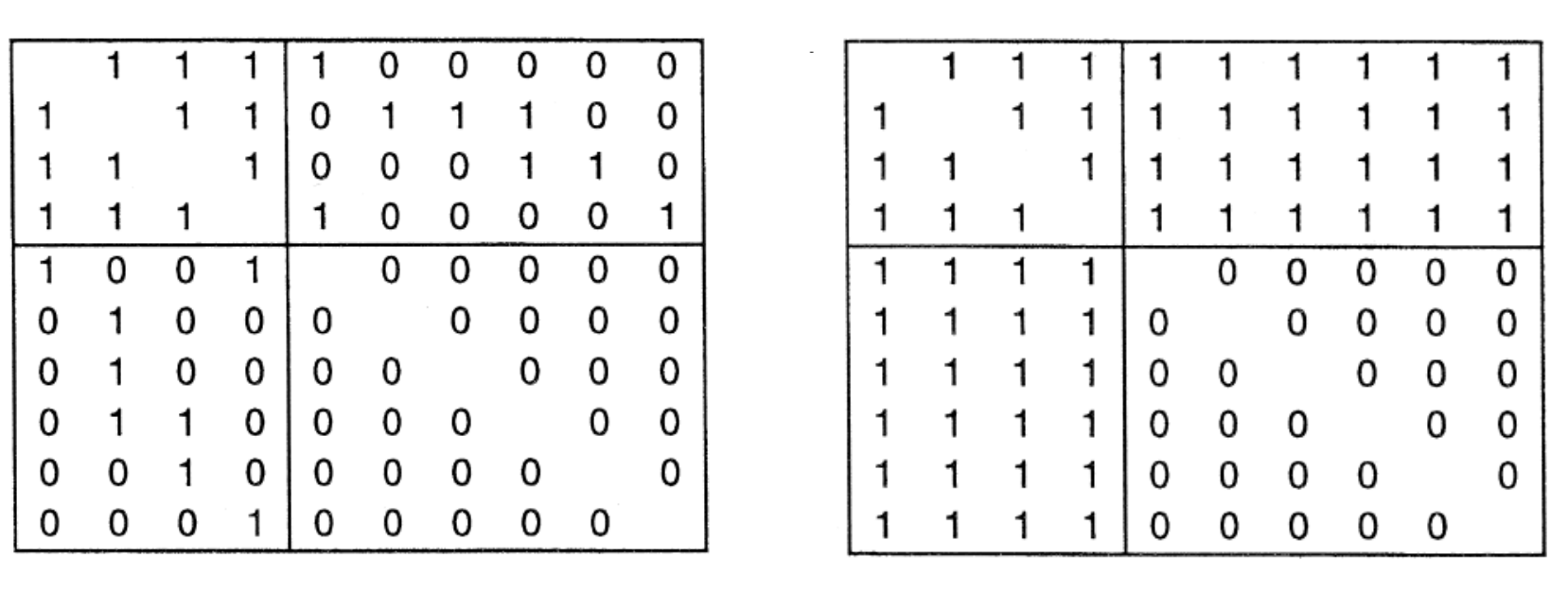}
\caption{Adjacency matrices of an undirected network with a (noisy) core-periphery structure (left) and an ideal core-periphery structure (right). Source: \citep{borgatti2000models}} 
\label{CP3}
\end{figure}

\citet{borgatti2000models}  first formalized the concept of core-periphery introducing the \emph{ideal core-periphery structure} (i.e. a fully-connected core and a periphery of nodes linked only with the core ones - see fig. \ref{CP3}) and measuring the deviation of an observed real network structure from it. In other words, the authors proposed to solve a maximization problem whose score function reads

\begin{equation}\label{Borg}
\Phi=\sum_i\sum_{j\neq i}a_{ij}\Delta_{ij}
\end{equation}
where $a_{ij}$ is the adjacency matrix of the network and $\Delta_{ij}$ represents the ideal core-periphery organization of a network of the same size. Since $\Delta = \delta^T\delta$, where $\delta$ is a boolean vector whose $i$-th entry is equal to $1$ if node $i$ belongs to the core and 0 if it does not, maximizing eq. \eqref{Borg} means finding a vector $\delta$ which maximizes the correlation between $\Delta_{ij}$ and $a_{ij}$. However, as the same authors explicitly stated, a significance test for the algorithm output is completely missing.

Along the same guidelines is the work by \citet{van2014finding}. They tested the goodness of three models in recovering the core-periphery structure of the DIN as defined by some axioms: the ER, the CM and the Barabasi-Albert (BA) model. As accuracy index, the authors used an error score counting the number of errors (i.e. the number of links to add/delete to recover the aforementioned axiomatic model) divided by the effective number of links. They found the error score characterizing both ER and BA models was too large for the DIN structure to be compatible with them; however, the error score was not significant under the CM, a result implying that the knowledge of the out- and in-degrees allows the core-periphery mesoscale organization to be recovered to a very good extent. This confirmed earlier results by \citet{lip2011fast} that, at least for the simplest specification of the error score given in \citet{borgatti2000models}, the core-periphery partition is completely determined by the degree sequence.\\

\begin{figure}[t!]
\centering
\includegraphics[width=\textwidth]{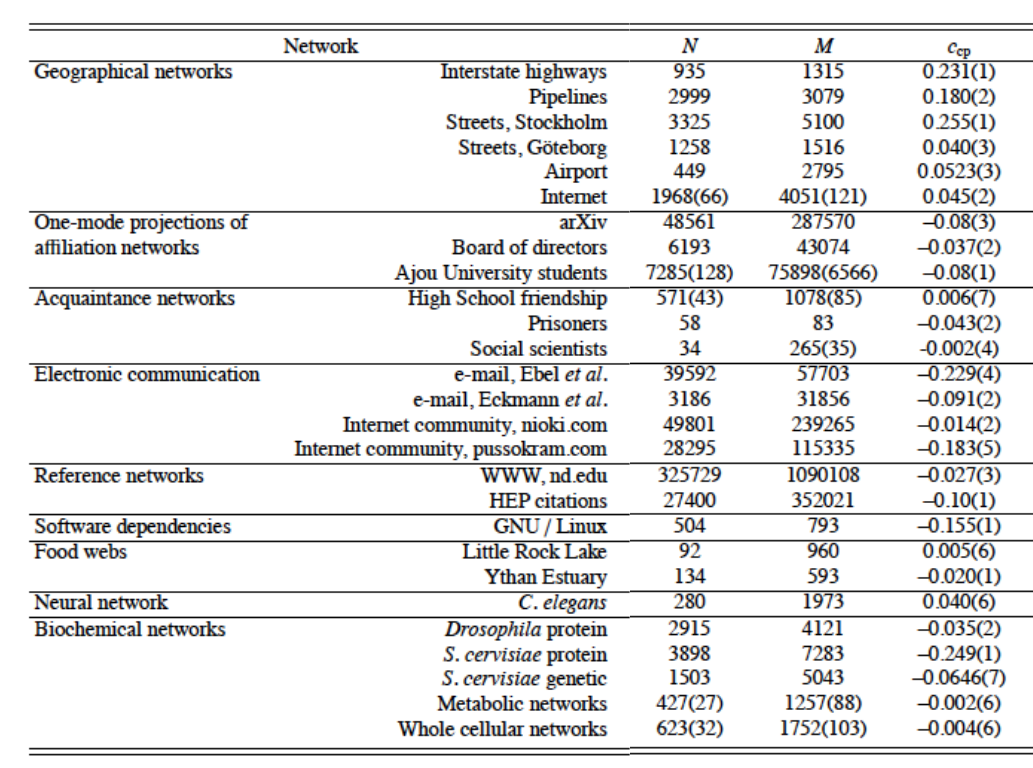}
\caption{Network size (number of nodes, $N$, and links, $M$) and core-periphery coefficient ($C_{cp}$) for different networks. Source: \citep{holme2005core}} 
\label{CP4}
\end{figure}

Let us now discuss the algorithms belonging to the second group, i.e. the ones not assuming the existence of an ideal core-periphery structure but comparing the observed topology with the outcome of a properly chosen benchmark model. The first model in this sense has been proposed by \citet{holme2005core} who introduced a generalization of the closeness centrality to be compared with a null model preserving the node degrees. The generalized closeness centrality refers to a subset $U$ of the set of nodes $V$ of a network

\begin{equation}\label{Holme}
C_{C}(U)=\left[\overline{\left(\overline{d(i,j)}_{j\in V\setminus\{i\}}\right)}_{i\in U}\right]^{-1}
\end{equation}
where $d(i,j)$ is the geodesic distance between nodes $i$ and $j$. The author searched for the optimization of the score function

\begin{equation}\label{Holme2}
C_{cp}=\frac{C_C[V_{core}]}{C_C[V]}-\left\langle\frac{C_C[V_{core}]}{C_C[V]}\right\rangle
\end{equation}
where the subset $U=V_{core}$ was taken to be the \emph{k-core}, representing the maximal subgraph of the network having minimum degree $k$ and maximal closeness, and the average was taken over the ensemble of networks having the same degree sequence as the observed one. Obviously, random graphs have (on average) $C_{cp}$ equal to 0; positive/negative values of $C_{cp}$ stand for over/under representation of the core-periphery structure. The authors find that the core-periphery structure clearly characterizes groups of networks (see table \ref{CP4}): geographically embedded networks often exhibit a positive $C_{cp}$ and this could be the effect of the optimization of temporal communications; on the other hand, social networks tend to show a slight negative coefficient. Upon considering the specific datasets under study, the authors concluded that the existence of modules, according to some kind of specialization, could imply the absence of a clearly-defined core.

\citet{della2013profiling} proposed an algorithm based on the standard random walk to identify the core-periphery structure of all kinds of networks, including the weighted ones. The authors associated to each network a \textit{core-periphery profile}, i.e. a discrete vector $\{\alpha_1,\alpha_2\dots\alpha_N\}$, with $N$ being the network size, which allows one to quantify to what extent a network is centralized and to associate to each node a measure of \textit{coreness}. Thanks to this definition it is possible to introduce an $\alpha$-dependent degree of \textit{peripheryness}, composed by nodes whose coreness is below a certain threshold~$\alpha$.

\begin{figure}[t!]
\centering
{\includegraphics[width=0.7\textwidth]{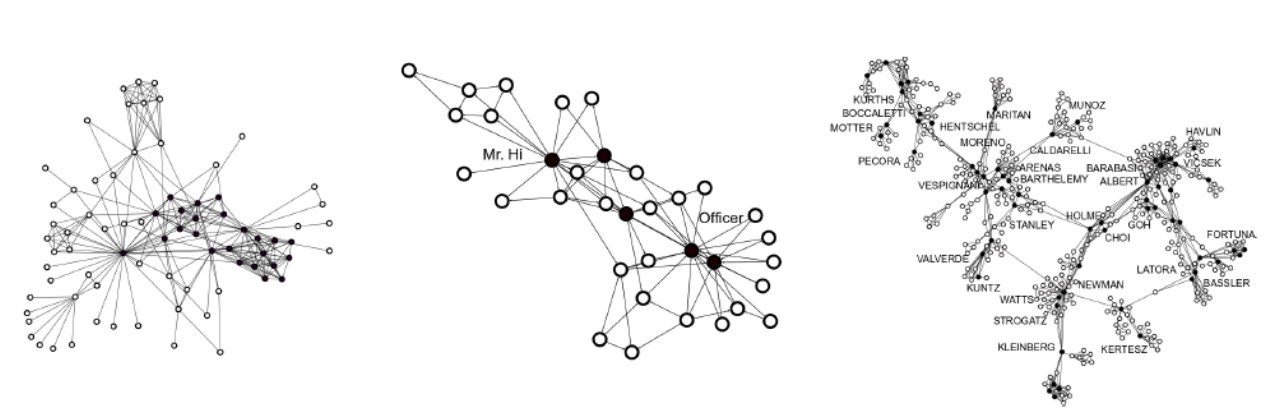}}
{\includegraphics[width=0.7\textwidth]{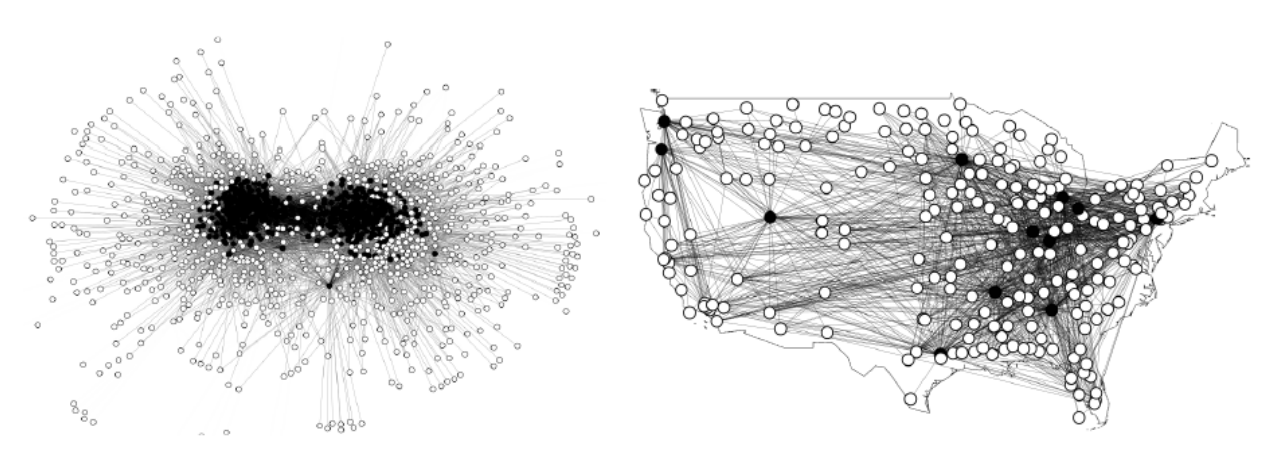}}
\includegraphics[width=0.7\textwidth]{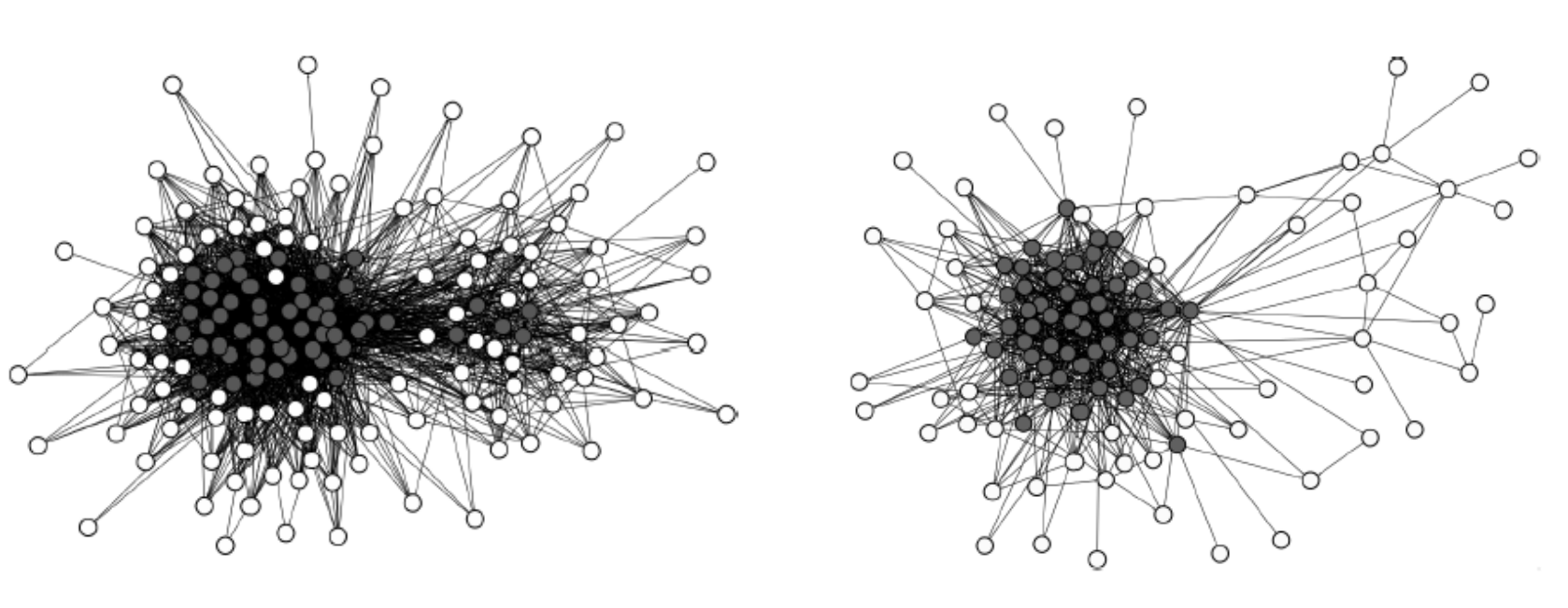}
\caption{Core-periphery structure of several networks detected using the multivariate extension of surprise. In the top panel, ``Les Miserables'' (left), the Zacary Karate Club (middle), the NetSci network of collaborations (right); in the middle pane: the US political weblog (left), the US airport network (right); in the bottom panel: eMID in January 2005 (left) and November 2009. Source: \citep{dejeude2019detecting}}
\label{CP8}
\end{figure}

Recently, \citet{dejeude2019detecting} proposed a benchmark model for core-periphery detection inspired by the definition of \emph{surprise}\footnote{The surprise is defined as the p-value of an hypergeometric distribution \citep{bifone2016surprise}.}. 
The idea is to detect \emph{bimodular structures}, such as the core-periphery one, by comparing the probability assigned to them by the \emph{Directed Erd\"os-Renyi Model} (DER) and by the SBM: finding the most statistically-significant partition ultimately means finding the partition that is least likely to be explained by the DER with respect to the SBM. As shown in fig.\ref{CP8}, this approach allowed identifying significant core-periphery structures in several real networks.

Similarly, \citet{kojaku2017finding} proposed a model to identify multiple core-periphery structures by extending the approach introduced in \cite{borgatti2000models} and using a random graph to test the significance of such a mesoscale organization. They also showed that the information encoded into the degree sequence always accounts for the organization of a network into a core and a periphery: therefore, it is not possible to use the CM to test the significance of such a mesoscale structure.\\

Lastly, let us discuss the approaches belonging to the third group. \citet{zhang2015identification} proposed a method to identify the generative model most likely to produce a given network partition and applied it to some synthetic systems and two empirical networks: Internet at the level of autonomous system and the US political blog-o-sphere \citep{adamic2005political}. The authors employed an expectation-maximization (EM) algorithm, assuming that the networks were generated by a SBM. Initially there are $N$ nodes, no links and two empty groups (the core and the periphery). Each node is randomly assigned to group 1 with probability $\gamma_1$ (and to group 2 with probability $\gamma_2 = 1-\gamma_1$). Then, each pair of nodes is connected with probability $p_{rs}$, where $r,s$ indicate the groups they belong to. Given the adjacency matrix $\mathbf{A}$, the likelihood that the network is generated by the model above is given by:

\begin{equation}\label{New}
P(\mathbf{A}|p,\vec{\gamma})=\sum_{g}P(\mathbf{A}|p,\vec{\gamma},g)P(g|\vec{\gamma}) = \sum_g\left[\prod_{i<j}p^{a_{ij}}_{g_ig_j} (1-p_{g_ig_j})^{1-a_{ij}}\prod_i\gamma_{g_i}\right]
\end{equation}
where again $g_i$ represents the group node $i$ belongs to and $\sum_g$ is the sum over all assignments of the nodes to groups. Upon maximizing the likelihood score function $\ln P(\mathbf{A}|p,\vec{\gamma})$ it is possible to determine the values of $\vec{\gamma}$ and $p_{rs}$.

Few years later, \citet{barucca2016disentangling,barucca2018organization} developed a SBM able to reproduce a bipartite or a core-periphery structure through a tuning parameter. They tested the e-MID structure to understand if it is better represented by a core-periphery or a bipartite structure. This work provided yet another evidence of the importance of degree heterogeneity and the information it carries about the system. Indeed, the authors found that while the SBM identifies a core-periphery structure for e-MID, the dcSBM highlights a bipartite organization.

\citet{dejeude2019reconstructing} focused on the DIN and tested the accuracy of a plethora of methods in reproducing its core-periphery structure: in particular, they compared the block models (SBM and dcSBM) with the DCM and the RCM, finding that the DCM always performs best. This result, in line with that of \citet{kojaku2018core}, further confirms the high informative role of node degrees in reproducing the core-periphery structure of a network.

\section{The bow-tie organization}

The first formal definition of a bow-tie network decomposition was given by \citet{yang2011bow}, even if its concept was firstly introduced by \citet{broder2000graph} for the study of the WTW. The bow-tie decomposition has, since then, been investigated both theoretically and empirically with applications in different fields: \citet{dill2002self} studied the WWW self-similarity; \citet{arasu2002pagerank} used the bow-tie structure as a model for the large scale structure of the WWW to test the PageRank algorithm; \citet{hirate2007web} studied the temporal evolution of bow-tie structures; \citet{zhang2007expertise} focused on the bow-tie structure of the Java Developer Forum to test some ranking algorithms for the expertise network; \citet{tanaka2005highly} showed that also metabolic networks could be characterized by such a mesoscale structure.

\begin{figure}[t!]
\centering
\includegraphics[width=\textwidth]{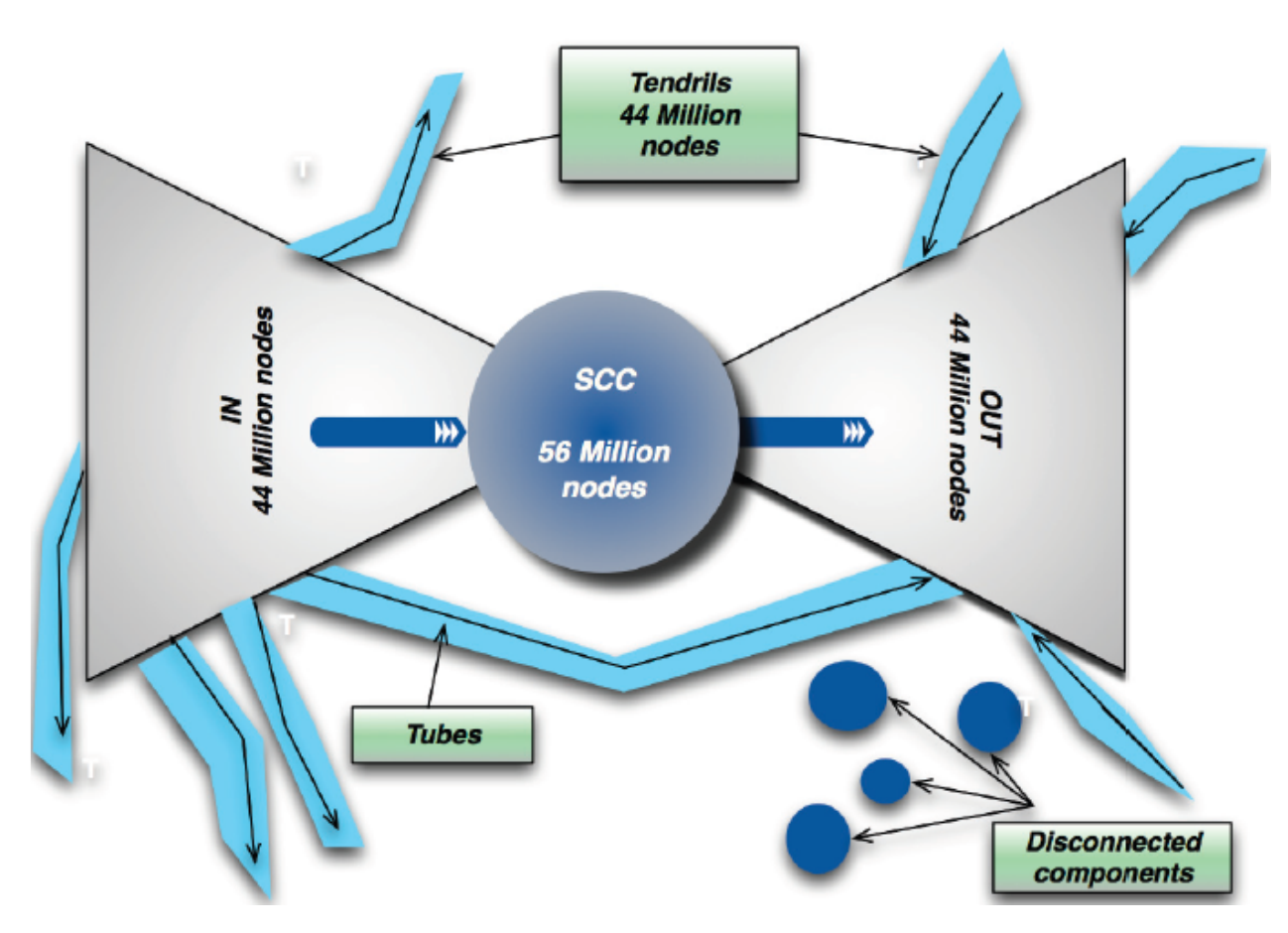}
\caption{A schematic representation of a bow-tie structure. It contains the SCC, IN and OUT components described in the main text. Moreover, there are (i) \textit{Tendrils} containing nodes reachable by the IN component or that can reach the OUT component without passing through the SCC one; (ii) \textit{tubes} allowing the passage from IN to OUT component without passing through the SCC one. Source: \citep{yang2011bow}. }
\label{BT}
\end{figure}

The definition of a bow-tie structure is strictly related with that of node \emph{reachability}. A node $i$ is reachable by a node $j$ if a path of consecutive links starting from node $j$ and ending to node $i$ exists. Using this concept, it is possible to fully describe a bow-tie structure as a composition of specific, not-overlapping network subgraphs \citep{broder2000graph}. We report here the definition of the three most relevant subgraphs (see fig. \ref{BT}):

\begin{itemize}
\item the Strongly Connected Component (SCC) contains all nodes reachable by any other node in the SCC;
\item the IN component (IN) contains all nodes not belonging to the SCC but from which all nodes in the SCC are reachable;
\item the OUT component (OUT) contains all nodes not belonging to the SCC but reachable from any other node in the SCC.
\end{itemize}

The adjacency matrix associated with a bow-tie structure can be rearranged in order to separate the three components described above:

\begin{equation}
\mathbf{A}=\begin{pmatrix}
\mathbf{A}^I & \mathbf{A}^> & \mathbf{0}\\
\mathbf{0} & \mathbf{A}^S & \mathbf{A}^{\gg}\\
\mathbf{0} & \mathbf{0} & \mathbf{A}^O
\end{pmatrix}
\end{equation}
where $\mathbf{A}^I$, $\mathbf{A}^S$ and $\mathbf{A}^O$ represent respectively the IN, SCC and OUT components, while $\mathbf{A}^>$ and $\mathbf{A}^{\gg}$ stand for the bipartite networks of their interactions.\\

The same considerations guiding the analysis of the core-periphery structure can be considered as valid also when analysing the bow-tie structure. Let us now revise the approaches that have been proposed to study it.

\citet{zhao2007bow} explored the gene-based metabolic network of 75 organisms, comparing their topological properties with a randomized counterpart preserving the degree of nodes and the number of bidirectional links. They found that the global bow-tie structure is still present in the randomized model, but with some differences: (i) the size of the SCC is smaller in the reconstructed network; (ii) the number of 2-cores is overestimated by the model and (iii) no 3-cores are detected in the random network. These results are in favour of a significant cliquish bow-tie topology characterizing metabolic networks that cannot be reconstructed by enforcing local constraints only.

\citet{vitali2011network} studied the network of trans-national corporations (TNCs) whose nodes are companies and (directed) links indicate that firm $i$ owns some share of firm $j$. The authors showed that the TNCs exhibit a bow-tie structure with a very small core of nodes and an OUT-component that is much bigger than the IN-component. Combining topology with control ranking, they find that the small core is densely connected, and a node randomly drawn from it is a the top holder with 50\% probability (which reduces to 6\% for the IN-component). The authors also state they did not perform any attempt to reconstruct the network using local information, as this operation would be meaningless in an economic system whose links represent the share of ownership among economic actors.

More recently, \citet{dejeude2019reconstructing} have investigated the problem of reconstructing the bow-tie structure of the WTW and the DIN using several ERG-based models, i.e. the ones defined by purely local information (out- and in-degrees of nodes) as well as those including the information about the membership of each node to a specific subgraph (i.e. SCC, IN, OUT). The WTW is characterized by a bow-tie structure where the OUT-component is completely missing and the size of SCC component increases over time together with the share of its reciprocated links. From an economic point of view this can be interpreted in terms of an ongoing globalization process, fostered by an increasing number of trade agreements. The comparison of the aforementioned models through the AIC and BIC criteria (see Appendix \ref{appc} for more details) allows concluding that the DCM is the best model to reconstruct the bow-tie structure of the WTW. This suggests that the information encoded into the degrees is enough to explain the peculiar mesoscale structure of the WTW.

\begin{figure}[t!]
\centering
\includegraphics[width=0.7\textwidth]{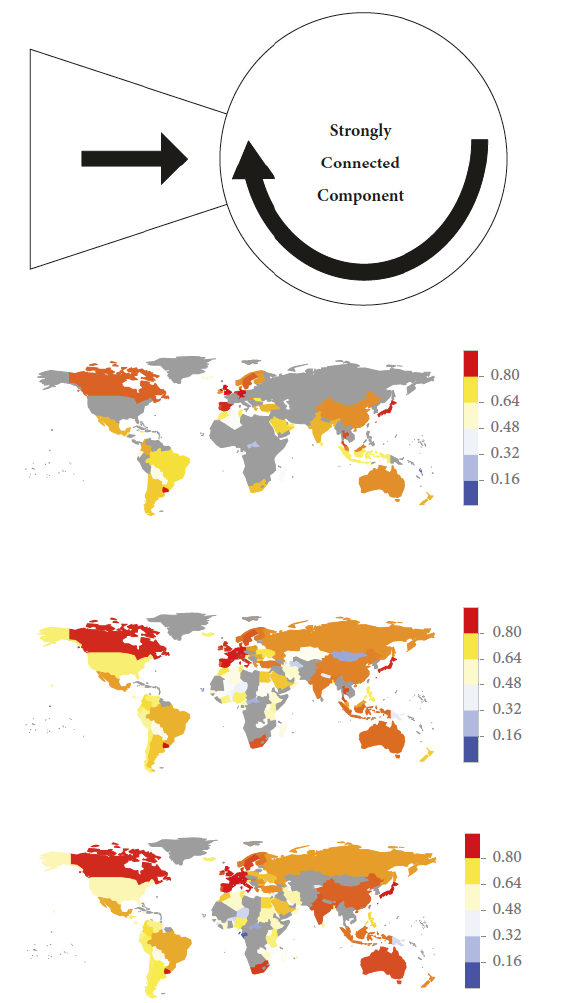}
\caption{Bow-tie structure of the WTW. In the bottom panels countries are colored according to their reciprocated degree (gray for countries belonging to the IN component). Source: \citep{dejeude2019reconstructing}.}
\label{BT5}
\end{figure}

Similarly, the DIN presents a bow-tie structure with three components (but without connections between the OUT- and the IN-component). The temporal evolution of the bow-tie structure is very informative about the ongoing structural organization of the system due to economic changes: the SCC size decreases before the 2008 crisis, while the size of the OUT- and IN-component appears as stable until they shrink around the crisis period. As for the WTW, the model accounting only for the degree heterogeneity is to be preferred to block models (except in specific periods, where it competes with the RCM): therefore, both the WTW and the DIN are characterized by a peculiar bow-tie structure that, basically, seems to be explained by local constraints.\\

Although the bow-tie structure characterizes several metabolic, technological and economic systems, little is known about its \emph{dynamical} origin. In order to investigate this aspect, \citet{zhang2007expertise} have studied the so-called \emph{community expertise network} (CEN), i.e. the network of online questions and answers among users. Following the idea that the user answering should have more knowledge than the one who asked the question, they built a network of expertise, and found an uneven bow-tie structure with half of users belonging to the IN-component (just asking question) and almost the same amount of people ($\simeq12\%$) belonging to the core (asking/answering) and to the OUT-component (just answering questions). The authors tried to reconstruct the network properties and its peculiar mesoscale structure by fixing a series of dynamical rules to generate a simulated network via Agent Based Models. They proposed two models differing in the choice of the user answering to a specific question: in the \emph{best preferred expert} model, the probability to reply increases exponentially with the difference between the expertise of the two users (asking/answering), while in the \emph{just better} one, the probability of answering just depends on having a slightly higher expertise than the user asking the question. They found a high similarity between the empirical bow-tie structure and the one of the network simulated via the first model while the core of network generated via the second model almost disappeared in favour of the tendrils.

\chapter{Network reconstruction at the microscale}
\label{chap4}

\section{The link prediction framework}

Network reconstruction at the microscale is the task of predicting specific individual links that are missing from the network. As mentioned in the introduction, there are many cases in which this is important. Take for instance biological networks, such as food webs, protein-protein interaction networks or metabolic networks \citep{barzel2013network,cannistraci2013link,kovacs2019network}. For these systems, whether a link between two nodes exists must be determined by field and/or laboratory experiments, which are usually very costly. Instead of blindly checking all possible interactions, focusing on those potential links that are most likely to exist can significantly reduce the experimental costs and speed the pace of uncovering the true network \citep{redner2008teasing}. Social network analysis also comes up against the missing data problem, therefore link prediction methods can be used to infer unknown relationships or collaborations \citep{liben-nowell2007link}. Also, the problem of {\em recommendation} in social networks or e-commerce applications is technically a link prediction task \citep{huang2005link,lu2012recommender}. 

In general, besides helping in analyzing networks with missing data, link prediction algorithms can be used to guess the links that may appear in the future. From this viewpoint, any model of evolving network corresponds in principle to a link prediction algorithm\footnote{As such, the correct prediction of future links has a self-reinforcing feedback effect in the ability to predict further new links \citep{li2018similairy}.} \citep{lu2015towards}.
On the other hand, link prediction techniques can be also used to identify ``spurious'' links resulting from inaccuracies in network data and, in general, to estimate the true structure of a network when data are noisy \citep{guimera2009missing,newman2018network}. For instance, protein-protein interaction networks are constructed on the basis of \emph{measured} interactions and not of \emph{actual} interactions, hence they can suffer not only from missing data but also from measurement errors. 
Another example is given by social friendship networks, which may be obtained from survey data (i.e. asking people who their friends are) that are affected by the different standards of participants on the concept of friendship itself. In these and many other cases, however, network data come with no error-assessment information of any kind \citep{peixoto2018reconstructing}.\\

The typical framework of link prediction takes as input the adjacency matrix $\mathbf{A}$ of a network, characterized by the set $E$ of \emph{observed links} and the set $U\setminus E$ of \emph{non-existent links} ($U$ is the set of all nodes pairs). We assume that there are some \emph{missing links} (or links that will appear in the future) in the set $U\setminus E$: the link prediction task is that of finding them out. Since by definition the missing links are unknown, in order to test a link prediction method the observed link set $E$ is partitioned into a \emph{training set} $E_T$ and a \emph{probe set} $E_P=E\setminus E_T$. The former is used to inform the prediction algorithm whereas the latter is used as the prediction target. For each \emph{non-observed link}, i.e. a link in the set $E_{N}=E_P \cup\:(U\setminus E)\equiv U\setminus E_T$, the link prediction algorithm provides a score (also known as \emph{reliability}) quantifying the likelihood of its existence. As we will see, the reliability scores can be used both to predict missing links (i.e. the non-existent links in the observed network having the highest reliability) and to identify possible spurious links (i.e. the observed links with the lowest reliability). \\

Link-prediction methods can be roughly classified into two main classes: \emph{similarity-based} and \emph{model-based}\footnote{These methods are also known in the literature as \emph{likelihood-based} or \emph{probabilistic} methods.} algorithms \citep{lu2011link}. While similarity-based methods rely on (a varying amount of) information about the network topology, model-based methods additionally assume the presence of some kind of organizing principle acting at the meso and/or macro-scale. In both cases, the key underlying assumption is that any two nodes are more likely to interact if they have a larger \emph{similarity}. Although ``similarity'' is quite an abstract concept and often depends on the specific context, it is typically proxied by the amount of direct or indirect paths between nodes \citep{martinez2017survey}. In the following, for simplicity we will only discuss the case of undirected binary networks.

\section{Similarity-based methods}

These methods represent the simplest approach for link prediction tasks. They assign a score $s_{ij}$ to each non-observed link of the network that is based on the \emph{structural similarity} of the two involved nodes $i$ and $j$. This similarity is computed using solely the knowledge of the network structure around $i$ and $j$, where ``around'' is measured in terms of \emph{graph distance} (i.e., the number of links that form a connection path between two nodes). Similarity-based methods are then classified as local, semi-local or global indices depending on how much the topological information used is distant from the target candidate link $i-j$. 

A first remark is in order here. A crucial advantage of local and semi-local methods with respect to global ones is that they can be implemented in a decentralized manner and can, thus, handle large-scale systems. Global indices, instead, require more computational power. Typically, the computational complexity of a similarity-based method increases exponentially with the maximum network distance used, while the prediction accuracy grows sub-linearly and eventually decreases. Indeed, when tested on real world-networks, global methods obtain the worst results because they make use of irrelevant information, while local techniques work surprisingly well, in turn suggesting that most of the useful information to predict links is local. However, the performance of each technique strongly depends on the structural properties of the network considered \citep{martinez2017survey}.

As a second remark, let us notice that similarity-based methods should be handled with care when detecting spurious connections. As we will see, by definition these methods tend to give high score to node pairs with many common connections, hence the \emph{weak ties} of the network (i.e. the links connecting different regions of the network) typically get low scores. However, treating weak ties as spurious links and removing them may have the undesired effect of breaking the network into disconnected components, thus destroying its functionality. A simple solution is to adjust the similarity score with some quantity such as the \emph{link betweenness} (the ratio of shortest paths passing through that link) \citep{zeng2012removing}. In this way, the weak ties that keep the network connected do not get the lowest similarity score and are not at risk of being removed.

\paragraph*{Local indices.} Local indices consider only the information about the first neighbors of each node (i.e., the nodes at graph distance equal to 1). Denoting by $\Gamma_i$ the set of neighbors of node $i$, with $k_i=|\Gamma_i|$ being the degree (number of neighbors) of $i$, the simplest index can be defined using the {\em preferential attachment} (PA) rule \citep{barabasi1999emergence}: connection probabilities are proportional to the degree of nodes. The score assigned by this method to a non-observed link between nodes $i$ and $j$ is thus:

\begin{equation}\label{s_pa}
s_{ij}^{\mbox{\tiny[PA]}}=k_i\cdot k_j.
\end{equation}

A step forward is made by considering the neighborhood structure of nodes $i$ and $j$. The idea is that any two nodes are more likely to be linked if they have many common neighbors (CN) \citep{newman2001clustering,kossinet2006effects}. The popular CN index is thus

\begin{equation}\label{s_cn}
s_{ij}^{\mbox{\tiny[CN]}}=|\Gamma_i\cap\Gamma_j|;
\end{equation}
note that $s_{ij}^{\mbox{\tiny [CN]}}$ is also equal to $(\mathbf{A}^2)_{ij}$, the number of different paths of length 2 connecting $i$ and $j$. Despite its simplicity, this measure performs surprisingly well on most real-world networks and can beat much more complicated approaches. Due to its success, many variations of this recipe do exist. For instance\footnote{Other examples are the Salton index (i.e. the cosine similarity), the S{\o}rensen index and the Hub Promoted/Depressed index \citep{ravasz2002hierarchical}, i.e.
\begin{equation}
s_{ij}^{\mbox{\tiny[Salton]}}=\frac{|\Gamma_i\cap \Gamma_j|}{\sqrt{k_i\cdot k_j}},\qquad 
s_{ij}^{\mbox{\tiny[S{\o}rensen]}}=\frac{2|\Gamma_i\cap \Gamma_j|}{k_i+k_j},\qquad
s_{ij}^{\mbox{\tiny[HPI]}}=\frac{|\Gamma_i\cap \Gamma_j|}{\min\{k_i,k_j\}}.
\end{equation}}, the Jaccard coefficient penalizes the nodes with many neighbors, whereas the Leicht-Holme-Newman index (LHN) \citep{leicht2006vertex} compares the CN score with the expected number of common neighbors under the PA model:

\begin{equation}\label{s_jc-lhn}
s_{ij}^{\mbox{\tiny[Jaccard]}}=\frac{|\Gamma_i\cap\Gamma_j|}{k_i+k_j},\qquad
s_{ij}^{\mbox{\tiny[LHN]}}=\frac{|\Gamma_i\cap \Gamma_j|}{k_i\cdot k_j}.
\end{equation}

More refined methods take into account the degrees of the common neighbors, thus assigning more weight to the less-connected nodes. These are the Adamic-Adar (AA) \citep{adamic2003friends} and the resource allocation (RA) \citep{zhou2009predicting} indices:

\begin{equation}\label{s_aa_ra}
s_{ij}^{\mbox{\tiny[AA]}}=\sum_{l\in\Gamma_i\cap\Gamma_j}\frac{1}{\ln k_l},\qquad s_{ij}^{\mbox{\tiny[RA]}}=\sum_{l\in\Gamma_i\cap\Gamma_j}\frac{1}{k_l}.
\end{equation}

The local na\"ive Bayes model (LNB) \citep{liu2011link}, instead, gives more weight to common neighbors with higher values of the clustering coefficient (we remind that the clustering coefficient $c_l$ of a node $l$ is the number of closed paths of length 3 involving that node, normalized by the maximum possible number of such paths, i.e. $c_l=(\mathbf{A}^3)_{ll}/[k_l(k_l-1)]$). By assuming that the clustering coefficients of the common neighbors are mutually independent, this index can be expressed as

\begin{equation}\label{s_lnb}
s_{ij}^{\mbox{\tiny[LNB]}}=\sum_{l\in\Gamma_i\cap\Gamma_j}\ln\left[\frac{c_l}{1-c_l}\right];
\end{equation}
notice that the different weighting procedures implemented by the AA-RA and LNB indices can be combined together in a straightforward way, for instance by multiplying term by term the corresponding sums \citep{liu2011link}.

\paragraph{Semi-local indices.} These indices rely on structural information up to the second neighbors of nodes $i$ and $j$ (i.e., the nodes at graph distance less or equal to 2). The local path (LP) index \citep{lu2009similarity} is defined as

\begin{equation}\label{s_lp}
s_{ij}^{\mbox{\tiny[LP]}}=(\mathbf{A}^2+\epsilon\mathbf{A}^3)_{ij}
\end{equation}
where $\epsilon<1$ is a free parameter. The LP index, thus, counts the number of paths connecting $i$ and $j$ of length 2 and 3, while penalizing the latter by the factor $\epsilon$. Clearly, LP reduces to CN when $\epsilon=0$. However, LP can also be extended to account for paths longer than three, i.e.

\begin{equation}\label{s_lp_n}
s_{ij}^{\mbox{\tiny[LP]}}=(\mathbf{A}^2+\epsilon\mathbf{A}^3+\epsilon^2\mathbf{A}^4+\dots+\epsilon^{n-2}\mathbf{A}^n)_{ij}
\end{equation}
where $n>2$ is the maximal path length considered. This index asks for more information and computational resources as $n$ increases. When $n\to\infty$, LP becomes equivalent to the Katz index \citep{katz1953new} (see below).

A related set of indices that exploit the local structure of the neighborhood is given by the LCP-based methods \citep{cannistraci2013link}. These indices rely on the idea that $i$ and $j$ are more likely to be connected if their common neighbors are members of a strongly-connected local community - an assumption known as the {\em local community paradigm} (LCP). The simplest LCP index is the CAR  similarity, a function of both the number of common neighbors and of the number of links between them (i.e. the number of links constituting the local community):

\begin{equation}\label{s_car}
s_{ij}^{\mbox{\tiny[CAR]}}=
s_{ij}^{\mbox{\tiny[CN]}}\cdot\sum_{l\in\Gamma_i\cap\Gamma_j}\frac{\gamma_l}{2}=
s_{ij}^{\mbox{\tiny[CN]}}\cdot\sum_{l\in\Gamma_i\cap\Gamma_j}\sum_{m\in\Gamma_i\cap\Gamma_j}\frac{a_{lm}}{2}
\end{equation}
where the factor $\gamma_l=\sum_{m\in\Gamma_i\cap\Gamma_j}a_{lm}$ counts how many neighbors of node $l\in\Gamma_i\cap\Gamma_j$ are also linked to both $i$ and $j$. Analogous modifications can be applied to the other similarity-indices discussed above - see \citep{cannistraci2013link}.

\begin{figure}[p]
\centering
\includegraphics[width=\textwidth]{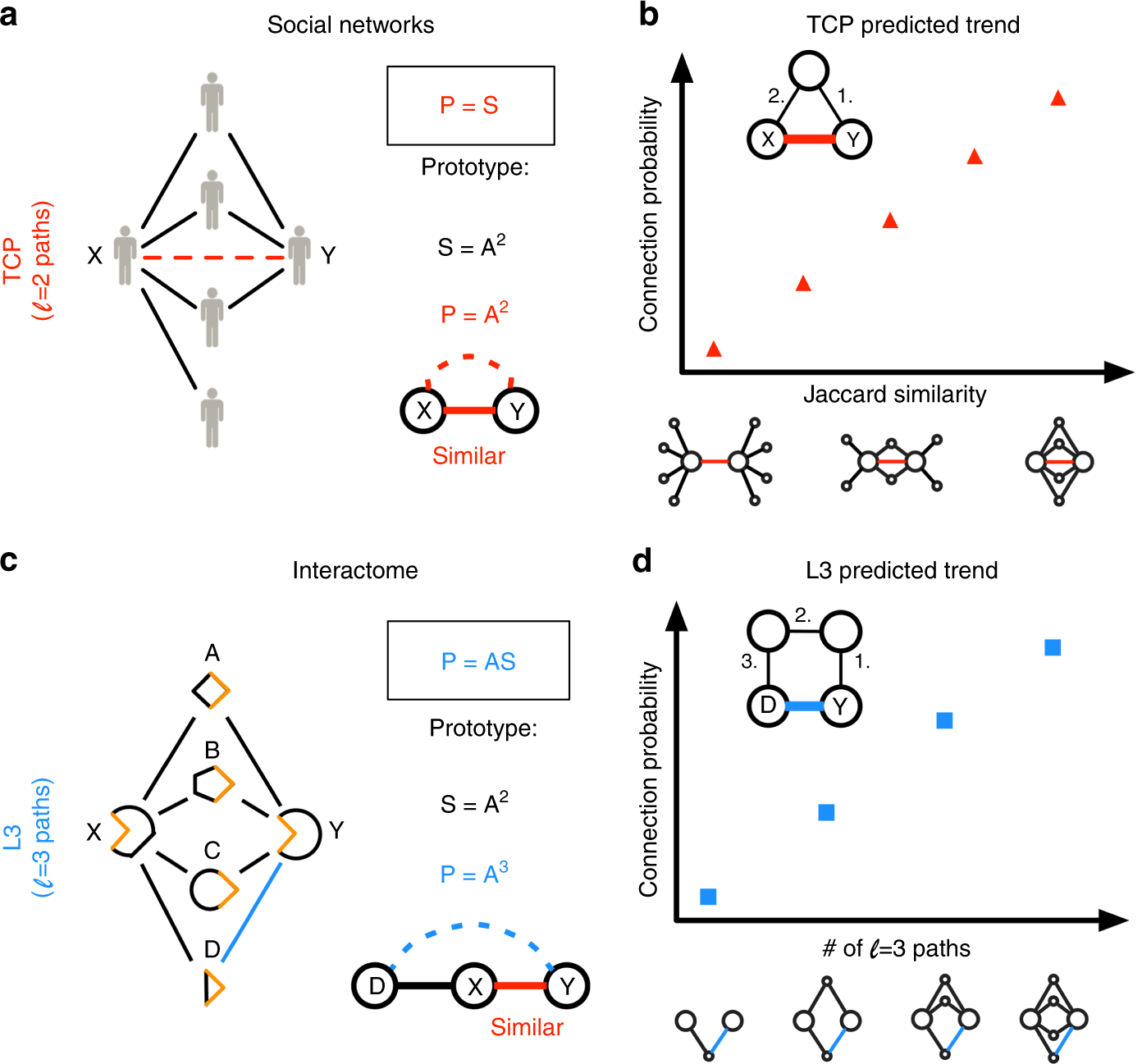}
\caption{Link prediction in social and biological networks. {\bf a)} In social networks, a large number of common friends (that is, paths of length two) implies a higher chance to become friends (red link between nodes X and Y). This is known as the Triadic Closure Principle (TCP). TCP predicts ($P$) links based on node similarity ($S$), quantifying the number of shared neighbors between each node pair ($\mathbf{A}^2$). {\bf b)} TCP implies that node pairs of high Jaccard similarity (a sample index based on paths of length two) are more likely to interact. {\bf c)} Protein interactions instead often require complementary interfaces. That is, two proteins X and Y sharing many neighbors will have similar interfaces, but typically this does not guarantee that X and Y directly interact with each other. Instead, Y might interact (blue link) with a neighbor of X (protein D). Such a link can be predicted through a Quadrangular Closure Principle, \ie, by using paths of length 3, since these paths identify similar nodes to the known partners ($P = \mathbf{A}S=\mathbf{A}^3$). {\bf d)} The two proteins Y and D are more likely to interact if they are linked by multiple paths of length three in the network. Source:  \citep{kovacs2019network}.}
\label{social-PPI}
\end{figure}

As we have seen up to now, most of the local similarity-based link prediction algorithms rely on the common neighbor idea, rooted in social network analysis, that the more common friends two individuals have, the more likely they know each other. This principle is known as {\em triadic closure} and leads to the key role of paths of length two. On the other hand, semi-local indices also rely on paths of length three, which can play a key role for link prediction in some contexts. Notably, this is the case of the protein-protein interaction networks (PPI): proteins turn out to interact not if they are similar to each other, but if one of them is similar to the partners of the other \citep{kovacs2019network}, see fig. \ref{social-PPI}. Moreover, PPI were shown to display a kind of LCP architecture where protein complexes are confined in topologically isolated network structures, which are often coincident with functional network modules that play an important role in molecular circuits \citep{muscoloni2018local}. Similar observations do apply to foodweb trophic relations and world trade network transitions \citep{muscoloni2018local}. In all these cases, similarity-based link prediction metrics have to rely on the {\em quadrangular closure} principle \citep{daminelli2015common}. For instance, a {\em degree-normalized L3 score} (DnL3) has been defined in the context of PPI by \citet{kovacs2019network}:

\begin{equation}\label{s_dnl3}
s_{ij}^{\mbox{\tiny[DnL3]}}=\sum_{l\neq i,m}\sum_{m\neq j}\frac{a_{il}a_{lm}a_{mj}}{\sqrt{k_l k_m}}=
\sum_{l\in\Gamma_i}\sum_{m\in\Gamma_j}\frac{a_{lm}}{\sqrt{k_l k_m}}
\end{equation}
a definition that can be extended to account for paths of length $n$ \citep{muscoloni2018local}. Note that the role of paths of length three is obvious in the case of bipartite networks: since in this case connections exist only across (and not within) two sets of nodes, by definition nodes belonging to different sets can be connected only by paths of odd length \citep{kunegis2010link} (see also the discussion at the end of the chapter).

\paragraph{Global indices.} Finally, global indices are computed by using the topological information encoded into the whole network. As mentioned above, the Katz index \citep{katz1953new} is the extension of the LP index that accounts for paths of any length, i.e.

\begin{equation}\label{s_katz}
s_{ij}^{\mbox{\tiny[Katz]}}=\sum_{m=1}^{\infty}\beta^m (\mathbf{A}^m)_{ij}=[(\mathbf{I}-\beta\mathbf{A})^{-1}-\mathbf{I}]_{ij}
\end{equation}
where the damping factor $\beta$ suppresses the exponential proliferation of longer paths in the network. To ensure convergence, $\beta$ must be smaller than the reciprocal of the largest eigenvalue of $\mathbf{A}$. Notice that the computational complexity of LP in an uncorrelated network is $O(N\avg{k}^n)$, converging to the complexity of the Katz index that is $O(N^3)$. Experimental evidence suggests that the optimal $n$ is positively correlated with the average shortest distance of the network \citep{lu2009similarity}.

Another similar index is the {\em local random walk} (LRW) index \citep{liu2010link}, defined as

\begin{equation}\label{s_lrw}
s_{ij}^{\mbox{\tiny[LRW]}}(t)=\sum_{t'=1}^t\left[q_i \pi_{ij}(t')+q_j \pi_{ji}(t')\right]
\end{equation}
where $q_i\propto k_i$ is the probability that the walker is initially on node $i$ and $\pi_{ij}(t')$ is the probability that this walker lands at node $j$ after $t'$ steps. Hence, this index measures the overall probability that a walker starting on either $i$ and $j$ will land to the other node within $t$ steps. Clearly, $t$ denotes the maximum length of the considered paths. For $t\to\infty$ the index reflects the steady-state behavior of the walker and is known under the name of {\em random walk with restart} (RWR) index \citep{tong2006fast}. Otherwise, adding a small amount of resistance on each link of the network (and a larger amount if the link has not been traveled yet), the index becomes the {\em random walk with resistance} (RWS) index \citep{lei2012novel}.

Two related variants are the SimRank \citep{jeh2002simrank} and the Global Leicht-Holme-Newman (GLHN) index \citep{leicht2006vertex}, both defined in a recursive way using the concept that two nodes are similar if their immediate neighbors are themselves similar. More quantitatively, 

\begin{equation}\label{s_sr-lhn'}
s_{ij}^{\mbox{\tiny[SimRank]}}=\frac{1}{k_ik_j}\sum_{l\in\Gamma_i}\sum_{m\in\Gamma_j}s_{lm}^{\mbox{\tiny[SimRank]}},\qquad s_{ij}^{\mbox{\tiny[GLHN]}}\simeq \frac{\alpha}{\lambda_1}\sum_l A_{il}\frac{k_l}{k_i}s_{lj}^{\mbox{\tiny [GLHN]}}.
\end{equation}
A different index is the Average Commute Time (ACT) index \citep{fouss2007random}, which builds on the idea that nodes $i$ and $j$ are similar if only a few steps are required to go from one to the other. It is defined as
 
\begin{equation}\label{s_act}
s_{ij}^{\mbox{\tiny[ACT]}}=\frac{1}{l_{ii}^++l_{jj}^+-2l_{ij}^+},
\end{equation}
where the denominator is proportional to the average commute time between $i$ and $j$ and is expressed using the elements of the pseudoinverse of the Laplacian matrix $\mathbf{L}^+$. 

The structural perturbation method (SPM) \citep{lu2015towards}, instead, relies on the spectrum of the adjacency matrix $\mathbf{A}$ (i.e. the set of its eigenvalues $\{\lambda_n\}_{n=1}^N$ and normalized eigenvectors $\{\mathbf{v}_n\}_{n=1}^N$) reflecting the network structural features. Indeed, any perturbation $\Delta\mathbf{A}$ of the network causes a shift of each eigenvalue that, at the first-order, is given by the projection of the perturbation on the corresponding (unperturbed) eigenvector: $\Delta\lambda_n\simeq \mathbf{v}_n^T(\Delta\mathbf{A})\mathbf{v}_n$. The perturbed matrix can, then, be approximated as $\tilde{\mathbf{A}}\simeq\sum_n(\lambda_n+\Delta\lambda_n)\mathbf{v}_n\mathbf{v}_n^T$. Given that the first-order spectral shifts induced by independent perturbations are strongly correlated, the entries of the approximated perturbed matrix can be taken as link prediction scores:

\begin{equation}\label{s_SPM}
s_{ij}^{\mbox{\tiny[SPM]}}=\left[\sum_n\Delta\lambda_n\mathbf{v}_n\mathbf{v}_n^T\right]_{ij}.
\end{equation}
Notice that if the perturbation does not significantly change the structural features of the network, the eigenvectors of the unperturbed and perturbed matrices do not change much: if this is the case, then $\tilde{\mathbf{A}}$ and $\mathbf{A}+\Delta\mathbf{A}$ become very close. The ``regularity'' of a network can then be measured using the similarity (or \emph{structural consistency}) between these two matrices, without any prior knowledge on the network organization \citep{lu2015towards}. The rationale behind SPM is then that a missing link is likely to exist if its appearance has only a small effect on the structural features of the network.

Actually this is an important observation that applies to all link prediction methods, which are designed to echo the fundamental organization and growth rules of complex networks: the precision of a link prediction algorithm tells us the extent to which the link formation process in the network can be explained by this algorithm. Missing links are, then, easy to predict if their addition causes little structural changes to the network, and hard to predict otherwise \citep{lu2015towards}.

\paragraph{Information theoretical methods.} We dedicate this section to approaches based on information theory which are more sophisticated that standard similarity-based indices but rest upon the same rationale. The mutual information (MI) index \citep{tan2014link} is defined as

\begin{equation}\label{s_MI}
s_{ij}^{\mbox{\tiny[MI]}}=-I\left[a_{ij}\middle\vert(\mathbf{A}^2)_{ij}\right]\equiv -I\left[a_{ij}\right]+I\left[a_{ij},(\mathbf{A}^2)_{ij}\right]
\end{equation}
namely the self-information\footnote{The self-information $I_i$ of an event $i$ whose probability is $p_i$ equals (minus) the logarithm of the occurrence probability of the event itself, i.e. $I_i=-\ln p_i$.} of the existence of a link between nodes $i$ and $j$ conditional on the presence of common neighbors. An explicit expression of the MI index can be derived in the case of uncorrelated networks \citep{tan2014link}.

This recipe can be easily extended to the case of multiple structural feature because the values of information brought by these features are additive. For instance, the neighbor set information (NSI) index \citep{zhu2015information} uses the structural information given by the common neighbors and by the links across the two neighbor sets

\begin{equation}\label{s_NSI}
s_{ij}^{\mbox{\tiny[NSI]}}=-I\left[a_{ij}\middle\vert(\mathbf{A}^2)_{ij}\right]-\lambda I\left[a_{ij}\middle\vert(\mathbf{A}^3)_{ij}\right]
\end{equation}
where $\lambda$ is a hybridization parameter. 

The path entropy (PE) index, instead, considers the self-information of all shortest paths between a node pair (with penalization to long paths) and can be expressed as \citep{xu2016link}

\begin{equation}\label{s_PE}
s_{ij}^{\mbox{\tiny[PE]}}=-I\left[a_{ij}\middle\vert\bigcup_{m>1}(\mathbf{A}^m)_{ij}\right]
\end{equation}
where the self-information of a path is approximated by the sum of the self-information of its constituent links. The idea behind this formulation is that paths with large self-information are critical substructures for the network and greatly reduce the self-information of the link between its end nodes. Note that MI, NSI and PE make use of an increasing amount of information, hence can be respectively classified within the local, semi-local and global category.

\section{Model-based methods}

\begin{figure}[t!]
\centering
\includegraphics[width=0.7\textwidth]{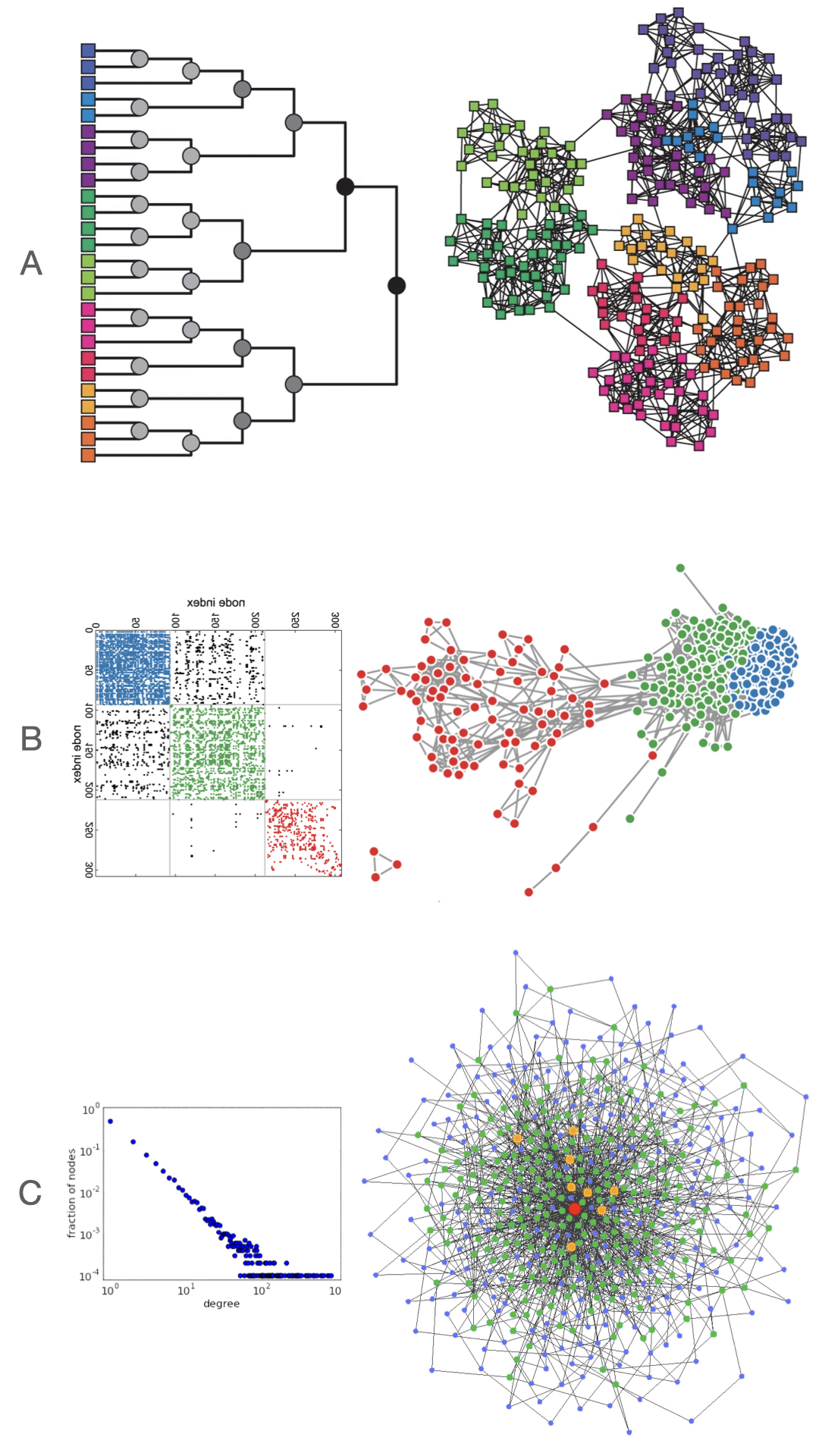}
\caption{Sample network structures at the basis of model-based methods. 
(A) Dendogram and hierarchical network structure. 
(B) Block-structured adjacency matrix and stochastic blockmodel network structure.
(C) Scale-free degree distribution and maximum-entropy network structure obtained by constraining the degree sequence. 
Figure adapted from \citep{clauset2008hierarchical,larremore2013network,oikonomou2006effects}.}
\label{structures}
\end{figure}

Methods belonging to this class build on a set of assumptions about the organizing principles of the network (see fig. \ref{structures}). 
The idea is to maximize the likelihood of the observed network under a given (probabilistic) parametric model, which in turns allows evaluating the likelihood of any non-existent link. 
From a practical viewpoint, the computational complexity of model-based methods is typically very high (much higher than that of similarity-based methods). 
In addition, they can be very accurate in predicting links but only when the underlying model properly describes the network and can infer its connection probabilities. 
Indeed, a given model-based method should represent the optimal link prediction strategy when the network is a direct realization of the model assumed \citep{perez2019predictability}. 
Otherwise, local methods can perform much better by mimicking the network growth process and thus acting as topological learning rules \citep{muscoloni2017localring}.

\paragraph{Hierarchical model.} This method is based on the evidence that many real-world networks are hierarchically organized, meaning that nodes can be divided into groups, further subdivided into groups of groups and so forth over multiple scales. Extracting such a hierarchy can, then, be useful to infer missing links. The hierarchical structure of a network can be represented as a dendrogram $D$ with $N$ leaves (corresponding to the nodes of the network) and $N-1$ internal nodes (representing the relationships among the descendant nodes in the dendrogram). The \emph{Hierarchical Random Graph Model} \citep{clauset2008hierarchical} assigns a probability $p_r$ to each internal node $r$, so that the connection probability of a pair of nodes (leaves) is equal to $p_{r'}$, where $r'$ is the lowest common ancestor of these two nodes. In order to find the hierarchical random graph $(D,\{p_r\})$ that best fits an observed real network $\mathbf{A}$, one assumes that all hierarchical random graphs are \emph{a priori} equally likely. Then, the likelihood that a given model $(D,\{p_r\})$ is the correct explanation of the data is

\begin{equation}
\mathcal{L}(D,\{p_r\})=\prod_rp_r^{E_r}(1-p_r)^{L_rR_r-E_r}
\label{likelihood-hierarchical}
\end{equation}
where $E_r$ is the number of links whose endpoints have $r$ as their lowest common ancestor in $D$ and $L_r$ and $R_r$ are, respectively, the number of leaves in the left and right subtrees rooted at $r$. Given a dendrogram $D$, the likelihood is maximized by the coefficients 

\begin{equation}
p_r=\frac{E_r^*}{L_rR_r},\:\forall\:r
\end{equation}
i.e. the fraction of potential links between the two subtrees of $r$ that actually appear in the network.  Markov Chain Monte Carlo (MCMC) method can be used to sample dendrograms with probability proportional to their likelihood. Once an ensemble of dendrograms is generated, the ensemble average of the connection probability for each pair of unconnected nodes represents the prediction score of the corresponding link.

\paragraph{Stochastic Block Model.} The SBM, which was introduced in the previous chapter, is an extremely popular network model: given a partition $\mathcal{M}$ of the network such that each node belongs to exactly one group, the model assumes that the connection probability for any two nodes belonging respectively to groups $r$ and $s$ can be indicated as $p_{rs}$. Hence, the likelihood of the observed network structure can be written as \citep{guimera2009missing}

\begin{equation}
\mathcal{L}(\mathbf{A}|\mathcal{M})=\prod_{r\leq s}p_{rs}^{L_{rs}}(1-p_{rs})^{N_{rs}-L_{rs}}
\label{likelihood-SBM}
\end{equation}
where $L_{rs}$ is the observed number of links between nodes in groups $r$ and $s$ (or within group $r$, in case $r=s$) and $N_{rs}$ is the maximum number of such links in a complete graph\footnote{$N_{rs}$ is the total number of node pairs $\binom{N_r}{2}$ whenever $r=s$, and $N_rN_s$ otherwise.}. Similarly to the previous case, the optimal set of $\{p_{rs}\}$ that maximizes the likelihood is $p_{rs}=L_{rs}^*/N_{rs}$. Under this framework, the \emph{reliability} of a link between nodes $i$ and $j$ (belonging respectively to groups $r$ and $s$) is:

\begin{equation}
R_{ij}\equiv\mathcal{L}(a_{ij}=1|\mathbf{A})=\frac{1}{Z}\sum_{\mathcal{M}}\left(\frac{l_{rs}+1}{N_{rs}+2}\right)e^{-H(\mathcal{M})},
\label{link_reliability}
\end{equation}
where $H(\mathcal{M})=\sum_{r\leq s}\left[\ln(N_{rs}+1)+\ln\binom{N_{rs}}{L_{rs}}\right]$ is a function of the partition and $Z=\sum_{\mathcal{M}}e^{-H(\mathcal{M})}$. As in the previous case, no prior knowledge on the true model is assumed, meaning that $p(\mathcal{M})$ is constant. 

Since the number of different partitions of $N$ elements grows faster than any finite power of $N$, summing over all partitions is not possible in practice. 
Therefore one can employ numerical MCMC sampling \citep{guimera2009missing} or other greedy stochastic sampling strategies \citep{liu2013correlations}.

\paragraph{Maximum-entropy models.} ERG models that reproduce the local features of the network, discussed in chapter 2, can be used for link prediction as well \citep{parisi2018entropy}. According to the Configuration Model (CM) -- built by maximizing the Shannon entropy of the network ensemble enforcing the node degrees as constraints -- the likelihood of the observed network is

\begin{equation}\label{likelihood-CM}
\mathcal{L}(\mathbf{A}|\vec{x})=\prod_{i<j}p_{ij}^{a_{ij}}(1-p_{ij})^{1-a_{ij}}
\end{equation}
where $p_{ij}=\frac{x_ix_j}{1+x_ix_j}$ is the probability that nodes $i$ and $j$ establish a connection. 
These connection probabilities $\{p_{ij}\}$ can be used as link prediction scores for the non-observed links of the network. Notice also that in the limit of sparse networks, the methods simplifies to $p_{ij}\simeq k_ik_j$, namely the preferential attachment method of eq. \eqref{s_pa}. The entropy-based approach to link prediction has been recently extended to account for non-linear constraints \citep{adriaens2020scalable}, a generalization that has been shown to outperform other competing algorithms.

\paragraph{Hamiltonian models.} As we have seen, the configuration model can be framed in statistical physics terms using the Hamiltonian $H(\mathbf{A})=\sum_i\theta_ik_i(\mathbf{A})$. The probability of the network $\mathbf{A}$, thus, becomes $P(\mathbf{A}|\vec{\theta})=\frac{e^{-H(\mathbf{A},\vec{\theta})}}{Z(\vec{\theta})}$ where $Z(\vec{\theta})$ is the partition function. Naturally, other choices of $H$ are possible: for instance, taking inspiration from the similarity-based indices relying on the short loops in the network, one can build an Hamiltonian of the form \citep{pan2016predicting}

\begin{equation}\label{HamC}
H(\mathbf{A})=-\sum_{m\le m_c} \beta_m \ln (\mbox{Tr}[\mathbf{A}^m])
\end{equation}
where the $m$-th term of the sum counts (approximately) the number of closed walks of length $m$ and the sum runs up to the maximum loop length $m_c$. The logarithm is used to rescale each term to the same magnitude, given that $\mbox{Tr}[\mathbf{A}^m]$ grows exponentially with the leading eigenvalue of the adjacency matrix. Differently from the CM, this formulation does not admit a closed-form solution for the partition function. This is a typical situation where maximum pseudo-likelihood methods have to be used to estimate the Lagrange multipliers $\{\beta_m\}$: in this case, one can replace the joint likelihood of the links of the network with the product over the conditional probability of each link, given the rest of the network; then, each non-observed link is scored by the conditional probability of adding it to the network.

\subsection{Hyperbolic latent space models}

Latent space network models assume the existence of an embedding space where the network nodes are located, such that connections are established with probabilities that decrease with the latent distance between nodes. Link prediction in this context boils down to ranking unconnected node pairs in order of increasing latent distances between them: the closer the two unlinked nodes in the latent space, the higher the probability of a missing link. 

The most popular model of this kind assumes that the latent space is hyperbolic; by doing so, it can reproduce the typical emerging properties of complex networks (sparsity, self-similarity and hierarchy, scale-freeness, small-worldness, and modular structure)~\citep{serrano2008self,krioukov2010hyperbolic,papadopoulos2012popularity,muscoloni2018nonuniform}. 
In particular, the {\em Hyperbolic Random Graphs} (HRG) model is defined on the two-dimensional hyperbolic disk of constant negative curvature $K=-1$, so that each node $i$ is identified by two hyperbolic coordinates: its radius $r_i$ and angle $\theta_i$. Therefore, the hyperbolic distance $x_{ij}$ between two nodes $i$ and $j$ is given by the hyperbolic law of cosines
\begin{equation}
\cosh x_{ij} = \cosh r_i \cosh r_j - \sinh r_i \sinh r_j \cos \Delta \theta_{ij}\label{eq:hypercos} 
\end{equation}
where $\Delta \theta_{ij} \simeq \theta_i - \theta_j$ is the angle between $i$ and $j$. In the HRG, $i$ and $j$ are connected with probability 
\begin{equation}
p\left(x_{ij}\right) = \left(1 + e^{\frac{x_{ij} - R}{2 T}}\right)^{-1}
\end{equation}
where the model parameters are the hyperbolic disk radius $R>0$ and the temperature $T \in [0,1)$. 

Link prediction with hyperbolic geometry is a two-step procedure. The first step, {\em network embedding}, consists is inferring node coordinates. 
The typical approach consists in finding the set of coordinates that maximize the likelihood that the network has been generated by a HRG  \citep{papadopoulos2015network,wang2016link}. 
Since node pairs are connected independently, the likelihood is given by
\begin{equation}
\mathcal{L}\left(  a_{ij}| \{r_i,\theta_i\},  T, R\right) = \prod_{i < j} \left[p\left(x_{ij}\right)\right]^{a_{ij}}  \left[1 -p\left(x_{ij}\right)\right]^{1 - a_{ij}};\label{eq:likelihood}
\end{equation}
alternative approaches are based on Laplacian eigenmaps~\citep{alanis-lobato2016efficient} and coalescent embedding \citep{muscoloni2017machine}. 
After coordinates have been inferred, hyperbolic distances can be used to find the most likely missing link candidates. 

Hyperbolic models have been found to make good link predictions only provided that node coordinates are inferred accurately \citep{kitsak2019linkprediction}. 
In this case, they work well also if if the fraction of missing links is high, and as the other model-based methods can effectively predict the nonlocal links (those between nodes that do not have any common neighbors). Link prediction performance can be further improved by computing the distances over the network topology, that is, ranking unconnected node pairs according to their hyperbolic shortest path length (the sum of the hyperbolic distances over the shortest path between these two nodes) \citep{muscoloni2018leveraging}.


\section{Network reconstruction from noisy data}

By postulating an underlying parametric model for the network under consideration, model-based methods have been widely used not only to predict individual missing or spurious link but also to reconstruct the more reliable network structure when the observed network data is noisy. Here we discuss the case of the SBM which is, by far, the most popular model in the literature. Given an observed network $\mathbf{A}$, the \emph{reliability} of an entire network $\tilde{\mathbf{A}}$ is the likelihood that $\tilde{\mathbf{A}}$ is the true network given the observation $\mathbf{A}$ (and the model belonging to the SBM family) \citep{guimera2009missing}. Analogously to eq. \eqref{link_reliability},

\begin{equation}
R(\tilde{\mathbf{A}})\equiv\mathcal{L}(\tilde{\mathbf{A}}|\mathbf{A})=\frac{1}{Z}\sum_{\mathcal{M}}e^{\sum_{r\leq s}\left[\ln\left(\frac{N_{rs}+1}{2N_{rs}+1}\right)+\ln\left(\frac{\binom{N_{rs}}{L_{rs}}}{\binom{2N_{rs}}{L_{rs}+\tilde{L}_{rs}}}\right)\right]}e^{-H(\mathcal{M})}.
\label{network_reliability}
\end{equation}
Finding the network $\tilde{\mathbf{A}}$ whose reliability is maximum is, however, computationally demanding. A simple alternative, greedy algorithm consists in evaluating the link reliability for all pairs of nodes and, then, iteratively removing the links with lowest reliability while adding non-links with high reliability. Each move is accepted only if the total network reliability increases.\\

A recent generalization of this approach consists in coupling the SBM generative process with a noisy measurement model, and performing Bayesian statistical inference of this joint model \citep{peixoto2018reconstructing}. Consider a true network $\tilde{\mathbf{A}}$ and a noisy observation of it, say $\mathbf{A}$. The inference framework to obtain $\tilde{\mathbf{A}}$ from $\mathbf{A}$ is based on:

\begin{itemize}
\item The network generating process. Using the SBM, a network is generated with probability
\begin{equation}
P(\tilde{\mathbf{A}}|\mathcal{M},\{p_{rs}\})=\prod_{r\leq s}p_{rs}^{\tilde{L}_{rs}}(1-p_{rs})^{\tilde{N}_{rs}-\tilde{L}_{rs}};
\end{equation}
\item The data generating process $P(\mathbf{A}|\tilde{\mathbf{A}},\mu,\nu)$ where $\mu$ is the probability of observing a missing link (i.e. a link that exist in $\tilde{\mathbf{A}}$ but not in $\mathbf{A}$) and $\nu$ is the probability of observing a spurious link (i.e. a link that exist in $\mathbf{A}$ but not in $\tilde{\mathbf{A}}$). Without any prior knowledge, these error rates lie anywhere in the unit interval. 
\end{itemize}

Under these assumptions, the final likelihood for the observed network $\mathbf{A}$ is identical to that of an effective SBM, i.e.

\begin{eqnarray}
P(\mathbf{A}|\mu,\nu,\mathcal{M},\{p_{rs}\})&=&\sum_{\tilde{\mathbf{A}}}P(\mathbf{A}|\tilde{\mathbf{A}},\mu,\nu)P(\tilde{\mathbf{A}}|\mathcal{M},\{p_{rs}\})\nonumber\\
&=&\prod_{r\leq s}{q_{rs}}^{L_{rs}}(1-{q_{rs}})^{N_{rs}-L_{rs}}
\end{eqnarray}
with $q_{rs}=(1-\mu-\nu)p_{rs}+\nu$ being an effective SBM-induced probability (scaled and shifted by the noise).

If the network partition $\mathcal{M}$ is known and if the number of modules is very small compared to the total number of nodes, the posterior distribution for $q_{rs}$ should be peaked around the maximum likelihood estimate $L_{rs}/N_{rs}$. In this case, computing the joint posterior distribution for $\mu$ and $\nu$ returns constraints implying that the inferred error rates are bounded by the maximum and minimum inferred connection probabilities \citep{peixoto2018reconstructing}:

\begin{equation}\label{eq:bounds}
\hat\nu\leq\min_{rs}\{L_{rs}/N_{rs}\},\qquad\hat\mu\leq 1-\max_{rs}\{L_{rs}/N_{rs}\};
\end{equation}
these bounds mean that $\mu$ ($\nu$) is small if we do (do not) observe many links between groups. Hence, as long as the inferred SBM probabilities are sufficiently heterogeneous (meaning that the observed network is sufficiently structured - besides being properly described by a SBM), the inferred error rates should be contained into narrow intervals. The key observation here is that the modifications induced by the error rates uniformly affect every link and non-link, thus with structured models we can exploit the observed correlations in the measurements to infer the underlying network (and even the error rates themselves).

Lastly, the reconstruction procedure consists of determining $\tilde{\mathbf{A}}$ from the posterior distribution $P(\tilde{\mathbf{A}}|\mathbf{A})=\frac{P(\mathbf{A}|\tilde{\mathbf{A}})P(\tilde{\mathbf{A}})}{P(\mathbf{A})}$, which defines an ensemble of possibilities for the underlying network that incorporates the amount of uncertainty resulting from the measurement. However, the marginal network probability $P(\tilde{\mathbf{A}}) = \sum_{\mathcal{M}}P(\tilde{\mathbf{A}}|\mathcal{M})P(\mathcal{M})$ involves an intractable sum over all possible network partitions; hence, instead of directly computing the posterior, one computes the joint posterior $P(\tilde{\mathbf{A}},\mathcal{M}|\mathbf{A}) \sim P(\mathbf{A}|\tilde{\mathbf{A}})P(\tilde{\mathbf{A}}|\mathcal{M})P(\mathcal{M})$, which defines simultaneously an ensemble of possibilities for the underlying network and its large-scale hierarchical modular organization (and involves only quantities that can be computed exactly). The original posterior can then be obtained by marginalization as $P(\tilde{\mathbf{A}}|\mathbf{A})  = \sum_{\mathcal{M}} P(\tilde{\mathbf{A}},\mathcal{M}|\mathbf{A})$ and sampled using MCMC methods.

The described setup is sufficiently general and can be used with any variant of the SBM, like the degree-corrected and the hierarchical ones, as well as different models for the noise. Additionally, this framework has been extended to situations in which several (possibly repeated) observations are available for a network \citep{newman2018network} and when explicit information on the measurement error is available \citep{peixoto2018reconstructing}, as well as to include data on dynamical processes taking place on the network \citep{peixoto2019network}.


\section{Quality metrics for link prediction}

Network reconstruction at the microscale is typically assessed against its ability to correctly predict the presence of individual links (i.e. the position of 1s in the binary adjacency matrix) and their absence (i.e. the position of 0s). From this purely topological perspective, a reconstruction algorithm can be seen as a binary classifier that determines if a given pair of unconnected nodes is linked or not. Hence, the evaluation metrics are those derived from the {\em confusion matrix} (see \citep{fawcett2006introduction} for an exhaustive treatment of the topic). Given a ``true'' binary matrix $\mathbf{A}$ and the reconstructed one $\hat{\mathbf{A}}$, for each pair of nodes we have four possible combinations:

\begin{itemize}
\item $a_{ij}=1$ and $\hat{a}_{ij}=1$: an existing link has been correctly predicted and we have a {\em true positive};
\item $a_{ij}=1$ but $\hat{a}_{ij}=0$: an existing link has been incorrectly predicted as missing and we have a {\em false negative};
\item $a_{ij}=0$ but $\hat{a}_{ij}=1$: a missing link has been incorrectly predicted as existing and we have a {\em false positive};
\item $a_{ij}=0$ and $\hat{a}_{ij}=0$: a missing link has been correctly predicted and we have a {\em true negative}.
\end{itemize}

The total number of events within these four categories, labeled respectively as TP, FN, FP and TN, are used to define various performance metrics, such as:

\begin{itemize}
 \item the {\em sensitivity} ({\em true positive rate}) or {\em recall}, i.e. the fraction of existing links that are correctly recovered
\begin{equation}
\mbox{TPR}=\frac{\mbox{TP}}{\mbox{TP}+\mbox{FN}};
\end{equation}
 \item the {\em specificity} ({\em true negative rate}), the fraction of non-existing links that are correctly recovered
\begin{equation}
\mbox{TNR}=\frac{\mbox{TN}}{\mbox{TN}+\mbox{FP}}.
\end{equation}
\end{itemize}

\begin{figure}[t!]
\centering
\includegraphics[width=\textwidth]{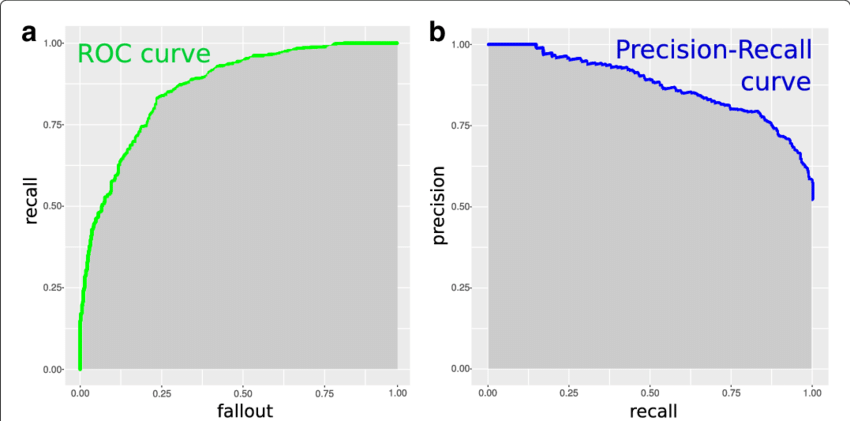}
\caption{(a) Example of ROC curve, defined by plotting the \emph{recall} score (or \emph{true positive rate}) VS the \emph{fallout} score (or \emph{false positive rate}). The grey region denotes the \emph{area under the ROC curve} (AUC) (b) Example of \emph{precision-recall} curve, with the \emph{precision} score on the vertical axis and the \emph{recall} score on the horizontal axis. The grey region denotes the \emph{area under the precision-recall curve} (AUPR). Source: \citep{chicco2017ten}.}
\label{AUC}
\end{figure}

A good classifier should be able to achieve high values for both these quantities. Unfortunately, this may be rather difficult. Indeed, a link prediction method characterized by a high discrimination threshold (meaning that only the candidate links with the largest reliability score are predicted as actually missing) likely generates many false negatives, thus achieving a low TPR but a large TNR. Instead if the discrimination thresholds is low, the method likely generates many false positives, thus achieving a low TNR (i.e. high {\em false positive rate} or {\em fallout}, defined as $\mbox{FPR}=1-\mbox{TNR}$) but a large TPR. 
Plotting the TPR against the FPR (i.e. \emph{recall} VS \emph{fallout}) as the discrimination threshold is varied generates the ROC (\emph{Receiver Operating Characteristic}) curve, a graphical way to illustrate the performance of a binary classifier (see fig. \ref{AUC}). Under this representation, the point $(0,1)$ corresponds to the perfect classifier, yielding neither false positives nor false negatives, whereas a random classifier lies along the line of no-discrimination, i.e. the diagonal between $(0,0)$ and $(1,1)$. The {\em area under the ROC curve} (AUC) has been largely used to quantify the overall performance of a classifier.

An equivalent formulation of AUC is the probability that the classifier ranks a random missing link (i.e., a link in $E_P$) higher than a random non-existing one (i.e., a link in $U\setminus E$). This is equivalent at carrying out the \emph{Mann-Whitney U test} computing the quantity

\begin{equation}
\mbox{AUC}=\frac{n'+\tfrac{1}{2}n''}{n}
\end{equation}
where $n$ is the number of times a missing link gets a higher score than a non-existent one, $n''$ is the number of ties in this comparison and $n$ is the total number of comparisons (i.e. the number of missing links times the number of non-existent links). One achieves the result $\mbox{AUC}=1/2$ if all links to be predicted get the same score and $\mbox{AUC}=1$ if all missing links occupy the top positions of the ranking.\\

The use of AUC has, however, been questioned since in link prediction applications it is common to deal with imbalanced network data, where one class (missing links) is over-represented with respect to the other (existing links). This happens because most real networks are sparse, with a number of links $L$ which is proportional to the number of nodes $N$ rather than to the number of node pairs, i.e. $O(N^2)$. Hence, when testing a link prediction scheme, the set of non-observed links will be dominated by the set of non-existing links (set $U\setminus E$). The situation is even more intricate because a non-existing link can result both from measuring the absence of the interaction or from not measuring the interaction at all, and typically network data come with no distinction between these two cases \citep{peixoto2018reconstructing}.

In any case, the result of the class imbalance is that correctly guessing non-existing links (the true negative instances) is far easier than correctly guessing existing links (the true positives) simply because of their absolute numbers. As such, weighting sensitivity and specificity equally (as AUC does) can lead to misleading conclusions on the performance of a link prediction method. A possible solution consists in replacing the ROC curve with the \emph{precision-recall} curve \citep{yang2015evaluating}, where the {\em precision} (or \emph{positive predicted value}) is defined as $\mbox{PPV}=\frac{\mbox{TP}}{\mbox{TP}+\mbox{FP}}$ (see fig. \ref{AUC}). This is because the optimization of the ROC curve tends to maximize both the TP (through the \emph{recall}) and the TN (through the \emph{fallout}), whereas the optimization of the precision-recall curve tends to maximize only the TP (through both the \emph{recall} and the \emph{precision}), without directly considering the TN (absent in both formulas).
An alternative recently-proposed solution consistis employing the \emph{Matthews correlation coefficient}

\begin{equation}
\mbox{MCC}=\frac{\mbox{TP}\cdot\mbox{TN}-\mbox{FP}\cdot\mbox{FN}}{\sqrt{(\mbox{TP}+\mbox{FP})\cdot(\mbox{TP}+\mbox{FN})\cdot(\mbox{TN}+\mbox{FP})\cdot(\mbox{TN}+\mbox{FN})}}
\end{equation}
which correctly takes into account the size of the confusion matrix elements \citep{chicco2017ten,chicco2020advantage}.

\chapter{Conclusions}
\label{chap5}

When studying social, economic and biological systems, one has often access to limited information about the structure of the underlying network. The need to compensate for the scarcity of data has led to the birth of a research field known as \emph{network reconstruction}. In the previous chapters we have discussed techniques to reconstruct networks at the macro-, meso- and micro-scale by either \emph{constraining} or \emph{targeting} very general network structures. We now provide a brief overview of recent research in the field that has remained outside this dissertation as well as of future perspectives.

\paragraph{Matrix completion.} A problem that is closely related to network reconstruction and link prediction is that of matrix completion: \emph{given a partially observed matrix $\mathbf{W}$, under which conditions is it possible to recover the whole matrix exactly?} While the problem is, by definition, underdetermined, much of the research in computational mathematics focuses on the case of low-rank matrices - where the rank $r$ is the number of linearly independent matrix columns. This case is of interest when a few variables are assumed to explain the structure of the matrix (e.g. in user-item ratings data, user preferences can often be described by a few factors). For a low-rank matrix the number of independent entries is $O(Nr)$, where $N$ is the dimension of $\mathbf{W}$, i.e. much smaller than the total number $N^2$. Thus, in this case a small number of observed elements could be sufficient for recovering it. 
Formally, the matrix completion problem can be stated as follows: finding the lowest rank matrix $\mathbf{X}$ that matches $\mathbf{W}$, over the set $E$ of observed links:

\begin{equation}
\underset {\mathbf{X}}{\min}\:\text{rank}(\mathbf{X})\quad\text{subject to}\:x_{ij}=w_{ij},\:\:\forall\:\:i,j\in E.
\end{equation}

While this is an NP-hard optimization problem, under proper assumptions on the sampling of the observed entries - and sufficiently many sampled entries - it can be proven to admit a unique solution, with high probability \citep{candes2009matrix}. A typical assumption is that the set of observed entries is sampled randomly - for instance, that each entry is observed with equal probability \citep{candes2010matrix}. Another important assumption concenrs the coherence of non-zero elements, which should be homogeneously distributed over the whole matrix. The minimum number of observed entities for the problem to be solvable has been estimated as of the order $O(nr\log n)$ \citep{candes2010power,xu2018minimal}. Various matrix completion algorithms have been proposed, including convex relaxation-based algorithm \citep{candes2009matrix}, gradient-based algorithm \citep{keshavan2010matrix} and alternating minimization-based algorithm \cite{jain2013low} - the approach varies depending on whether the rank of the matrix is known or not. We refer the interested reader to \citet{nguyen2019low} for a recent, accessible review of the field.

\paragraph{Link prediction for directed and weighted networks.} Specifically for the link prediction topic, directed networks are not easy to deal with. Since in this case paths do depend on links directionality, the simple triadic closure principle at the basis of the local similarity-based indices has to be modified to take into account the whole family of triadic motifs \citep{alon2007network}. Hence, for instance, the likelihood of a link $i\to j$ will be in general different from the likelihood of $j\to i$.
The case of weighted networks is even more complicated, since no clear indication exists on how link weights should contribute to the similarity or likelihood metrics. Additionally, contradictory results have been obtained on whether the strong ties \citep{murata2007link} or the weak ties \citep{lu2010link} are more important to successfully predict links. An even harder problem is to predict the weights of the non-observed links - a simple solution being to set weights of missing links proportional to their similarity scores \citep{zhao2015prediction}. In general, there are just a few techniques that generalize to these two more complex scenarios - notable cases being the SBM \citep{aicher2014learning} and the ERG-based models \citep{parisi2018entropy}.

\paragraph{Reconstructing valued networks.} A big challenge is the network reconstruction task for \emph{valued} networks, where links can have a \emph{categorical} meaning. An example is provided by signed social networks where links can be positive or negative: in this case, network reconstruction methods have to take into account the structural balance theory (which, for instance, implies that {\em the enemy of my enemy is my friend}) \citep{leskovec2010predicting}, thus requiring additional constraints to the local ones we discussed.

\paragraph{Reconstructing temporal networks.} On a different direction, the temporal evolution of links occurrences can be used to deal with the network reconstruction of temporal networks. Attempts have been done to encode the link persistence as a constraint, in order to explain the bursts of activity characterizing social networks. On the side of single link reconstruction, the evidence that older events (links) are in general less relevant to future links than recent ones can be directly incorporated into the similarity indices \citep{tylenda2009towards}. Additionally, algorithms may benefit from the fact that in temporal networks, node attract links depending not only on their structural importance but also on their current level of activity \citep{wang2017link}. Moreover, when temporal data have varying periodic patterns, tensor-based techniques turn out to be particularly effective \citep{dunlavy2011temporal}.

\paragraph{Reconstructing multiplex networks.} In multiplex networks the same set of nodes have different interaction patterns across various layers (like users interacting over multiple social networking platforms). Some attempts have been done to extend the Exponential Random Graph framework to multiplex networks: the resulting null models, however, are based on independent layers (i.e. are factorized, the factors representing single-layer null models), a characteristic that make them suitable to be used as benchmarks and not as proper reconstruction models (a notable exception is \citep{menichetti2014correlations}). For what concerns the link prediction topic, algorithms based on {\em meta-paths} (i.e. paths across the various layers) can be used to predict links \citep{sun2012when,jalli2017link}, since the likelihood of a link increases when the ending nodes have high neighborhood similarity over multiple layers \citep{hristova2016multilayer,hajibagheri2016holistic}. Multilayer mixed-membership SBM can be used in this case to develop model-based prediction methods \citep{debacco2017community}.

\paragraph{Reconstruction beyond networks.} Very recently, the ERG framework has been extended to \emph{simplicial complexes}, namely generalized network structures where interactions take place among more than two nodes at once. Although proper simplicial reconstruction techniques have not been developed yet, algorithms for link prediction can be implemented by extending the triadic closure principle as simplicial closure mechanisms \citep{benson2018simplicial}.

\paragraph{Reconstruction beyond Shannon entropy.} As discussed in chapter 2, the ERG formalism is based on the maximization of the Shannon entropy. However, it is in principle possible to employ non-Shannon functionals such as those belonging to the \emph{Cressie-Read family of divergences} or \emph{non-extensive functionals} such as the Renyi and the Tsallis entropy. Although of great interest from a purely mathematical perspective, solving the constrained maximization of such non-standard functionals may be problematic, the main challenges being that of 1) properly extending the likelihood maximization principle and 2) finding a suitable procedure for sampling the ensemble induced by the chosen functional. For a complete overview on the topic, see \citep{squartini2018reconstruction}.

\paragraph{Macroscale reconstruction: recent case studies.} Recently, ERG-based reconstruction techniques have been applied in the context of cryptocurrencies, e.g. for reconstructing the Bitcoin Lightning Network (BLN) representing transactions between users \citep{lin2020lightning}. Interestingly, for this network enforcing the out- and in-degree sequences (hence, using the DCM) is not enough to reproduce some centrality indicators: quantities like betweenness and eigenvector centralities are severely underestimated. This finding implies that the BLN is growing following an increasingly centralized fashion, thus becoming increasingly more fragile to failures and attacks (in particular, those aiming at splitting the network into separate components). The aforementioned analysis, however, concerns only the binary topological structure, and more work needs to be done on the weighted counterpart. Aside from the purely financial applications, a class of systems that has recently gained attention is that of \emph{interfirm networks}, i.e. networks whose nodes are \emph{firms} and whose links are \emph{buying/selling relationships} between them - a construction following the so-called \emph{supply chains} \citep{uchida2015interfirm,watanabe2015economics,goto2017estimating,ozaki2019modeling}. An even more interesting class of systems is that of bipartite bank-firm networks - where a link between a bank $i$ and a firm $j$ indicates that $i$ has lent money to $j$: representations like these allow the effects of shocks at the interface between the financial and the economic system to be properly understood \citep{huang2013cascading,poledna2018identifying}.

Reconstructing these kinds of networks ultimately aims at identifying the minimal amount of information needed to reproduce the so-called \emph{value chains}. This, in turn, opens up the possibility of generating realistic economic scenarios, to be used for testing the effects of disruptive economic events such as the crisis induced by the recent Covid-19 pandemics. The resilience of these reconstructed production networks could be, then, measured via the economic analogues of the regulatory stress tests; as their financial counterparts, they are expected to be very sensitive to the network features of the underlying system \citep{ramadiah2020network}.


\backmatter

\appendix
\chapter{Reconstructing bipartite networks}
\label{appb}

By definition, the adjacency matrix $\mathbf{A}$ of a bipartite network features two empty diagonal blocks of dimension $N_1\times N_1$ and $N_2\times N_2$, where  $N_1$ and $N_2$ are the number of nodes in the two sets. Hence a sufficient representation of a bipartite network is the $N_1\times N_2$ \emph{biadjacency matrix} $\mathbf{B}$, which is the off-diagonal block of the matrix $\mathbf{A}$. 
The generic element $b_{i\alpha}$ of the biadjacency matrix equals 1 if nodes $i$ and $\alpha$ are connected, and 0 otherwise. 
Latin and Greek characters are used here to indicate the two sets forming the bipartite graph.

Such a representation allows generalizing the CM to reconstruct bipartite networks as well. In particular, it is possible to consider the ensemble of undirected, binary, bipartite networks $\mathcal{B}$ and solve the maximization problem of eq. (\ref{cons}) under the constraints summed up by the Hamiltonian

\begin{equation}
H(\mathbf{B})=\sum_i\theta_ik_i(\mathbf{B})+\sum_\alpha\psi_\alpha h_\alpha(\mathbf{B})
\end{equation}
where $k_i(\mathbf{B})=\sum_\alpha b_{i\alpha}$ and $h_\alpha(\mathbf{B})=\sum_ib_{i\alpha}$ are the degrees of the nodes belonging to the two sets defining a bipartite network. The probability of a network in the ensemble is given by:

\begin{equation}
P(\mathbf{B})=\prod_i\prod_\alpha p^{b_{i\alpha}}_{i\alpha}(1-p_{i\alpha})^{1-b_{i\alpha}}
\end{equation}
where $p_{i\alpha}=\frac{x_iy_\alpha}{1+x_iy_\alpha}$ stands for the probability that a link exists between node $i$ and node $\alpha$; analogously to the CM, the parameters can be numerically determined by solving the system

\begin{eqnarray}
k_i(\mathbf{B}^*)&=&\sum_{\alpha}\frac{x_iy_\alpha}{1+x_iy_\alpha}=\langle k_i\rangle_\text{BiCM},\:\forall\:i\\
h_\alpha(\mathbf{B}^*)&=&\sum_{i}\frac{x_iy_\alpha}{1+x_iy_\alpha}=\langle h_\alpha\rangle_\text{BiCM},\:\forall\:\alpha
\end{eqnarray}
that defines the \emph{Bipartite Configuration Model} (BiCM) \citep{saracco2015randomizing}. Notice that the system of equations defining the BiCM is formally analogous to that of defining the DCM. If \emph{directed} bipartite networks are considered, instead, the system of equations to be solved becomes more complicated \citep{dejeude2019reconstructing}.\\

Starting from the BiCM recipe we can immediately define a bipartite version of the dcGM:

\begin{equation}\label{bib}
p_{i\alpha}^\text{dcGM}=\langle b_{i\alpha}\rangle_\text{dcGM}=\frac{zs_it_\alpha}{1+zs_it_\alpha},\:\forall\:i,\alpha
\end{equation}
where $s_i(\mathbf{V})=\sum_\alpha w_{i\alpha}$ and $t_\alpha(\mathbf{V})=\sum_iw_{i\alpha}$ are the strengths of the nodes belonging to the two sets and $\mathbf{V}$ represents the weighted biadjacency matrix of the considered bipartite network; for what concerns the estimation of weights, instead, the recipe reads

\begin{equation}\label{biw}
\langle w_{i\alpha}\rangle_\text{dcGM}=\frac{s_it_\alpha}{W},\:\forall\:i,\alpha.
\end{equation}

The model defined by eqs. (\ref{bib}) and (\ref{biw}) is also known as \emph{Enhanced Capital-Asset Pricing Model} (ECAPM) \citep{squartini2017stock}.

\chapter{Model selection: a quick look at AIC and BIC}
\label{appc}

In a context such that of network reconstruction, a tool is needed to find out the model best fitting a given data set. A possible criterion is based on the concept of \textit{information} and rests upon Fisher's idea that the best model is characterized by the smallest amount of information loss. The latter is computed via the Kullback-Leibler (KL) divergence, accounting for the information lost when the reality, $f$, is approximated by a model, $g$:

\begin{equation}\label{inf}
I(f,g)=\int f(x)\ln\left(\frac{f(x)}{g(x|\vec{\theta})}\right)dx
\end{equation}
(here for simplicity we consider just the unidimensional case). Hence, finding the best model means finding the model $g$ minimizing $I(f,g)$. Information criteria are nothing else that estimations of the K-L information loss. Akaike \citep{akaike1974new} found an estimator of $I(f,g)$ reading

\begin{equation}\label{AIC}
\text{AIC}=-2\mathcal{L}(\hat{\theta}|\text{data})+2K
\end{equation}
where $\mathcal{L}$ is the maximum of the log-likelihood function of model $g$ (notice that the vector of parameters has been estimated through the available data) and $K$ is the number of parameters defining the model itself (a quantity introduced to prevent overfitting). For two, or more, competing models, the optimal choice is the one minimizing AIC\footnote{When the sample size is small, compared to the number of parameters defining a model, a corrected version of AIC must be considered, reading

\begin{equation}\label{AIC2}
\text{AICc}=-2\mathcal{L}(\hat{\theta}|\text{data})+2K+\frac{2K(K+1)}{n-K-1}
\end{equation}
where $n$ is the sample size.}. Based on the same idea, the Bayesian Information Criterion (BIC) provides an estimation of $I(f,g)$ reading \citep{schwarz1978estimating}

\begin{equation}\label{BIC}
\text{BIC}=-2\mathcal{L}(\hat{\theta}|\text{data})+K\ln n;
\end{equation}
also in this case, the optimal model is the one minimizing BIC.

\endappendix

\bibliographystyle{cambridgeauthordate}
\bibliography{Bibliography}\label{refs}

\end{document}